# Long-wave instability of stratified two-phase MHD channel flow

Arseniy Parfenov, Alexander Gelfgat, and Neima Brauner

School of Mechanical Engineering, Faculty of Engineering, Tel-Aviv University, Tel-Aviv 6997801, Israel

## Abstract

Instability of a stratified two-phase MHD parallel flow between two infinite plates is addressed. We examine the effect of the transverse magnetic field on the base flow and long wave instability of a two-layer system consisting of conductive liquid and non-conductive gas. Both perfectly insulating and perfectly conducting boundaries are considered. To capture the behavior at small but finite wavenumbers, the conventional first-order long-wave stability analysis is extended to higher order terms. Using mercury-air system as a representative test case, the results demonstrate distinct and non-similar base flow and disturbance profiles, as well as different stability maps for insulating versus conducting boundaries. The stability diagrams reveal a non-monotonic influence of the magnetic field on flow stability, showing that, in addition to its expected stabilizing effect, the field can also induce destabilization under certain conditions. Inspection of the disturbance profiles indicates that despite the strong damping of mercury flow by the magnetic field, interaction of the two fluids at the interface and the shear-induced instabilities in the gas layer dominate and can lead to flow destabilization as the magnetic field strength increases.

## 1. Introduction

Multiphase flow systems, where one or more of the phases are electrically conducting, are encountered in various fields, including nuclear and petroleum industries, metallurgical processes, geophysics and MHD power generation (e.g., Kim et al., 2017; Wang et al., 2022; Lu et al., 2022; Saraswat et al., 2025). In many applications, an external magnetic field is applied to such flows as means of control, giving rise to a broad range of multiphase MHD flow problems addressed in the literature. Examples are two-phase liquid metal magnetohydrodynamics generators (Dune 1980, Inoue et al., 1994; Lee et al., 2022; Wang et al., 2024a,b), transport of fluids in microscale systems (Bau et al., 2003; Kabbani et al., 2008, Khan et al., 2023), and magnetic field-driven micropumps (Yi et al., 2002; Weston et al., 2010).

To control and optimize the targeted process, one needs to know how the two-phase flow pattern and its stability change with the magnetic field and other governing parameters. Because the two phases differ in density, stratified flows are frequently observed in horizontal and inclined conduits. Studies that modeled the steady two-phase stratified MHD flow are available in the literature (e.g., Shail, 1973; Lohrasbi and Sahai; 1988, Malashetty and Umavathi 1997; Nikodijević et al., 2011; Parfenov et al., 2024). The stability of electrically non-conducting two-phase flows is quite well studied (see Barmak et al., 2016; Gelfgat et al., 2020; Barmak et al., 2024 and references therein). The same can be said about stability of the single-phase Hartmann flow (see, e.g., Potter et al., 1973; Takashima, 1996; Krasnov et al., 2004 and the references therein). The Squire's transformation was used to show that two-dimensional motions are more unstable than three-dimensional ones (e.g., Hughes and Tobias, 2001; Shankar and Shivakumara, 2018). However, we found only one experimental study (He et al., 2024) that addressed stability of a two-phase air-ferrofluid slug flow. The effect of magnetic field on the stability of two-phase stratified MHD flows has not yet been explored.

It should be noted that, while in the single-phase shear flows the primary transition is not predicted by the linear stability theory, in two-phase electrically neutral flows the primary instability does take place in accordance with the linear stability predictions. This has been confirmed by the experiments (Nezihovski et al., 2022; Nezihovski et al., 2025) and attributed to the addition of the interfacial instability mechanisms to the shear induced ones. It can be expected that under the influence of a magnetic field, the onset of instability will also align with linear stability theory, though this remains to be investigated.

In this study we address the problem of stability of a two-phase stratified MHD flow in channels, assuming that the lower and heavier phase is electrically conducting (e.g., mercury), while the upper lighter phase is not conducting (e.g., air). Following previous studies of non-conducting stratified two-phase flows, we focus here on the long-wave instability of the two-phase MHD flows between two infinite plates, and extend results of (Kushnir et al., 2014) to MHD effects.

Evaluating the base flow, we noticed that electric properties of the flow boundaries produce a non-trivial effect. In single-phase Hartmann flow (Hartmann, 1937), the flow is defined either by a prescribed pressure gradient, or by a prescribed mass flux through the channel. While electrically insulating, partially conducting, or perfectly conducting channel walls influence the amplitude of the base flow, they do not alter the shape of the Hartmann velocity profile (Müller and Bühler, 2001). However, in two-phase stratified flow, the phases with different conductivities are affected unevenly, so that velocity profile shapes are qualitatively different for different electric conductivity of the channel walls. Moreover, the position of the interface for a prescribed flow rate ratio is also affected. This leads to qualitatively different base flows, for instance, when comparing perfectly insulating versus perfectly conducting lower walls. Therefore, stability analysis must be carried out separately for each boundary condition scenario.

Since the governing equations for the flow and its linear stability have been already discussed in previous studies (e.g., Potter et al., 1973; Takashima, 1996; Krasnov et al., 2004), as well as in textbooks (e.g., Müller and Bühler, 2001), we revisit them in the Supplementary materials 1 and 2, and Appendix A. Our focus is on the modification introduced by the Lorenz force, as well as the changes in the base flow and its stability characteristics resulting from the electromagnetic boundary conditions at the walls and the fluids' interface. A particularly important result is the validity of the Squire transformation, originally proven for non-conducting two-phase flow in (Barmak et al., 2017), which we extend here for the two-phase MHD flow. This allows us to restrict our analysis to two-dimensional disturbances.

Stability of non-conducting two-phase flows with respect to long wave disturbances was studied in the limit of vanishing (dimensionless) wavenumber $k \to 0$ (see, e.g. Kushnir et al., 2014 and references therein). It was shown in (Barmak et al., 2016) that numerical solution of the whole problem for finite wavenumbers allows one to reproduce the analytical long wave results of (Kushnir et al., 2014) with the choice of $k = 10^{-5}$. In this study we propose to construct an asymptotic expansion of solution of the linearized stability problem for powers of small $k$, such that the first two terms will correspond to the previously used technique, while added terms allow for an accurate analytical solutions at slightly larger wavenumbers, reaching $k = 10^{-3}$ in some of the considered cases. To the best of our knowledge, such an asymptotic expansion has not been previously considered in long wave stability analysis. We show that the analytical and numerical solutions at small wavenumbers are in good agreement and validate each other.

The results of the analysis are demonstrated for mercury and air, as lower and upper fluids, respectively. Two cases of perfectly conducting or perfectly insulating lower boundary are considered. Since air is a perfect insulator, electric properties of the upper boundary are irrelevant. The estimated value of the magnetic Prandtl number for mercury is of the order of $10^{-7}$, which allows us to neglect infinitesimally small disturbances of the induced magnetic field in the linear stability problem. The results presented include the characteristics of the base flow, stability maps and profiles of the most unstable disturbances. It is shown that already within the long wave model, the stability results are quite complicated and are noticeably different for the cases of perfectly insulating or perfectly conducting lower boundary. Furthermore, dependence of the critical superficial velocities on the magnetic field is not monotonic, so that we can observe stabilization at lower magnetic fields and destabilization at larger ones. Despite significant differences in the stability characteristics obtained for the perfectly insulating and perfectly conducting boundaries, the profiles of the streamwise velocity disturbances appear similar, thus suggesting a similarity in the underlying routes to the instability in both cases. Recalling to our previous results for instability of this configuration in the absence of the magnetic field, we attribute the observed similarities to a similar interactions between the interfacial and shear instability mechanisms, while the differences in stability characteristics are attributed to noticeably different base flow profiles.

## 2. Formulation of the problem

We consider isothermal steady parallel laminar stratified flow of two immiscible fluids between two infinite horizontal plates, with a plane and smooth interface between the phases. The flow is driven by an axial pressure gradient, which is constant and the same in both layers (e.g., Barmak et al., 2016). The fluids are affected by the imposed constant transverse magnetic field. While we are interested primarily in the configuration with heavy conducting lower phase and lighter non-conducting upper phase, the problem formulation is given for a general case. The layout of the problem is presented in Figure 1.

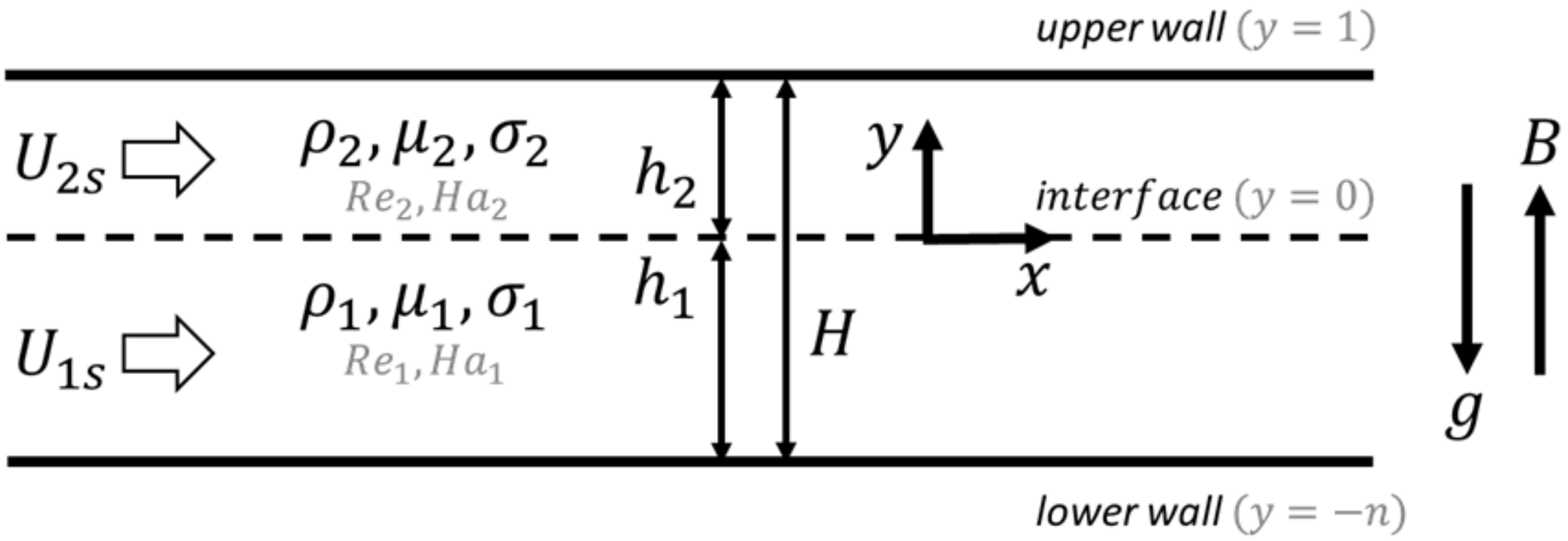

*Figure 1. Configuration of stratified two-phase flow between two parallel plates. The plates are shown by bold solid lines. The two phases are separated by an interface shown by dash line.*

We denote phases' densities, dynamic viscosities, and electric conductivities as $\rho_j$, $\mu_j$, and $\sigma_j$ respectively, where index $j = 1$ corresponds to the lower phase, and $j = 2$ to the upper one. The channel height is $H$, and the heights of the layers are $h_j$; gravitational acceleration is $g$, surface tension coefficient is $\gamma$, and the magnitude of the imposed magnetic field is $B_0$. The walls of the channel are assumed thin, with a small thickness, $t_w \ll H$, and the external space is assumed to be non-conducting. Following (Müller and Bühler, 2001), the problem is formulated for an arbitrary walls conductivity $\sigma_w$, while in the present paper we consider only two cases of zero and of infinite conductivity.

To preserve consistency with previous works on two-phase flow stability (Barmak et al., 2016; Yih, 1967; Yiantsios, 1988), the problem is rendered dimensionless using $h_2$ as the length scale, and velocity at the interface as a velocity scale $U_*$. The overall notation is similar to (Barmak et al., 2016). We introduce dimensionless parameters as listed in Table 1.

*Table 1. Dimensionless parameters*

| | |
|---|---|
| $n = h_1/h_2$ | -- layer height ratio |
| $m = \mu_1/\mu_2$ | -- viscosity ratio |
| $r = \rho_1/\rho_2$ | -- density ratio |
| $Re_j = \rho_j U_* h_2 / \mu_j$ | -- Reynolds numbers (j = 1,2) |
| $Ha_j = B_0 h_2 \sqrt{\sigma_j / \mu_j}$ | -- Hartmann numbers (j = 1,2) |
| $Re_{mj} = \mu_{mj}\sigma_j U_* h_2$ | -- magnetic Reynolds numbers (j = 1,2) |
| $Pr_{mj} = \mu_{mj}\mu_j\sigma_j/\rho_j$ | -- magnetic Prandtl numbers (j = 1,2) |
| $Fr_2 = U_*^2 / g h_2$ | -- Froude number |
| $We_2 = \rho_2 h_2 U_*^2 / \gamma$ | -- Weber number |

Note that the Hartmann numbers are defined using the length scale $h_2$, so that the governing equations do not contain $n$ and match with the previous works on the effect of magnetic field (Takashima, 1996; Lock, 1955). The dimensionless coordinate $y$ changes in the range $[-n; 1]$, where $y = -n$ and $y = 1$ correspond to the lower and upper boundaries respectively, and the interface is located at $y = 0$.

### *2.1. Base flow*

We assume the base (undisturbed) flow to be steady and plane-parallel. Under these assumptions, the reduced problem involves only axial components of the velocity and of the induced magnetic field, which depend only on $y$.The full derivation is presented in Supplementary material 1.

Following (Müller and Bühler, 2001), we represent dimensionless velocity $\boldsymbol{V}_j = U_j(y)\boldsymbol{e}_x$ and dimensionless magnetic field (scaled by the magnitude of imposed magnetic field) $\boldsymbol{B} = \boldsymbol{e}_y + B_j(y)\boldsymbol{e}_x$. The induced component of the magnetic field $B_j(y)$ is further rescaled as $B_j(y) = \frac{Re_{mj}}{Ha_j}\mathfrak{B}_j(y)$, where $Re_{mj} = \mu_{mj}\sigma_j U_* h_2$ is the magnetic Reynolds number, and $\mu_{mj}$ is the magnetic permeability. Then the equations for $U_j, \mathfrak{B}_j(y)$ become

$$\begin{gathered} \frac{\mu_j}{\mu_2}\left(U_j'' + Ha_j\mathfrak{B}_j'\right) = -G = const, \\ \mathfrak{B}_j'' + Ha_j U_j' = 0; \end{gathered} \quad \begin{cases} y \in [-n; 0], & j = 1 \\ y \in [0; 1], & j = 2 \end{cases} \tag{2.1}$$

with no-slip boundary conditions

$$U_1 \underset{y=-n}{=} 0, \qquad U_2 \underset{y=1}{=} 0, \tag{2.2}$$

continuity of the velocity and tangent stresses at the interface

$$U_1 \underset{y=0}{=} U_2 \underset{y=0}{=} 1, \tag{2.3}$$

$$m\, U_1' \underset{y=0}{=} U_2', \tag{2.4}$$

the conditions on the interface for $\mathfrak{B}_j$ are (note that the coefficients here appear due to rescaling; also note that $mHa_1^2/Ha_2^2 = \sigma_1/\sigma_2$):

$$mHa_1\mathfrak{B}_1 \underset{y=0}{=} Ha_2\mathfrak{B}_2, \tag{2.5}$$

$$\mathfrak{B}_1'/Ha_1 \underset{y=0}{=} \mathfrak{B}_2'/Ha_2; \tag{2.6}$$

boundary conditions at the walls for $\mathfrak{B}_j$ are obtained similar to (Kushnir et al., 2014):

$$C_1^w\mathfrak{B}_1' - \mathfrak{B}_1 \underset{y=-n}{=} 0, \qquad C_2^w\mathfrak{B}_2' + \mathfrak{B}_2 \underset{y=1}{=} 0, \tag{2.7}$$

where $C_j^w = \frac{t_w \sigma_w}{h_2 \sigma_j}$. In particular, $C_j^w = 0$ for non-conducting walls, and $C_j^w = \infty$ for perfectly conducting walls. The constant $G$ in (2.1) represents the dimensionless pressure gradient:

$$G = -Re_2 \frac{\partial P_j}{\partial x} = const. \tag{2.8}$$

In the equations above, we assumed $Ha_j > 0$, and the cases of $Ha_j = 0$ can be obtained by taking a limit $Ha_j \to 0$. For instance, if $\sigma_1 > 0,\ \sigma_2 = 0$, then $\mathfrak{B}_2 = 0$ and $\mathfrak{B}_1(0) = 0$. The general solution, as well as the expression for the flow rates in the case of one conducting phase, can be found in the Supplementary material 2. Note that the solution for the base flow is valid for all $Re_{mj}$, while the linear stability problem is formulated below only for small magnetic Reynolds numbers. Note also, that the relation between *n* and the flow rate ratio is affected by the conductivity of the channel walls, which is further discussed in the Results section.

## *2.2. Linear stability problem*

The linear stability problem is reduced to an eigenvalue problem defined by the Orr-Sommefeld equation and is similar to the one considered in the previous works (e.g., Barmak et al., 2016), with the addition of the electromagnetic Lorenz force. The linear problem is derived under assumption that the magnetic Reynolds number $Re_m$ is small compared to $Re$, as it was assumed also in (Lock, 1955; Chandrasekhar, 1953; Stuart, 1954) in linear stability analysis for single phase flow of the conducting layer. This assumption is equivalent to assuming small magnetic Prandtl numbers, since $Pr_{mj} = Re_{mj}/Re_j$. The latter assumption is often preferred because the magnetic Prandtl number is a material property (Table 1) and is independent of the characteristic velocity. The resulting problem, evaluated for small $Re_m$, allows for the Squire's transformation (see Appendix A), so that only two-dimensional disturbances of horizontal velocity $u_j$, vertical velocity $v_j$, pressure $p_j$, and the interface $\eta$ have to be considered. The disturbances are expressed in the form

$$\left(\phi_j(x,y), p_j(x,y), \eta(x)\right) = \left(\phi_j(y), p_j(y), \eta\right) exp[\, ik(x-ct)], \tag{2.9}$$

where $k$ is a wave number, $c$ is a phase velocity and $\phi$ is a stream function:

$$u_j = \frac{\partial \phi_j}{\partial y} = \phi_j', \qquad v_j = -\frac{\partial \phi_j}{\partial x} = -ik\, \phi_j. \tag{2.10}$$

The problem for $\phi_j$, $\eta$ and $c$ is derived from the linearized Navier-Stokes equations (see Appendix A). It consists of Orr-Sommerfeld equations for both phases

$$\phi_j'''' - 2k^2 \phi_j'' + k^4 \phi_j - Ha_j^2 \phi_j'' = ik\, Re_j\left[\left(U_j - c\right)\left(\phi_j'' - k^2\phi_j\right) - \phi_j U_j''\right]; \begin{cases} y \in [-n; 0], & j = 1 \\ y \in [0; 1], & j = 2 \end{cases}, \tag{2.11}$$

no-slip boundary conditions

$$\phi_1(-n) = \phi_1'(-n) = 0, \qquad \phi_2(1) = \phi_2'(1) = 0, \tag{2.12}$$

continuity of the vertical velocity and the kinematic condition

$$\phi_1 \underset{y=0}{=} \phi_2 \underset{y=0}{=} \left(c - U_j\right)\eta, \tag{2.13}$$

continuity of the axial velocity

$$\phi_1' + U_1'\eta \underset{y=0}{=} \phi_2' + U_2'\eta\,, \tag{2.14}$$

continuity of the shear stress

$$m[\phi_1'' + U_1''\eta + k^2\phi_1] \underset{y=0}{=} \phi_2'' + U_2''\eta + k^2\phi_2, \tag{2.15}$$

and the jump of normal stress due to non-zero deformation of the interface:

$$\begin{aligned} & m\,(\phi_1''' - 3k^2\phi_1' - Ha_1^2\phi_1') - (\phi_2''' - 3k^2\phi_2' - Ha_2^2\phi_2') + \\ & + ik\,Re_2(c - U_1)[\,r(\phi_1' + U_1'\eta) - (\phi_2' + U_2'\eta)] \underset{y=0}{=} ik\,Re_2((r-1)Fr_2^{-1} + k^2\,We_2^{-1})\eta. \end{aligned} \tag{2.16}$$

Note that the Lorenz force produces additional terms with $Ha_j$ in eqs. (2.11) and (2.16), which also affects the problem indirectly by altering the base flow $U_j$. It is noticed also, that the walls conductivity affects the base flow, so that the stability results for perfectly conducting and perfectly insulated walls are noticeably different.

## 3. Long-wave asymptotic solution

While the general problem can be solved numerically as, e.g., in (Barmak et al., 2016), an asymptotic limit of small wavenumber $k$ allows for an analytical solution. The asymptotic solution procedure is similar to that presented in (Kushnir et al., 2014) for non-conducting fluids. The procedure outlined below is extended for the present magnetohydrodynamic formulation, as well as for a larger number of terms in asymptotic series. We start with introducing several amendments in the formulation of the linear stability problem, and then present the solution in the form of asymptotic expansion. For convenience, we introduce the following coefficients:

$$K_F \overset{\text{def}}{=} \frac{g\,\rho_2^2\,h_2^3}{\mu_2^2} \;\Rightarrow\; Fr_2^{-1} = K_F\,Re_2^{-2}, \tag{3.1}$$

$$K_W \overset{\text{def}}{=} \frac{\gamma\,\rho_2 h_2}{\mu_2^2} \Rightarrow\; We_2^{-1} = K_W Re_2^{-2}, \tag{3.2}$$

$$\vartheta_j \overset{\text{def}}{=} \frac{\rho_j\mu_2}{\rho_2\mu_j} \Rightarrow\; Re_j = \vartheta_j Re_2. \tag{3.3}$$

So that only $Re_2$ explicitly depends on the velocity scale $U_*$, which, as noted above, is taken as the velocity at the interface. Thus,

$$U_1(0) = U_2(0) = 1. \tag{3.4}$$

Note that for another velocity scale $\widetilde{U}_* = \alpha\,U_*$, values $Re_2$, $U$, and $c$ are rescaled as $\widetilde{Re}_2 = \alpha Re_2$, $\widetilde{U} = \frac{1}{\alpha}\,U$ and $\tilde{c} = \frac{1}{\alpha}c$.

As the interface is disturbed, its amplitude is non-zero, $\eta \neq 0$. However, this amplitude, as well as the amplitudes of other disturbances, cannot be resolved within the linear stability analysis. We normalize the eigenvector $(\phi_1, \phi_2, \eta)$ so that the amplitude of the disturbed interface is $\eta = 1$. This turns the eigenvalue problem into a system of equations, which is linear with respect to the stream function $\phi_j$.

For simplicity, we rewrite several terms in boundary conditions at the interface ($y = 0$) using the following substitutions:

In (2.14), using (2.4): $U_2' - U_1' = (m-1)U_1'$;

In (2.15), using (2.13, 2.3): $m\phi_1 - \phi_2 = (m-1)(c-1)$;

In (2.16), using (2.4, 2.14): $m\phi_1' - \phi_2' = (m-1)(\phi_1' + U_1')$;

In (2.16), using (2.14): $r(\phi_1' + U_1') - (\phi_2' + U_2') = (r-1)(\phi_1' + U_1')$.

Where appropriate, we collect all the terms that vanish when $k \to 0$. The above stability problem can then be rewritten as

$$\begin{aligned}\phi_j'''' - Ha_j^2\phi_j'' = (ik)Re_2\vartheta_j\big[\big(U_j - c\big)\big(\phi_j'' + (ik)^2\phi_j\big) - \phi_j U_j''\big] - \\ - (ik)^2\big[\, 2\,\phi_j'' + (ik)^2\phi_j\big] \overset{\text{def}}{=} F_j(y);\end{aligned} \qquad \begin{cases} y \in [-n; 0], & j = 1, \\ y \in [0; 1], & j = 2, \end{cases} \tag{3.5}$$

$$\phi_1(-n) = \phi_1'(-n) = 0, \;\; \phi_2(1) = \phi_2'(1) = 0, \tag{3.6}$$

$$\phi_1 - \phi_2 \underset{y=0}{=} 0, \tag{3.7}$$

$$\phi_1' - \phi_2' \underset{y=0}{=} (m-1)U_1' \overset{\text{def}}{=} G^v, \tag{3.8}$$

$$m\phi_1'' - \phi_2'' \underset{y=0}{=} (U_2'' - m\,U_1'') + (ik)^2\,(m-1)(c-1) \overset{\text{def}}{=} G^s, \tag{3.9}$$

$$\begin{aligned} m\,(\phi_1''' - Ha_1^2\,\phi_1') - (\phi_2''' - Ha_2^2\,\phi_2') \underset{y=0}{=} - 3(ik)^2(m-1)(\phi_1' + U_1') - \\ - (ik)Re_2\,(r-1)(c-1)(\phi_1' + U_1') + (ik)Re_2^{-1}\left((r-1)K_F - (ik)^2\,K_W\right) \overset{\text{def}}{=} G^n. \end{aligned} \tag{3.10}$$

The kinematic condition (2.13) becomes:

$$c = 1 + \phi_1(0). \tag{3.11}$$

Here we denoted the right-hand side of equations (3.5, 3.8-3.10) as $F_j(y; c, \phi_j)$, $G^v$, $G^s(c)$ and $G^n(c, \phi_1)$ respectively.

## *3.1. Basis functions*

To modify the problem further, we introduce functions $\xi_j$ as the solutions of the following homogeneous 4th-order ODEs

$$\xi_j'''' - Ha_j^2\,\xi_j'' = 0, \tag{3.12}$$

which satisfy following conditions at the interface

$$\xi_j(0) = \xi_j'(0) = \xi_j''(0) = 0, \;\; \xi_j'''(0) = 1. \tag{3.13}$$

These functions possess the following properties:

The functions $\xi_j$, $\xi_j'$, $\xi_j''$ and $\xi_j'''$ are linearly independent solutions of (3.12), and also

$$\begin{gathered}\left(\frac{d^2}{dy^2} - Ha_i^2\right)\xi_i = y, \\ \xi_i(0) = \xi_i'(0) = \xi_i''(0) = 0, \;\; \xi_i'''(0) = 1, \\ \xi_i^{(4)}(0) = 0, \;\; \xi_i^{(5)}(0) = Ha_i^2, \quad \xi_i^{(6)}(0) = 0. \end{gathered} \tag{3.14}$$

The base flow $U_j$ can be represented using $\xi_j$ (see Supplementary material 2):

$$U_j = \xi_j''' + C_j^u\xi_j'' + B_j^u\xi_j'. \tag{3.15}$$

Defining $I_j(y)$ as:

$$I_j(y) \overset{\text{def}}{=} \int_0^y \xi_j(y - s)\cdot F_j(s)ds\,, \tag{3.17}$$

it satisfies

$$\left(\frac{d^4}{dy^4} - Ha_j^2\frac{d^2}{dy^2}\right)I_j(y) = F_j(y),$$
$$I_j(0) = I_j'(0) = I_j''(0) = I_j'''(0) = 0. \tag{3.18}$$

It is also convenient to introduce another solution of the nonhomogeneous equation as:

$$f_j(y) = A_j^f\xi_j + B_j^f\xi_j' + I_j(y), \tag{3.18a}$$

which satisfies

$$f_j'''' - Ha_j^2 f_j'' = F_j(y),$$
$$f_j(0) = f_j'(0) = 0, \qquad f_1(-n) = f_1'(-n) = 0, \qquad f_2(1) = f_2'(1) = 0; \tag{3.18b}$$

Hence, the coefficients $A_j^f$, $B_j^f$ are found from

$$\begin{cases} A_1^f\xi_1(-n) + B_1^f\xi_1'(-n) + I_1(-n) = 0, \\ A_1^f\xi_1'(-n) + B_1^f\xi_1''(-n) + I_1'(-n) = 0; \end{cases} \begin{cases} A_2^f\xi_2(1) + B_2^f\xi_2'(1) + I_2(1) = 0, \\ A_2^f\xi_2'(1) + B_2^f\xi_2''(1) + I_2'(1) = 0. \end{cases} \tag{3.18c}$$

The explicit expressions for $\xi_j$ are:

$$\begin{cases} Ha_j = 0: & \xi_j(y) = y^3/6; \\ Ha_j > 0: & \xi_j(y) = \left(\sinh(Ha_j y) - Ha_j y\right)/Ha_j^3. \end{cases} \tag{3.19}$$

Considering (3.14) to (3.19), the linear stability problem can be presented as an integral equation

$$\phi_j = A_j\xi_j + B_j\xi_j' + C_j\xi_j'' + D_j\xi_j''' + f_j(y); \tag{3.20}$$

with the following boundary conditions, at the channel walls:

$$\begin{cases} \xi_1(-n)A_1 + \xi_1'(-n)B_1 + \xi_1''(-n)C_1 + \xi_1'''(-n)D_1 = 0, \\ \xi_1'(-n)A_1 + \xi_1''(-n)B_1 + \xi_1'''(-n)C_1 + Ha_1^2\xi_1''(-n)D_1 = 0, \\ \xi_2(1)A_2 + \xi_2'(1)B_2 + \xi_2''(1)C_2 + \xi_2'''(1)D_2 = 0, \\ \xi_2'(1)A_2 + \xi_2''(1)B_2 + \xi_2'''(1)C_2 + Ha_2^2\xi_2''(1)D_2 = 0; \end{cases} \tag{3.21}$$

and at the interface:

$$\begin{gathered} D_1 - D_2 = 0, \\ C_1 - C_2 = G^v, \\ m(B_1 + Ha_1^2 D_1) - (B_2 + Ha_2^2 D_2) = G^s - (mB_1^f - B_2^f), \\ mA_1 - A_2 = G^n - \left(mA_1^f - A_2^f\right). \end{gathered} \tag{3.22}$$

Also, from the kinematic condition (3.11):

$$c = 1 + D_1. \tag{3.23}$$

The above system of equations (3.21-3.22) for $A_j, B_j, C_j, D_j$,can be presented in a matrix form

$$[\![\mathrm{S}]\!]\boldsymbol{a} = \boldsymbol{b}, \tag{3.24}$$

where

$$\boldsymbol{a} = (A_1, B_1, C_1, D_1, A_2, B_2, C_2, D_2)^T, \tag{3.25}$$

$$\boldsymbol{b} = \left(0,0,0,0,0, G^v, G^s - (mB_1^f - B_2^f), G^n - (mA_1^f - A_2^f)\right)^T, \tag{3.26}$$

and

$$
[\![\mathrm{S}]\!] = \begin{pmatrix}
\xi_1(-n) & \xi_1'(-n) & \xi_1''(-n) & \xi_1'''(-n) & & & & \\
\xi_1'(-n) & \xi_1''(-n) & \xi_1'''(-n) & Ha_1^2\xi_1''(-n) & & & & \\
 & & & & \xi_2(1) & \xi_2'(1) & \xi_2''(1) & \xi_2'''(1) \\
 & & & & \xi_2'(1) & \xi_2''(1) & \xi_2'''(1) & Ha_2^2\xi_2''(1) \\
 & & & 1 & & & & -1 \\
 & & 1 & & & & -1 & \\
 & m & & mHa_1^2 & & -1 & & -Ha_2^2 \\
m & & & & -1 & & &
\end{pmatrix}; \quad (3.27)
$$

Denoting

$$
\boldsymbol{a}^v = [\![\mathrm{S}]\!]^{-1}\boldsymbol{e}_6, \quad \boldsymbol{a}^s = [\![\mathrm{S}]\!]^{-1}\boldsymbol{e}_7, \quad \boldsymbol{a}^n = [\![\mathrm{S}]\!]^{-1}\boldsymbol{e}_8. \tag{3.28a}
$$

where $\boldsymbol{e}_j$ are the unit vectors in $R^6$ (i.e. $[\![\mathrm{S}]\!]^{-1}\boldsymbol{e}_j$ is the $j$-th column of the matrix $[\![\mathrm{S}]\!]^{-1}$), and noting that $[\![\mathrm{S}]\!]$, as well as $\boldsymbol{a}^v$, $\boldsymbol{a}^s$, $\boldsymbol{a}^n$, depend only on $n, Ha_1, Ha_2$, the coefficients in (3.25) can be presented as

$$
\boldsymbol{a} = G^v\boldsymbol{a}^v + \left(G^s + B_2^f - mB_1^f\right)\boldsymbol{a}^s + \left(G^n + A_2^f - mA_1^f\right)\boldsymbol{a}^n. \tag{3.28b}
$$

If $F_j$, $G^v$, $G^s$, $G^n$ in eqs. (3.8-3.10) are known (e.g., for $k = 0$), eq. (3.20) would become the explicit expression for $\phi_j$.

## 3.2. Asymptotic solution

Considering $k \to 0$, assume:

$$
c = c_0 + (ik)c_1 + (ik)^2c_2 + \dots, \tag{3.29}
$$

$$
\phi_j = \phi_{j;0} + (ik)\phi_{j;1} + (ik)^2\phi_{j;2} + \dots . \tag{3.30}
$$

Note, that here we use powers of $(ik)$ instead of $k$ to define $c_m$, $\phi_{j;m}$.

### 3.2.1. Zero order terms

Substituting expansions (3.29-3.30) into the problem (3.5-3.10) and collecting all terms of the zero power of $(ik)$, results in the following equations and boundary conditions

$$
\phi_{j;0}'''' - Ha_j^2\phi_{j;0}'' = 0, \tag{3.31}
$$

$$
\phi_{1;0}(-n) = \phi_{1;0}'(-n) = 0, \quad \phi_{2;0}(1) = \phi_{2;0}'(1) = 0, \tag{3.32}
$$

$$
\begin{gathered}
\phi_{1;0} - \phi_{2;0} \underset{y=0}{=} 0, \\
\phi_{1;0}' - \phi_{2;0}' \underset{y=0}{=} (m-1)U_1', \\
m\phi_{1;0}'' - \phi_{2;0}'' \underset{y=0}{=} U_2'' - m\,U_1'', \\
m\left(\phi_{1;0}''' - Ha_1^2\phi_{1;0}'\right) - \left(\phi_{2;0}''' - Ha_2^2\phi_{2;0}'\right) \underset{y=0}{=} 0,
\end{gathered} \tag{3.33}
$$

$$
c_0 = 1 + \phi_{1;0}(0). \tag{3.34}
$$

Since eq. (3.31) is homogeneous, we can represent $\phi_{j;0}$ as

$$
\phi_{j;0} = A_{j;0}\xi_j + B_{j;0}\xi_j' + C_{j;0}\xi_j'' + D_{j;0}\xi_i''', \tag{3.35}
$$

so that the solution is found as:

$$\left( A_{1;0}, B_{1;0}, C_{1;0}, D_{1;0}, A_{2;0}, B_{2;0}, C_{2;0}, D_{2;0}\right)^T = \boldsymbol{a}_0,$$
$$\boldsymbol{a}_0 = (m-1)U_1'(0)\boldsymbol{a}^v + (U_2''(0) - m\,U_1''(0))\boldsymbol{a}^s, \tag{3.36}$$
$$c_0 = 1 + D_{1;0}.$$

In the absence of magnetic field $U_2'' = mU_1''$, whereby (3.36) yields:

$$c_{0|\{Ha_j=0\}} = 1 + \frac{2mn^2(n+1)(m-1)U_1'(0)}{n^4 + 4n^3m + 6n^2m + 4nm + m^2}. \tag{3.37a}$$

By substituting $U_1'$, an expression presented in (Kushnir et al., 2014) is obtained:

$$c_{0|\{Ha_j=0\}} = 1 + \frac{2n(m-1)(m-n^2)}{n^4 + 4n^3m + 6n^2m + 4nm + m^2}. \tag{3.37b}$$

In the case where only the lower phase is conductive ($Ha_1 \coloneqq Ha > 0, Ha_2 = 0$), we have:

$$c_{0|1ph} = 1 + \frac{A^{1ph}\cdot\left(U_2''(0) - m\,U_1''(0)\right) + B^{1ph}\cdot(m-1)U_1'(0)}{D^{1ph}}, \tag{3.38a}$$

where

$$A^{1ph} = (mHa^2 + 12)(\cosh(Ha\,n) - 1) - 6Ha\,n\sinh(Ha\,n), \tag{3.38b}$$

$$B^{1ph} = mHa\,[(6n+4)Ha\,(\cosh(Ha\,n) - 1) - 6(\sinh(Ha\,n) - Ha\,n)], \tag{3.38c}$$

$$D^{1ph} = (12(m-1)Ha^2 - mHa^4 - 24)(\cosh(Ha\,n) - 1) + $$
$$+ mHa^2(mHa^2 + 12n)\cosh(Ha\,n) + \tag{3.38d}$$
$$+\left(Ha^2(6n+4m) - 12(m-n)\right)Ha\sinh(Ha\,n);$$

For the case of perfectly insulating lower wall:

$$U_1'(0)_{\{ins\}} = \frac{mHa\sinh(Ha\,n) - 2n(\cosh(Ha\,n) - 1)}{2m(n+1)\,(\cosh(Ha\,n) - 1)}, \tag{3.39a}$$

$$\left(U_2''(0) - mU_1''(0)\right)_{\{ins\}} = \frac{2Ha(n-m)\sinh(Ha\,n) - (mHa^2+4)\,(\cosh(Ha\,n) - 1) + 2nm\,Ha^2}{2(n+1)(\cosh(Ha\,n) - 1)}; \tag{3.39b}$$

and for perfectly conductive lower wall:

$$U_1'(0)_{\{cnd\}} = \frac{mHa^2\cosh(Ha\,n) - 2(\cosh(Ha\,n) - 1)}{m\left(Ha\sinh(Ha\,n) + 2(\cosh(Ha\,n) - 1)\right)}, \tag{3.39c}$$

$$\left(U_2''(0) - mU_1''(0)\right)_{\{cnd\}} = -mHa^2. \tag{3.39d}$$

*3.2.2. First order terms*

Considering now terms with the first power of $(ik)$, results in a problem for $\phi_{j;1}$ and $c_1$. The Orr-Sommerfeld equations read:

$$\phi_{j;1}'''' - Ha_j^2\phi_{j;1}'' = Re_2\vartheta_j\left[\left(U_j - c_0\right)\phi_{j;0}'' - \phi_{j;0}U_j''\right], \tag{3.40}$$

The no-slip boundary conditions at the walls are:

$$\phi_{1;1}(-n) = \phi_{1;1}'(-n) = 0, \qquad \phi_{2;1}(1) = \phi_{2;1}'(1) = 0, \tag{3.41}$$

and the conditions at the interface are:

$$\phi_{1;1} - \phi_{2;1} \underset{y=0}{=} 0,$$
$$\phi'_{1;1} - \phi'_{2;1} \underset{y=0}{=} 0,$$
$$m\phi''_{1;1} - \phi''_{2;1} \underset{y=0}{=} 0, \tag{3.42}$$
$$m(\phi'''_{1;1} - Ha_1^2\phi'_{1;1}) - (\phi'''_{2;1} - Ha_2^2\phi'_{2;1}) \underset{y=0}{=} - Re_2(r-1)(c_0-1)(\phi'_{1;0} + U'_1) + Re_2^{-1}(r-1)K_F .$$

Since the right-hand side of the above problem consists of two parts proportional to $Re_2$ and $Re_2^{-1}$, respectively, the solution is represented as

$$\phi_{j;1} = Re_2^{-1}\phi_{j;1;-1} + Re_2\phi_{j;1;1},$$
$$c_1 = Re_2^{-1}c_{1;-1} + Re_2 c_{1;1}. \tag{3.43}$$

The solution for $\phi_{j;1,-1}$ is written as

$$\phi_{j;1,-1} = A_{j;1,-1}\xi_j + B_{j;1,-1}\xi'_j + C_{j;1,-1}\xi''_j + D_{j;1,-1}\xi'''_j, \tag{3.44}$$

where the coefficients are to be found from the set of B.Cs, whereby:

$$\left(A_{1;1,-1}, B_{1;1,-1}, C_{1;1,-1}, D_{1;1,-1}, A_{2;1,-1}, B_{2;1,-1}, C_{2;1,-1}, D_{2;1,-1}\right)^T = \boldsymbol{a}_{1,-1},$$
$$\boldsymbol{a}_{1,-1} = (r-1)K_F\boldsymbol{a}^n, \tag{3.45}$$
$$c_{1,-1} = D_{1;1,-1}.$$

Note that this part of the solution is proportional to $K_F$, defined in (3.1), and vanishes in the zero-gravity case. As suggested in (Kushnir et al., 2014), to compute an approximate solution for a given $k$, the term $(r-1)K_F$ can be replaced with $[(r-1)K_F + k^2 K_W]$, which will allow one to account for the surface tension effect without considering the higher order terms of $c_2$ and $c_3$.

The solution for $\phi_{j;1,1}$ in the r.h.s. of (3.43) is represented as:

$$\phi_{j;1,1} = A_{j;1,1}\xi_j + B_{j;1,1}\xi'_j + C_{j;1,1}\xi''_j + D_{j;1,1}\xi'''_j + f_{j;1,1}(y), \tag{3.46}$$

where

$$f_{j;1,1}(y) = A^f_{j;1,1}\xi_j + B^f_{j;1,1}\xi'_j + I_{j;1,1}(y),$$
$$I_{j;1,1}(y) = \int_0^y \xi_j(y-s)\cdot F_{j;1,1}(s)ds, \tag{3.47}$$
$$F_{j;1,1}(y) = \vartheta_j\left[(U_j - c_0)\phi''_{j;0} - \phi_{j;0}U''_j\right].$$

The coefficients $A^f_{j;1,1}$ and $B^f_{j;1,1}$ are found from

$$\begin{cases} A^f_{1;1,1}\xi_1(-n) + B^f_{1;1,1}\xi'_1(-n) + I_{1;1,1}(-n) = 0, \\ A^f_{1;1,1}\xi'_1(-n) + B^f_{1;1,1}\xi''_1(-n) + I'_{1;1,1}(-n) = 0; \end{cases} \begin{cases} A^f_{2;1,1}\xi_2(1) + B^f_{2;1,1}\xi'_2(1) + I_{2;1,1}(1) = 0, \\ A^f_{2;1,1}\xi'_2(1) + B^f_{2;1,1}\xi''_2(1) + I'_{2;1,1}(1) = 0; \end{cases} \tag{3.48}$$

and

$$\left(A_{1;1,1}, B_{1;1,1}, C_{1;1,1}, D_{1;1,1}, A_{2;1,1}, B_{2;1,1}, C_{2;1,1}, D_{2;1,1}\right)^T = \boldsymbol{a}_{1,1},$$
$$\boldsymbol{a}^f_{1;1} = \left(mB^f_{1;1,1} - B^f_{2;1,1}\right)\boldsymbol{a}^s + \left(mA^f_{1;1,1} - A^f_{2;1,1}\right)\boldsymbol{a}^n,$$
$$\boldsymbol{a}_{1;1} = -(r-1)D_{1;0}\left(C_{1;0} + C^u_1\right)\boldsymbol{a}^n - \boldsymbol{a}^f_{1;1}, \tag{3.49}$$
$$c_{1,1} = D_{1;1,1}.$$

Since $Im(c) = kc_1 - k^3c_3 + \cdots$, the term $c_1$ determines the long-wave stability when $k \to 0$. The other terms can be used to calculate asymptotic $c$ for a given value of $k > 0$. The corresponding procedure is described in Appendix B. Results obtained by including terms with higher order of $k$ are presented in the Results section.

# 4. Results and discussion

Considering the number of governing dimensionless parameters involved ($n, m, r, Re_2, Fr_2, Ha_j$, see Table 1), to enable a meaningful parametric study, we perform all the calculations for two well defined fluids, mercury and air, in a channel of fixed height, $H = 0.02$m. Although all the evaluations and computations are carried out for dimensionless equations, we report the dimensional results. In our opinion, this allows for a better understanding of the results, as well as the ranges of parameters considered. Specifically, the range of magnetic field strength considered is 0 to 0.2T, which for mercury corresponds to $Ha = B_0\, H \sqrt{\sigma_1 / \mu_1}$ values up to 103 in a channel of 0.02 m. The magnetic Prandtl number, $Pr_{m1} = Re_{m1}/Re_1 = \mu_{m1}\nu\sigma,\ \ \mu_{m1} \approx \mu_o = 1.256 \times 10^{-6}$) is $1.383 \times 10^{-7}$. The position of the interface is represented by the heavier (conducting) fluid holdup, $h = h_1/H = n/(n+1)$, which is the fraction of the channel cross-section occupied by mercury.

*Table 2. Problem parameters*

| | | |
|---|---|---|
| Channel height | | $\mathrm{H} = 0.02$ m |
| Surface tension | | $\gamma = 0.45$ N/m |
| Gravitational acceleration | | $\mathrm{g} = 9.81\ \mathrm{m/s^2}$ |
| Mercury | density | $\rho_1 = 1.35 \cdot 10^4\ \mathrm{kg/m^3}$ |
| | dynamic viscosity | $\mu_1 = 1.49 \cdot 10^{-3}$ Pa s |
| | electric conductivity | $\sigma_1 = 10^6\ 1/\,\Omega \cdot \mathrm{m}$ |
| Air | density | $\rho_2 = 1\ \mathrm{kg/m^3}$ |
| | dynamic viscosity | $\mu_2 = 1.8 \cdot 10^{-5}$ Pa s |
| | electric conductivity | $\sigma_2 = 0$ |

| Dimensionless parameters | |
|---|---|
| $m = 82.7777$ | $r = 13500$ |
| $Re_1 = 181208 \times (1-h)U_*$ | $Re_2 = 1111.11 \times (1-h)U_*$ |
| $Re_{m1} = 0.02514 \times (1-h)U_*$ | $Re_{2m} = 0$ |
| $Pr_{m1} = 1.38735 \cdot 10^{-7}$ | $Pr_{m2} = 0$ |
| $Ha_1 = 518.1278 \times (1-\mathrm{h})B_0$ | $Ha_2 = 0$ |
| $Fr_2 = 5.0968 \times U_*^2/(1-h)$ | $K_F = 2.4222 \cdot 10^5 \times (1-h)^3$ |
| $We_2 = 0.0444 \times (1-h)U_*^2$ | $K_W = 2.7777 \cdot 10^7 \times (1-h)$ |

## *4.1. Base flow*

In case of a single-phase MHD flow, it is known (see, e.g., Müller and Bühler, 2001) that the conductivity of the walls only affects the amplitude of the base flow, i.e. the relation between flux and pressure gradient. However, the shape of the velocity profile, known as the Hartmann profile, remains the same. Therefore, by using the maximal (or mean velocity) as the velocity scale, the same dimensionless results are obtained, irrespective of the walls conductivity.

In the two-phase flow case, the holdup and the pressure gradient are defined by the prescribed ratio of the volumetric (dimensional) flow rates of the specified fluids, $\widetilde{Q}_1$, and $\widetilde{Q}_2$. However, the conductivity of the walls affects the holdup (i.e., the interface location), and consequently, the shape of the base flow velocity profile depends on the walls conductivity. Obviously, when only the lower phase is conducting (e.g., mercury-air flow), only the conductivity of the lower wall affects the flow configuration. This is illustrated in Figure 2 (a,b), which shows the dependence of the holdup on the ratio $Q_1/Q_2$ (=$\widetilde{Q}_1/\widetilde{Q}_2$) and the external magnetic field strength for two limit cases of the lower wall conductivity, i.e., perfectly insulating (a) and perfectly conducting (b) lower wall. As shown, the dependence of the holdup on the ratio $Q_1/Q_2$ and magnetic field strength is different in these two cases. Frames (c,d) additionally show ratios of layer thicknesses (i.e. $h_1$ and $h_2$), yielding an additional comparison of the conductive and insulating wall cases. Both $h_{cnd}/h_{ins}$ and $(1-h_{cnd})/(1-h_{ins})$ tend to unity for small and large values of $Q_1/Q_2$.

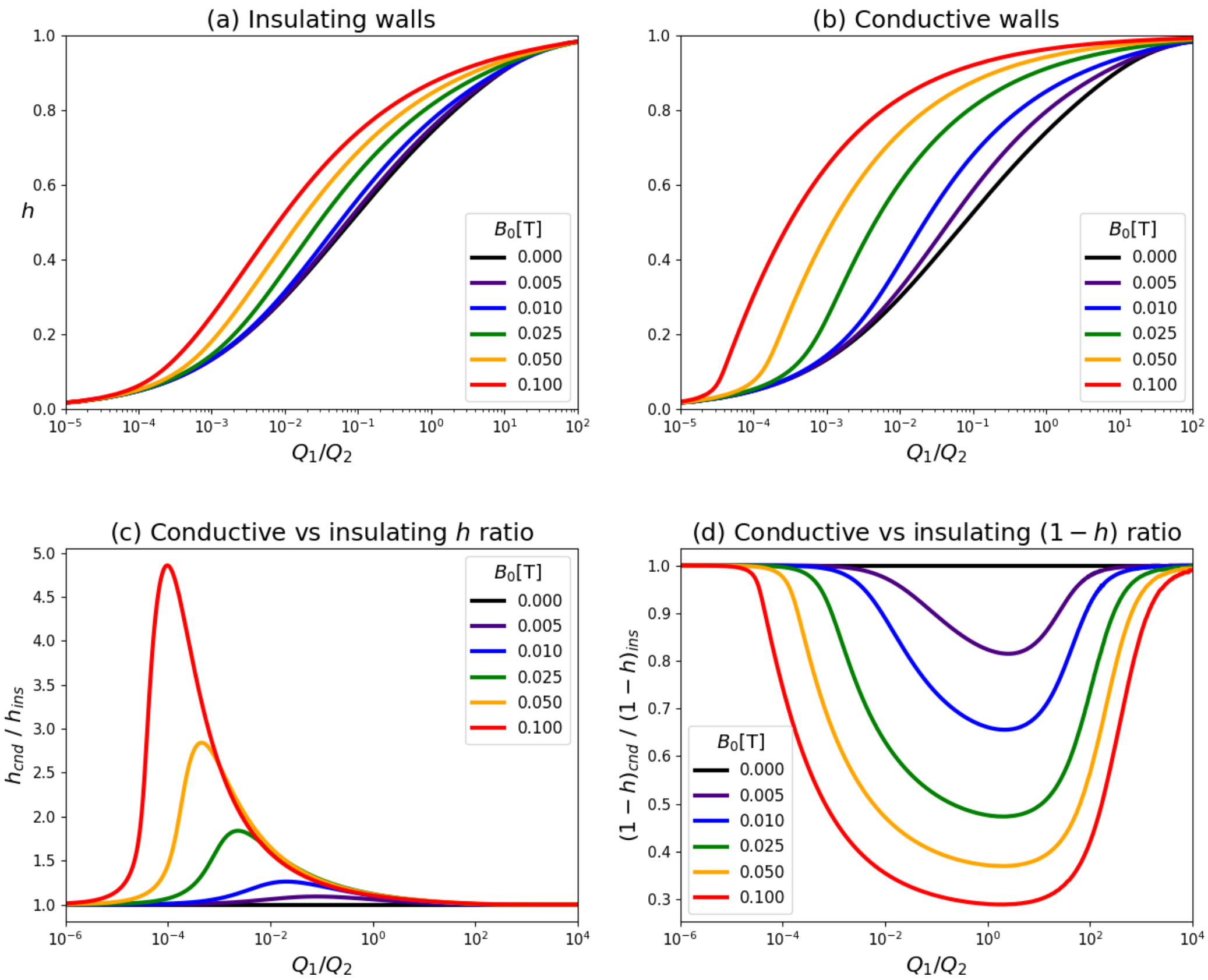

*Figure 2. (a,b): Holdup h versus volumetric flux ratio $Q_1/Q_2$ for base flow with insulating (a) and with perfectly conducting (b) lower wall. (c,d): ratios of layers' thicknesses [mercury-air]*

The influence of wall conductivity and magnetic field strength on the holdup and velocity profiles is illustrated in Figure 3(a,b) for cases with insulating and perfectly conducting bottom walls ($Q_1/Q_2 = 1$). The corresponding induced magnetic field distributions, scaled by $B_0$ and the magnetic Reynolds number, are presented in Figure 3(c,d). As established in previous studies (e.g., Müller and Bühler, 2001), the wall conductivity affects the electric current distribution in the conducting phase and thereby modifies the Lorentz force. In the present two-phase configuration, the Lorentz force acts solely within the lower, conducting layer, reducing its velocity, while the upper, non-conducting gas layer remains unaffected. Comparison of Figures 3(c) and 3(d) shows that for the same Ha, the damping effect of the Lorentz force, which is represented by the negative gradient of the induced magnetic field ($\mathfrak{B}_j'$, see Eq. 2.1), is considerably stronger when the lower wall is perfectly conducting. Consequently, the conductive phase exhibits lower velocity, and a larger holdup (i.e., flow area) is required to maintain its volumetric flow rate (see Figures 3a and 3b). The reduced gas flow area results in significantly higher gas velocity when the bottom wall is conducting. As noted by Parfenov et al. (2024), at sufficiently strong magnetic fields, the conducting liquid can occupy nearly the entire cross-section, while a thin gas layer may act as a lubricant when $Q_1/Q_2$ is large (i.e., at low gas flow rates).

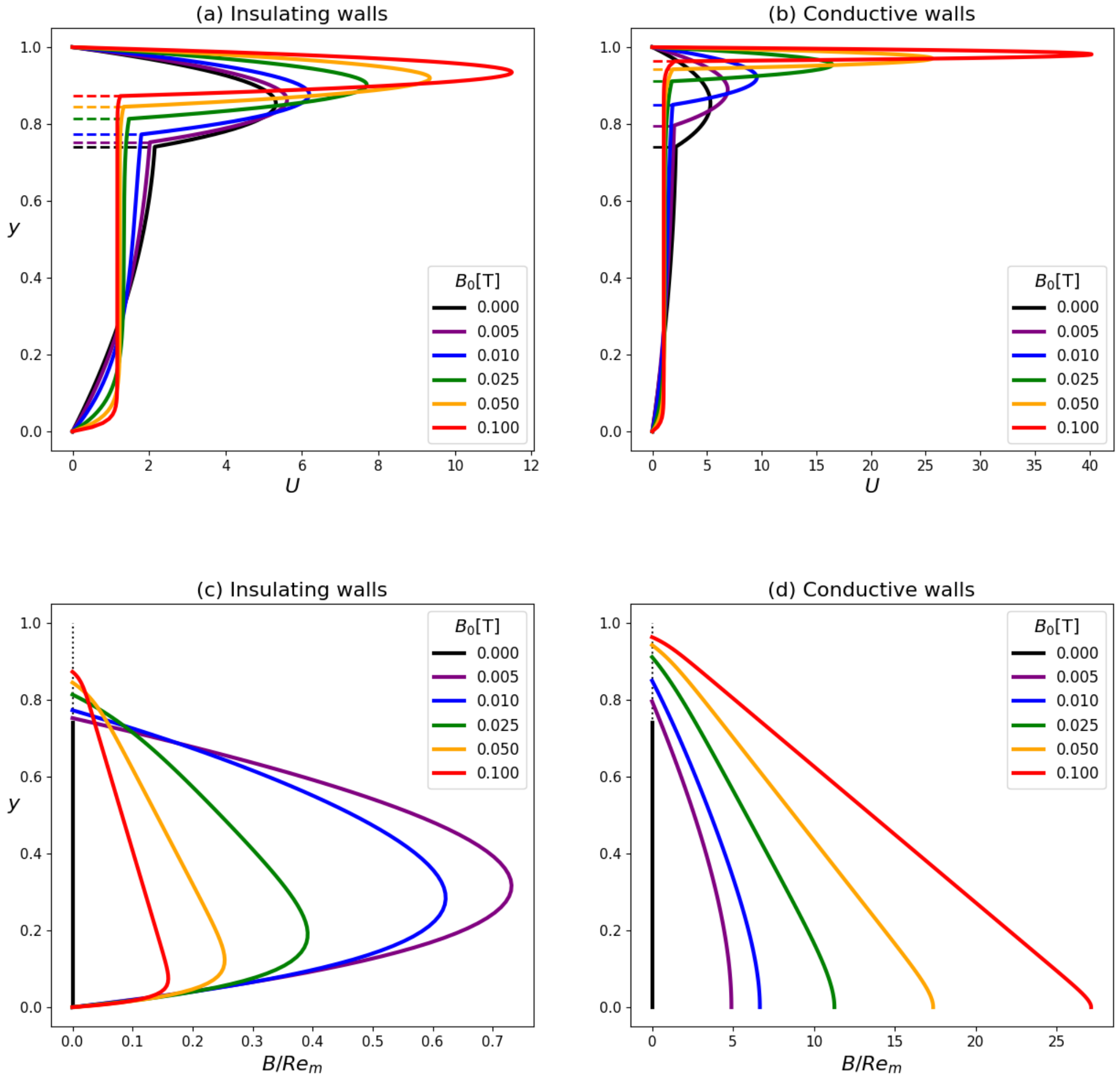

*Figure 3. Dimensionless base flow velocity profiles (a,b) for $Q_1/Q_2 = 1$ and respective induced magnetic field components (c,d) with insulating (a,c) and with perfectly conducting (b,d) lower wall. The velocity is scaled by $U_{1s} = \tilde{Q}_1/H$. [mercury-air]*

### *4.2. Eigenvalues*

In the first order approximation for small values of $k$, the eigenvalue $c$ is represented as

$$c \approx c_0 + ik\left( c_{1,1} \cdot Re_2 + c_{1,-1}/ Re_2\right), \tag{4.1}$$

where the values $c_0$, $c_{1,1}$, $c_{1,-1}$ are real numbers which are obtained for given $B_0$ and $h$ by the procedure described in the previous section. The real part of $c$ corresponds to the wave speed (normalized by the interfacial velocity) of the disturbance. It does not depend on $Re_2$ and its value is determined by the zero-order solution, $c_0$. The long-wave stability of the flow is determined by the sign of $Im(c)$, where a growing disturbance corresponds to $Im(c) > 0$.

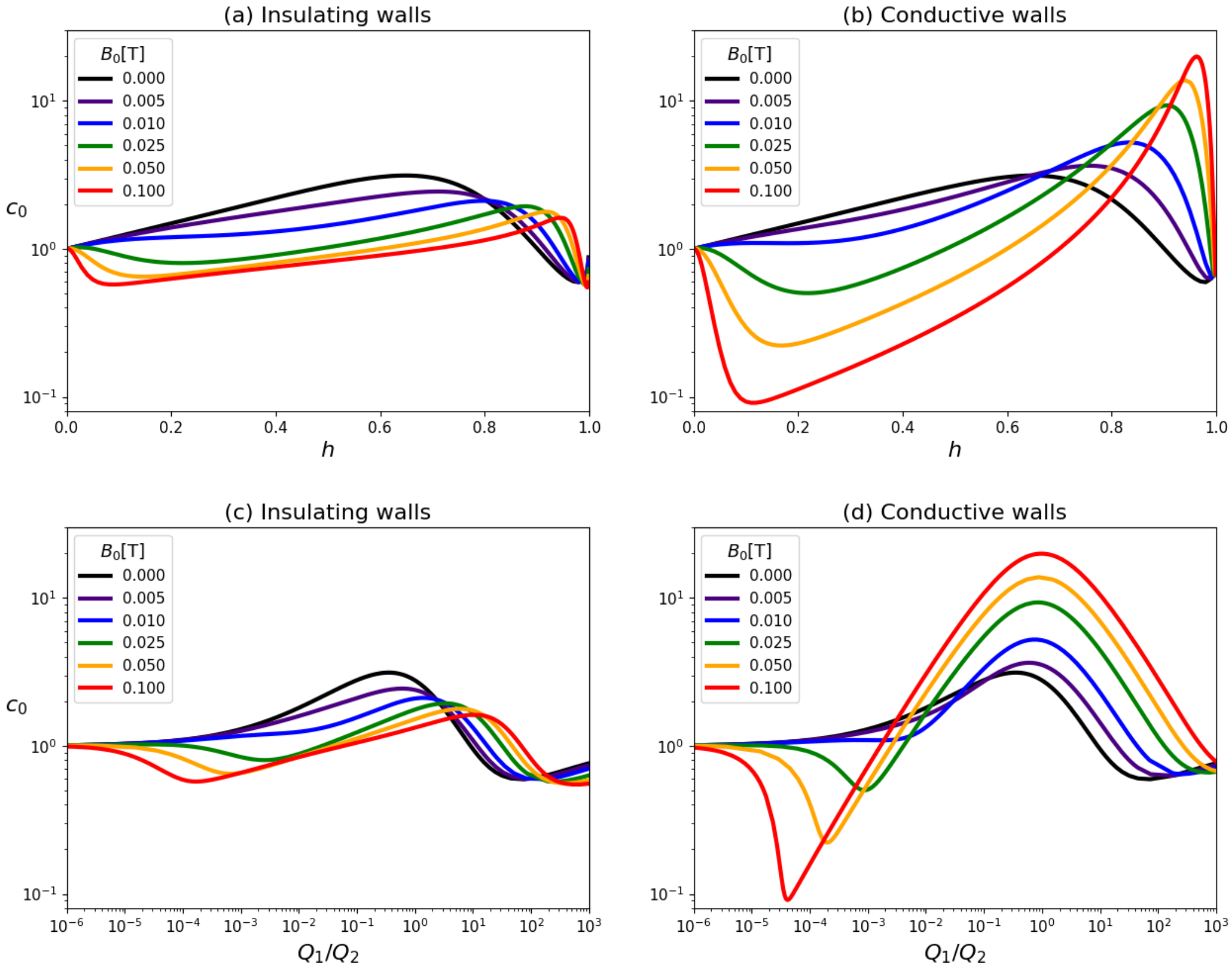

*Figure 4. Wave speed $c_0$ versus holdup $h$ (a,b) and flux ratio $Q_1/Q_2$ (c,d) for different values of the magnetic field $B_0$ in a channel with insulating (a,c) and perfectly conducting (b,d) lower wall. [mercury-air]*

Figure 4 shows the dependence of $c_0$ on holdup $h$ and on the flux ratio $Q_1/Q_2$ for different values of $B_0$. Note that according to (3.34), the conductivity of the lower wall affects the value of $c_0$ only through the interface position and the values of $U_j'$ and $U_j''$ at the interface. This results in two quite unexpected observations. First, the phase velocity is noticeably affected by the wall conductivity. Second, with the increase of magnetic field, the wave speed decreases at small ratios $Q_1/Q_2$, increases at larger ratios, and then again decreases at very large flow rate ratios. The minima and maxima associated with these trends are several times larger in the case of conductive walls than in the case of insulating walls. Furthermore, in the case of insulating wall, the wave speed is of the order of the interfacial velocity (i.e., its dimensionless value is of the order of unity) over a wide range of flow rates and $B_0$, and the maximum value of $c_0$ decreases with $B_0$. On the other hand, with a conducting lower wall, the maximal $c_0$ increases with $B_0$, and reaches values which are much higher than the interface velocity when $Q_1/Q_2 > 0.1$, and much lower than the interface velocity for small $Q_1/Q_2$. In a channel with an insulating wall, the maximum of $c_0$ is shifted to higher $Q_1/Q_2$ and $h$ with the increase of $B_0$. In the case of conducting wall, the maximal $c_0$ is obtained at similar values of $Q_1/Q_2$ (a slight shift is observed between $B_0 = 0$ and $B_0 = 0.005$ T), which, unlike the insulating wall case, corresponds to larger holdups for larger $B_0$.

The growth rate of the perturbation is determined by the imaginary part of $c$, which is proportional to $k$, and is unaffected by the zero-order solution. It consists of two parts: $kc_{1,1}Re_2$ and $kc_{-1,1}/Re_2$.

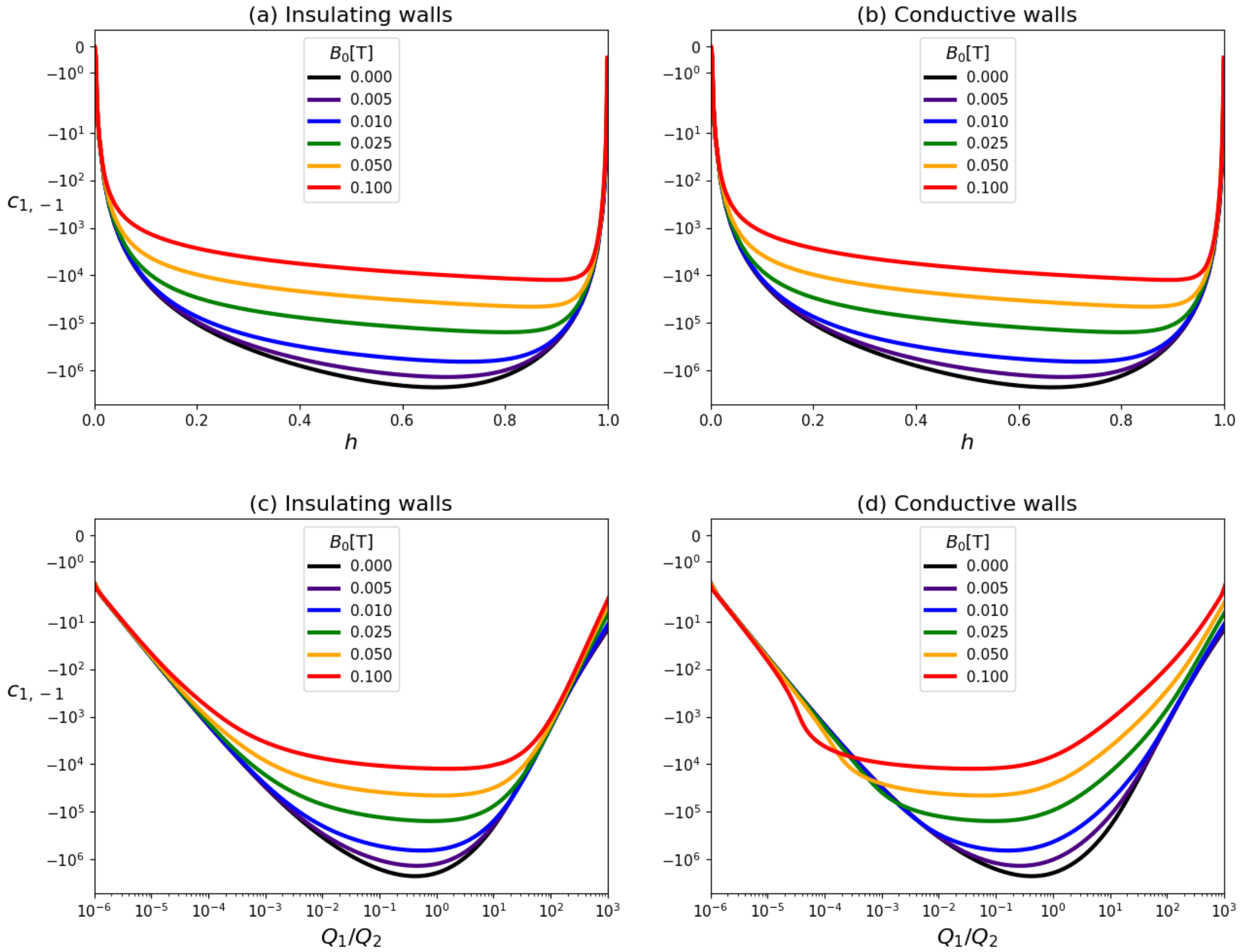

*Figure 5. Values of $c_{1,-1}$ versus holdup $h$ (a,b) and flux ratio $Q_1/Q_2$ (c,d) for different values of the magnetic field $B_0$ in a channel with insulating (a,c) and perfectly conducting (b,d) lower wall. [mercury-air]*

Figure 5 shows how $c_{1,-1}$ depends on $h$ and $Q_1/Q_2$ for different $B_0$. It is always negative, and its absolute values decrease with increasing the magnetic field in all the cases except at very small $Q_1/Q_2$ ratios in the frame 5d. Frames (a) and (b) are identical, since according to (3.45), for a given holdup, $c_{1,-1}$ is independent of the other details of the base flow, hence does not depend on the wall conductivity. The effect of different holdups obtained for specified flow rate ratio and either conducting or insulating lower wall, is evident when comparing frames (c) and (d). Since the values of $c_{1,-1}$ are always negative, the instability onset is possible only when values of $c_{1,1}$ are positive and large enough. Note that $c_{1,-1}$ is proportional to $g$ and to $(r-1)$, i.e., it vanishes in the zero-gravity case or for equal densities of the two fluids, in which case the stability is determined solely by $c_{1,1}$.

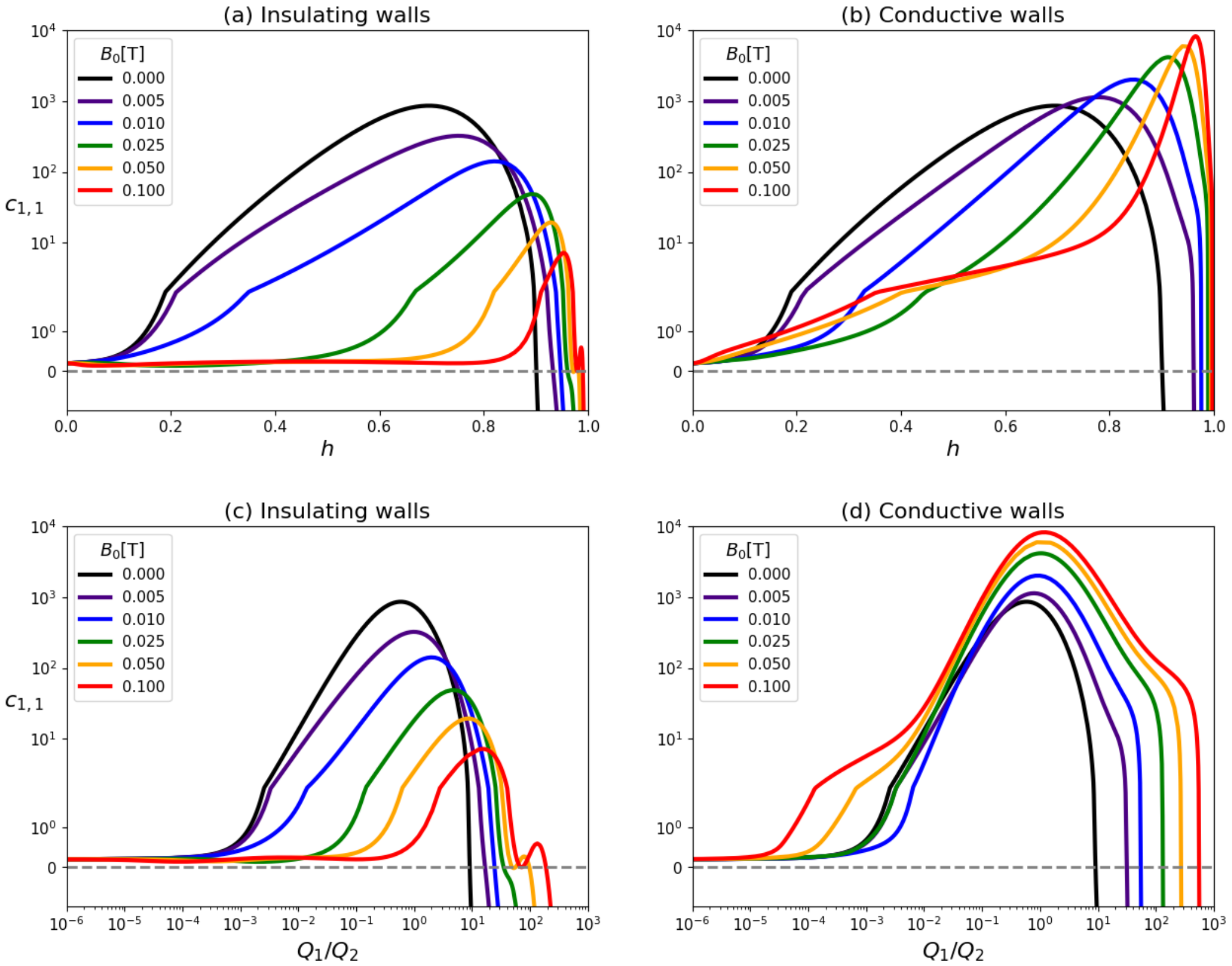

*Figure 6. Values of $c_{1,1}$ versus holdup $h$ (a,b) and flux ratio $Q_1/Q_2$ (c,d) for different values of the magnetic field $B_0$ in a channel with insulating (a,c) and perfectly conducting (b,d) lower wall. [mercury-air]*

The dependences of $c_{1,1}$ on the holdup and the flow rates ratio are shown in Figure 6. Note that according to eq. (4.1), $c_{1,1}$ dominates the growth rate at large $Re_2$, and that for small and moderate values of the flux ratio and holdup its value is positive, which indicates on a tendency of losing stability. As $Q_1/Q_2$ (or $h$), increases, the sign of $c_{1,1}$ changes from positive to negative, indicating the possibility of flow stabilization. The values of $Q_1/Q_2$ (and the corresponding holdup) at which the sign change occurs increase with $B_0$, indicating that the system tends to stabilize as the mercury flow intensifies and the air flow weakens. In the case of insulating lower wall, the sign of $c_{1,1}$ changes several times for $B_0 >\approx 0.04$ T. To illustrate that, the intervals corresponding the sign changes are enlarged in Figure 7. Similar to $c_0$, with the increase of $B_0$, the maximum of $c_{1,1}$ shifts to larger $h$, while the maximal value decreases for an insulating wall and increases for a conducting lower wall.

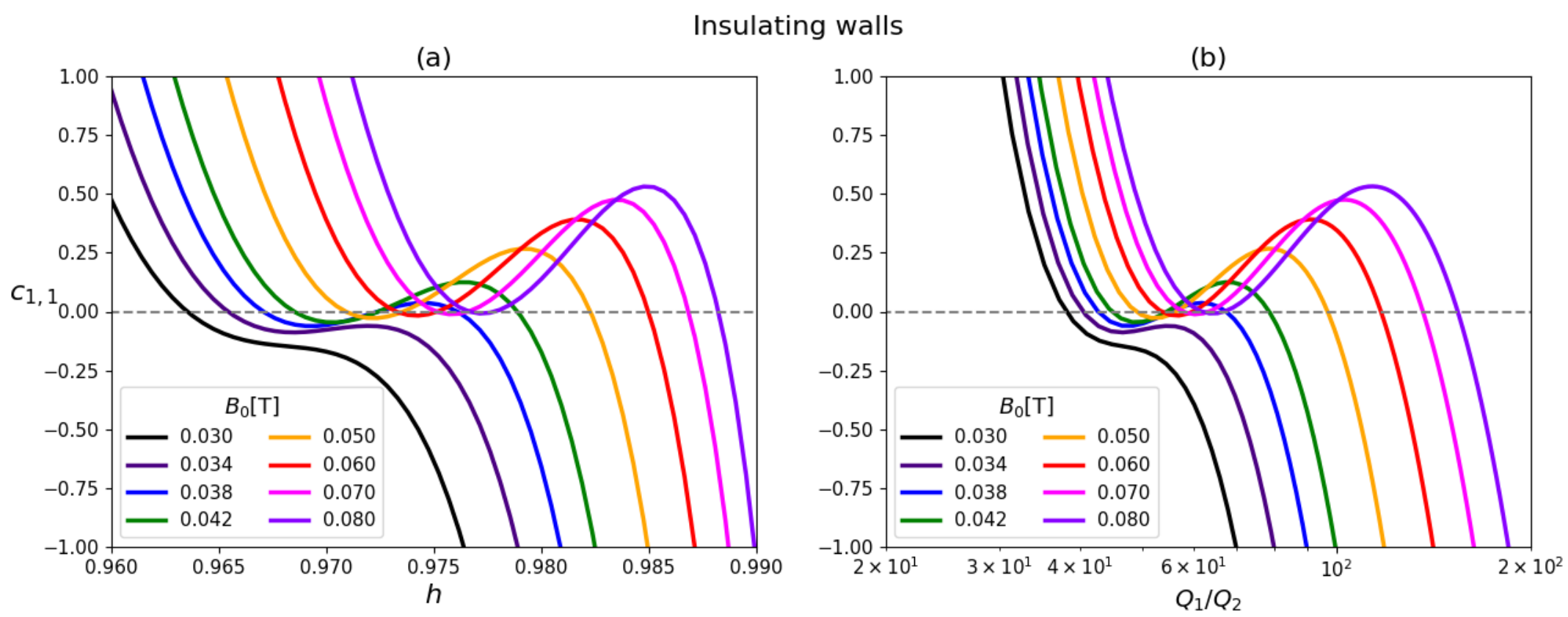

*Figure 7. Values of $c_{1,1}$ versus holdup $h$ (a) and flux ratio $Q_1/Q_2$ (b) for different values of magnetic field $B_0$. [mercury-air]*

The range of wave numbers for which the first-order asymptotic approximation remains valid can be assessed by comparing the predicted values of $c$ with those obtained numerically (Table 3). The numerical values are computed using the Chebyshev collocation method (description of the numerical method can be found in (Peyret, 2002)), similar to the approach described in (Barmak et al., 2016). As shown in Table 3 for a perfectly conducting lower boundary, the asymptotic and numerical results begin to diverge for $k = 10^{-5}$. Similar behavior is observed for a perfectly insulating boundary.

*Table 3. Asymptotic vs numerical c (mercury-air, conducting lower boundary, $h = 0.5$)*

| $B_0$[T] | $Re_2$ | k | c asymptotic | c numerical |
|---|---|---|---|---|
| 0 | 1 | $10^{-6}$ | 2.6694 − 1.4265E+00 i | 2.6697 − 1.4266E+00 i |
| | | $10^{-5}$ | 2.6694 − 1.4265E+01 i | 2.7016 − 1.4381E+01 i |
| | 10 | $10^{-6}$ | 2.6694 − 1.4059E-01 i | 2.6697 − 1.4060E-01 i |
| | | $10^{-5}$ | 2.6694 − 1.4059E+00 i | 2.7011 − 1.4165E+00 i |
| | 100 | $10^{-6}$ | 2.6694 + 6.5523E-03 i | 2.6692 + 6.5488E-03 i |
| | | $10^{-5}$ | 2.6694 + 6.5523E-02 i | 2.6557 + 6.2277E-02 i |
| 0.01 | 1 | $10^{-6}$ | 1.7708 − 4.4239E-01 i | 1.7749 − 4.4429E-01 i |
| | | $10^{-5}$ | 1.7708 − 4.4239E+00 i | 1.7767 − 4.4461E+00 i |
| | 10 | $10^{-6}$ | 1.7708 − 4.4025E-02 i | 1.7749 − 4.4212E-02 i |
| | | $10^{-5}$ | 1.7708 − 4.4025E-01 i | 1.7767 − 4.4243E-01 i |
| | 100 | $10^{-6}$ | 1.7708 − 2.2602E-03 i | 1.7749 − 2.2515E-03 i |
| | | $10^{-5}$ | 1.7708 − 2.2602E-02 i | 1.7758 − 2.2488E-02 i |
| 0.1 | 1 | $10^{-6}$ | 0.3420 − 7.1629E-03 i | 0.3410 − 7.1677E-03 i |
| | | $10^{-5}$ | 0.3420 − 7.1629E-02 i | 0.3410 − 7.1677E-02 i |
| | 10 | $10^{-6}$ | 0.3420 − 6.8344E-04 i | 0.3410 − 6.8389E-04 i |
| | | $10^{-5}$ | 0.3420 − 6.8344E-03 i | 0.3410 − 6.8389E-03 i |
| | 100 | $10^{-6}$ | 0.3420 + 2.6010E-04 i | 0.3410 + 2.6043E-04 i |
| | | $10^{-5}$ | 0.3420 + 2.6010E-03 i | 0.3410 + 2.6043E-03 i |

## *4.3. Stability boundary*

The long-wave stability of the flow is determined by the sign of $Im(c)$, which is obtained from (4.1):

$$Im(c) = k\left( c_{1,1} \cdot Re_2 + c_{1,-1} / Re_2 \right). \tag{4.2}$$

As mentioned above, positive $Im(c)$ corresponds to instability. As shown above (Figure 5) the values of $c_{1,-1}$ are always negative in the presence of gravity ($g > 0$ and $r > 1$), and vanish in zero-gravity case. Therefore, when $c_{1,1} < 0$, the flow is stable with respect to long-wave perturbations for any $Re_2$. When $c_{1,1} > 0$, the flow is unstable for large enough $Re_2$, or for any $Re_2$ in the zero-gravity case. Thus, the zero-gravity stability boundaries are determined by the roots of the equation $c_{1,1} = 0$, which are presented in Fig. 8. In the case of a conducting lower wall, there is only one root corresponding to each $B_0$. However, for an insulating wall, multiple sign changes occur, consistent with the behavior shown in Figure 7. As shown in Figure 8, the critical holdup $h_{cr}$ (i.e. upper boundary, such that $c_{1,1} = 0$ for $h = h_{cr}$, and $c_{1,1} < 0$ for any $h > h_{cr}$), as well as the corresponding critical flow ratio $(Q_1/Q_2)_{cr}$, increase with the magnetic field. A steeper increase takes place in the conducting wall case. Thus, in the limiting case of zero gravity, with increasing the magnetic field at a constant superficial velocity of air, the critical mercury superficial velocity grows. Note that in this case all the neutral stability curves approach a linear dependence at large $B_0$.

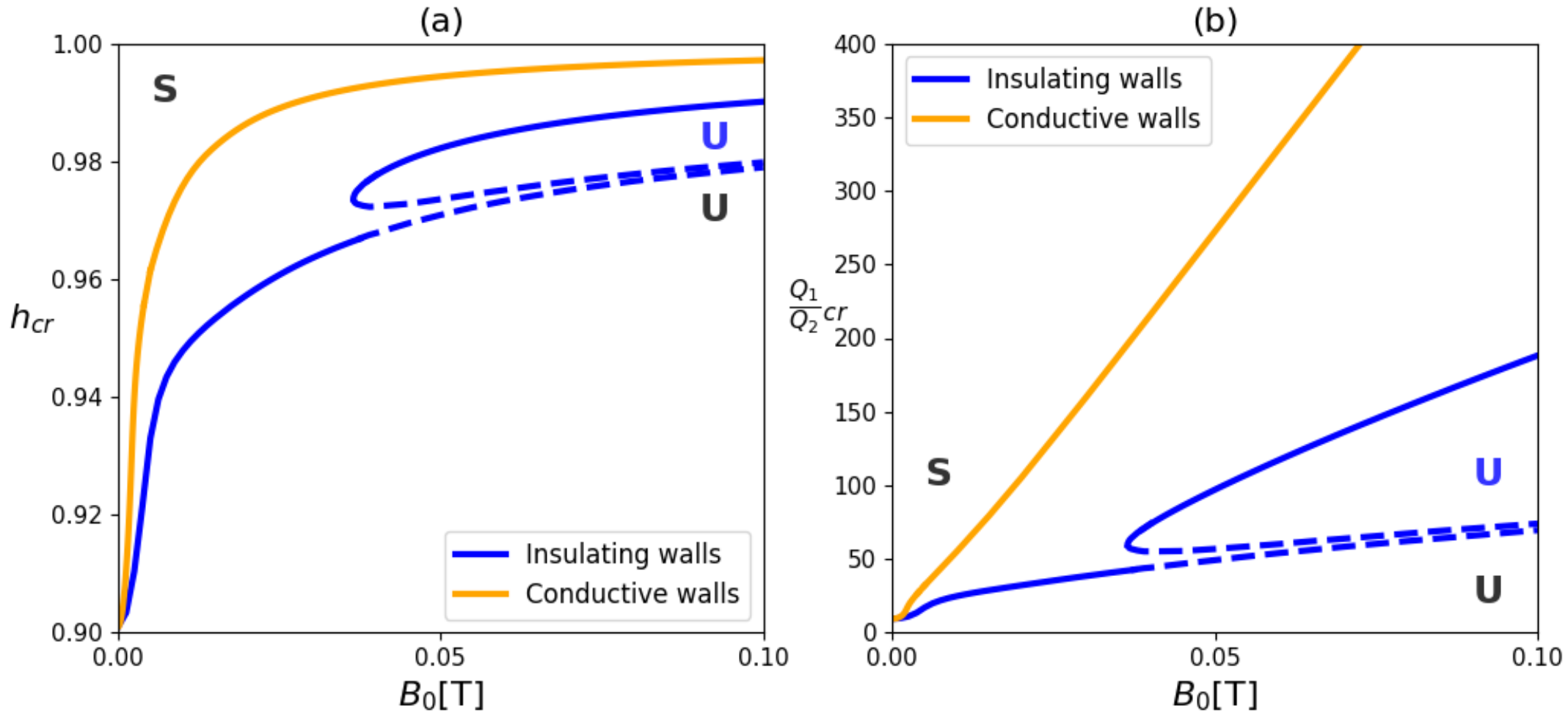

*Figure 8. Stability boundaries for zero-gravity case: critical holdup h (a) and flow rate ratio $Q_1/Q_2$ (b), which correspond to $c_{1,1} = 0$, versus magnetic field $B_0$. For $B_0 = 0$, $h_{cr} \approx 0.9$ and $(Q_1/Q_2)_{cr} \approx 9.1$. [mercury-air]*

It is known that for $B_0 = 0$ the neutral stability conditions also correspond to equal average velocities of the phases $U_{1,avg} = U_{2,avg}$ (where $U_{1,avg} = Q_1/h$ and $U_{2,avg} = Q_2/(1-h)$), which is also associated with zero shear stress at the interface (i.e. $U_j'(0) = 0$) (Kushnir et al, 2014). This is not the case when the flow is under a magnetic field effect. Frame (a) of Figure 9 shows the $U_{1,avg}/U_{2,avg}$ ratio at the neutral stability boundary, indicating that the ratio of the critical phases' average velocities increases with $B_0$. The lines corresponding to either $c_{1,1} = 0$, to equal average velocities, or to zero interfacial shear stress are plotted on the frame (b). As can be seen, they only coincide in the absence of the magnetic field.

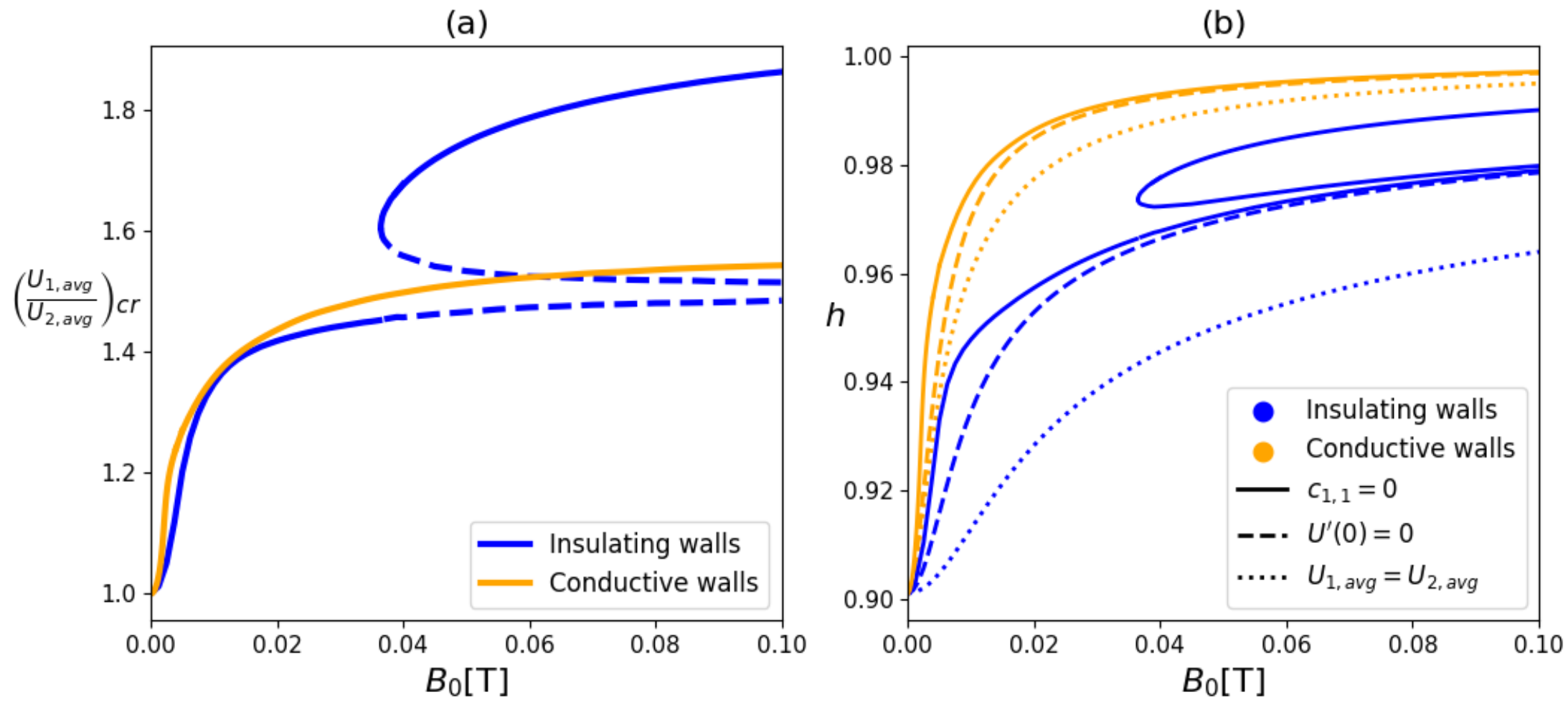

*Figure 9. [mercury-air](a): Critical $U_{1,avg}/U_{2,avg}$ vs $B_0$. (b): Values of h corresponding to neutral stability (solid lines), equal average velocities (dotted lines), and zero interfacial shear (dashed lines) stress vs $B_0$ for both perfectly insulating (blue) and perfectly conductive (orange) walls.*

For non-zero gravity, when $c_{1,-1} < 0$ and $c_{1,1} > 0$ , the long-wave stability is determined by $Re_2$: $Im(c) > 0$ when $Re_2 > Re_{2,cr}$, where critical $Re_{2,cr}$is found from:

$$c_{1,1} \cdot Re_{2,cr} + c_{1,-1}/ Re_{2,cr} = 0 \Rightarrow Re_{2,cr} = \sqrt{-c_{1,-1}/c_{1,1}}. \quad (4.3)$$

In the following, we report the critical superficial velocities, which are commonly used in two-phase flow studies to represent the stability boundaries of the flow configuration considered (e.g., Barmak et al., 2016). We refer to the (critical) mercury superficial velocity, $\mathrm{U}_{1s} = \widetilde{\mathrm{Q}}_1/\mathrm{H}$, defined as a volumetric flow rate ($Q_1$) normalized by the channel height:

$$U_{1s} = \frac{\widetilde{\mathrm{Q}}_1}{H} = \frac{1}{H}\int_{-h_1}^{0} V_x(\tilde{y})d\tilde{y} = \frac{1}{H}\cdot\frac{h_1}{n}\cdot U_*\int_{-n}^{0} U(y)dy = \frac{Re_2\mu_2}{H\,\rho_2}\int_{-n}^{0} U(y)dy. \quad (4.4)$$

The critical superficial velocity $U_{1s,cr}$ if found from (4.4) when $Re_2 = Re_{2,cr}$. Also, the corresponding (critical) air superficial velocity is found as $U_{2s} = \widetilde{\mathrm{Q}}_2/H = U_{1s}Q_2/Q_1$.

Figures 10 and 11 show the dependence of $U_{1s,cr}$ (Fig. 10) and $U_{2s,cr}$ (Fig. 11) on $B_0$ for fixed holdups (a,b) and for fixed flow rate ratios (c,d). The variations of $U_{1s,cr}$ and $U_{2s,cr}$ with the holdup and flow rate ratio at fixed values of the magnetic field are presented in Fig. 12(a,b), 13(a,b) and Fig. 12(c,d), 13(c,d), respectively. The main observation that follows from all these stability diagrams is the non-monotonic dependence of the critical superficial velocities on the magnetic field strength, which is quite counter-intuitive. One would expect that the critical velocity $U_{1s,cr}$ increases with the increasing the magnetic field strength, similar to the single-phase scenario. However, in Figure 10 a non-monotonic behavior is most commonly observed, with $U_{1s,cr}$ increasing up to a certain $B_0$, and then decreasing with the further strengthening of the magnetic field. In some cases, when the walls are perfectly conductive (b,d) and for large enough holdups ($\geq 0.6$) or $Q_1/Q_2$ ($\geq 1.0$), $U_{1s,cr}$ decreases monotonically for the whole range of $B_0$. Also, the initial decrease of $U_{1s,cr}$ (from infinite to finite values) for the cases of $Q_1/Q_2 = 10$ in Figure 10, frames (c,d), is due to the zero-gravity destabilization, discussed previously. As can be seen from Figure 8b, this flow rate ratio corresponds to the zero-gravity stability region, when $B_0 = 0$ and, as discussed above, the flow is stable. Then, with the increase of $B_0$, the flow destabilizes.

The complementary data on the corresponding critical superficial gas velocity (Figures 11 and 13) show that for both fixed holdup or fixed flow-rate ratio, the critical superficial velocity in the gas phase approaches a constant value as $B_0$ increases. This suggests that the instability originates in the non-conducting gas phase, which becomes dominant when the velocity in conducting phase is strongly suppressed by the magnetic field. In this case, one might expect the onset of instability at high magnetic field strength to occur at a constant Reynolds number of the air flow, $e_{air} = U_{2,avg}\rho_2 h_2/\mu_2 = U_{2s}\rho_2 H/\mu_2$, which is independent of $h$. However, this is not observed. The critical $Re_{air}$ for different holdups in Figure 11 converges to distinct values as $B_0$ increases ($Re_{air} \approx 4 \cdot 10^3$ for $h = 0.8$ up to $\approx 27 \cdot 10^3$ for $h = 0.2$). It is worth noting that these limiting values are not affected by the walls' conductivity (as shown by frames (a,b) of Figure 11).

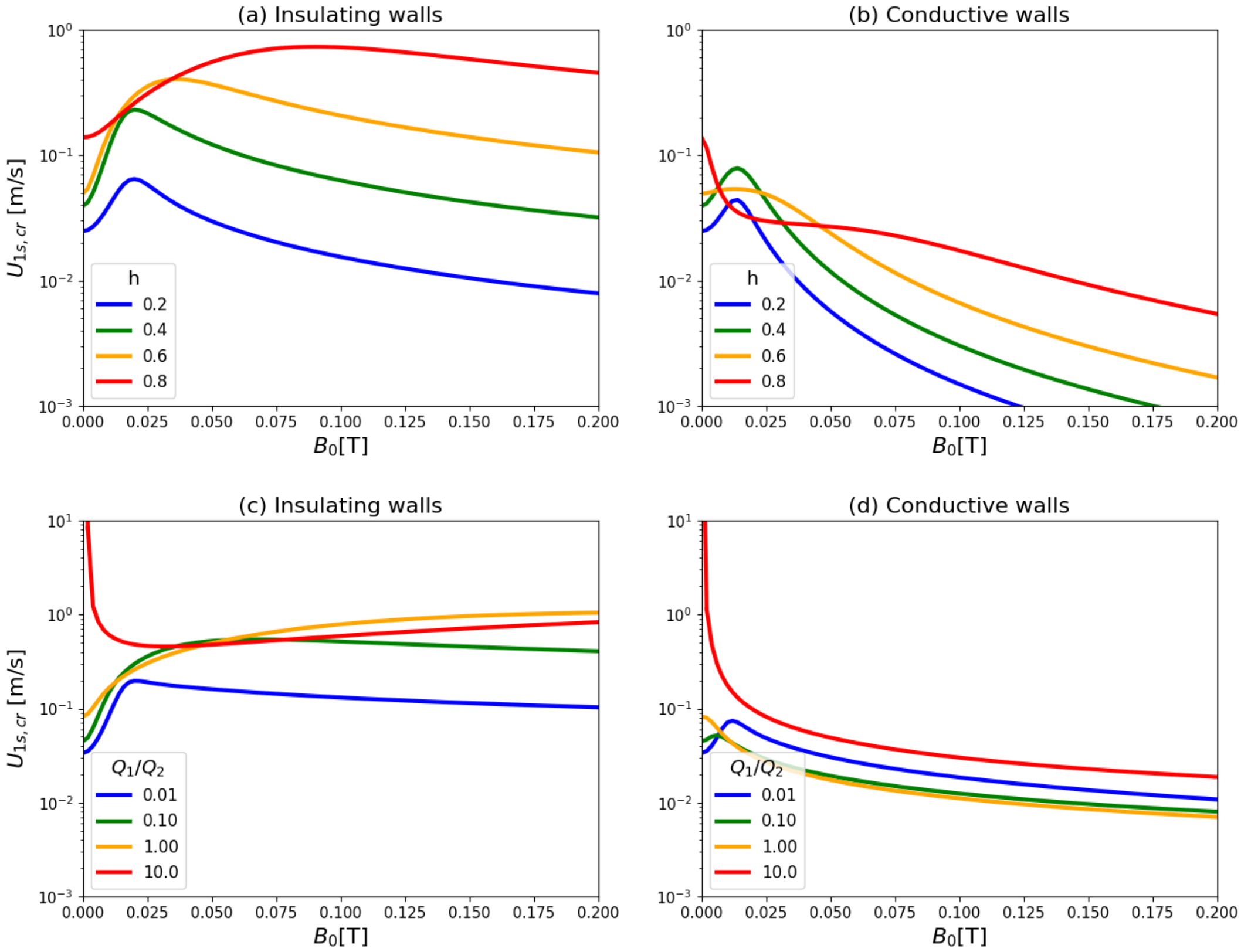

*Figure 10. Critical superficial velocity $U_{1s,cr}$ versus magnetic field $B_0$ for fixed values of holdup h (a,b) and for fixed values of flux ratio $Q_1/Q_2$ (c,d), in a channel with insulating (a,c) and perfectly conducting (b,d) lower wall. [mercury-air]*

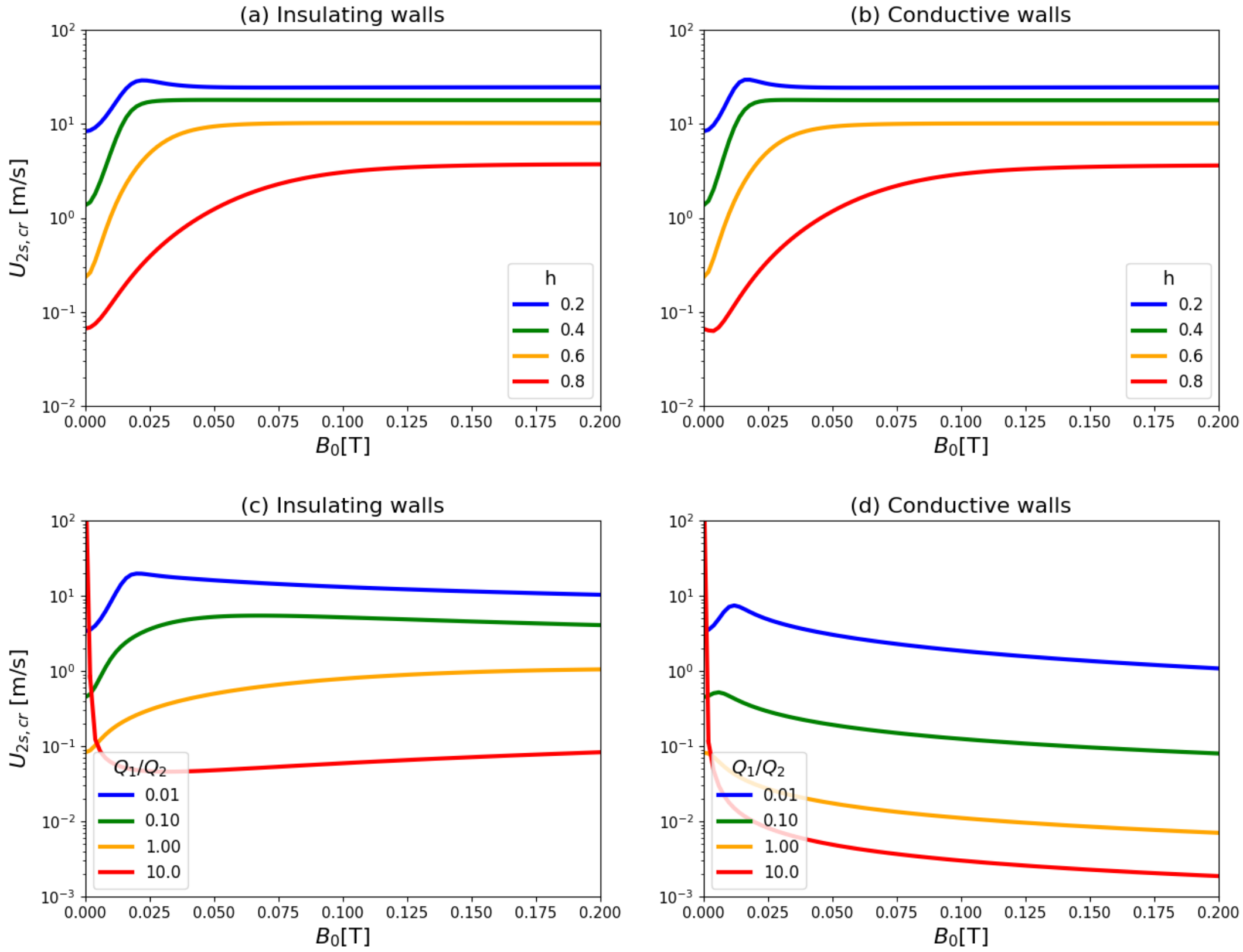

*Figure 11. Critical superficial velocity $U_{2s,cr}$ versus magnetic field $B_0$ for fixed values of holdup $h$ (a,b) and for fixed values of flux rate $Q_1/Q_2$ (c,d), in a channel with insulating (a,c) and perfectly conducting (b,d) lower wall. [mercury-air]*

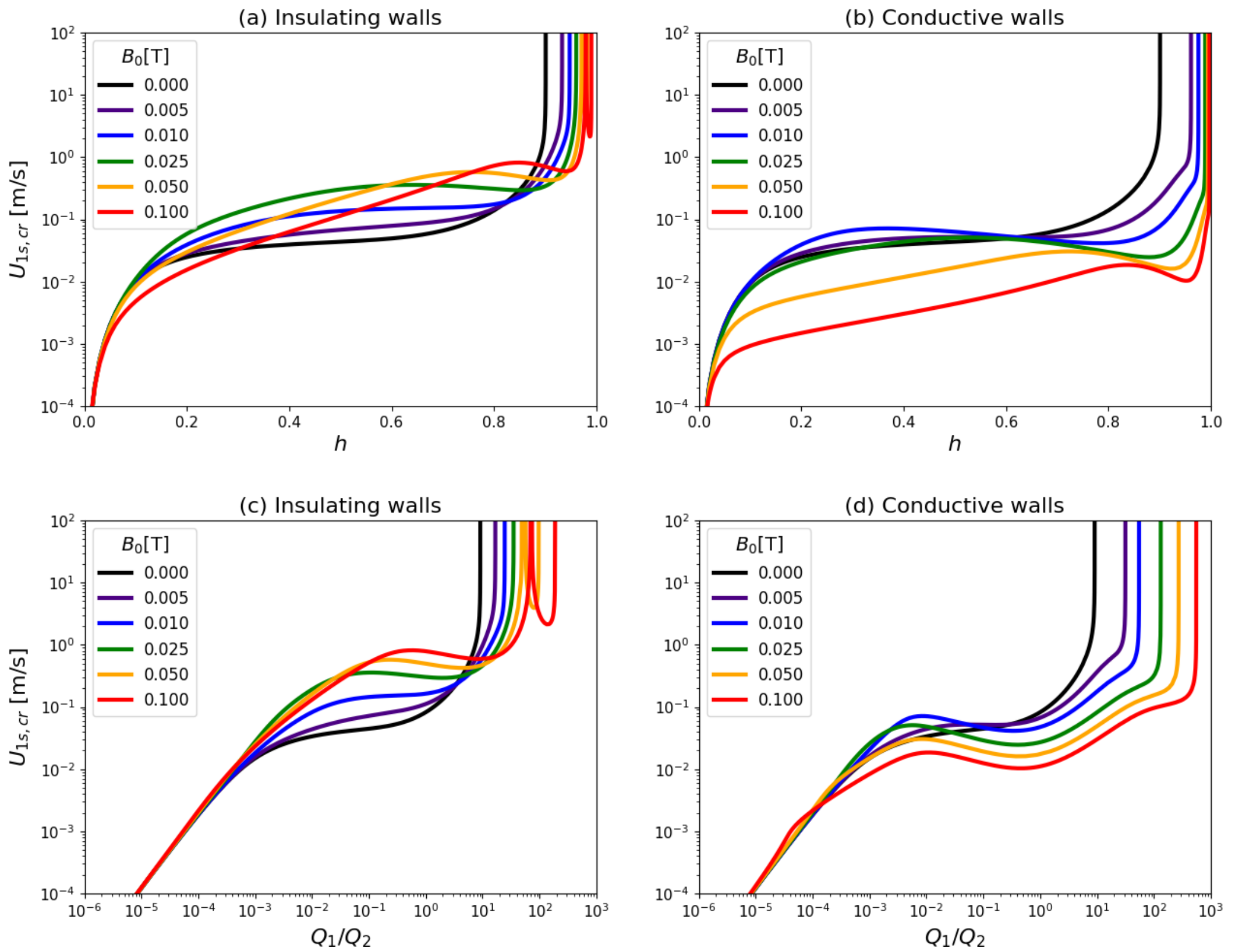

*Figure 12. Critical superficial velocity $U_{1s,cr}$ versus holdup $h$ (a,b) and flux ratio $Q_1/Q_2$ (c,d) for different values of the magnetic field $B_0$ in a channel with insulating (a,c) and perfectly conducting (b,d) wall. [mercury-air]*

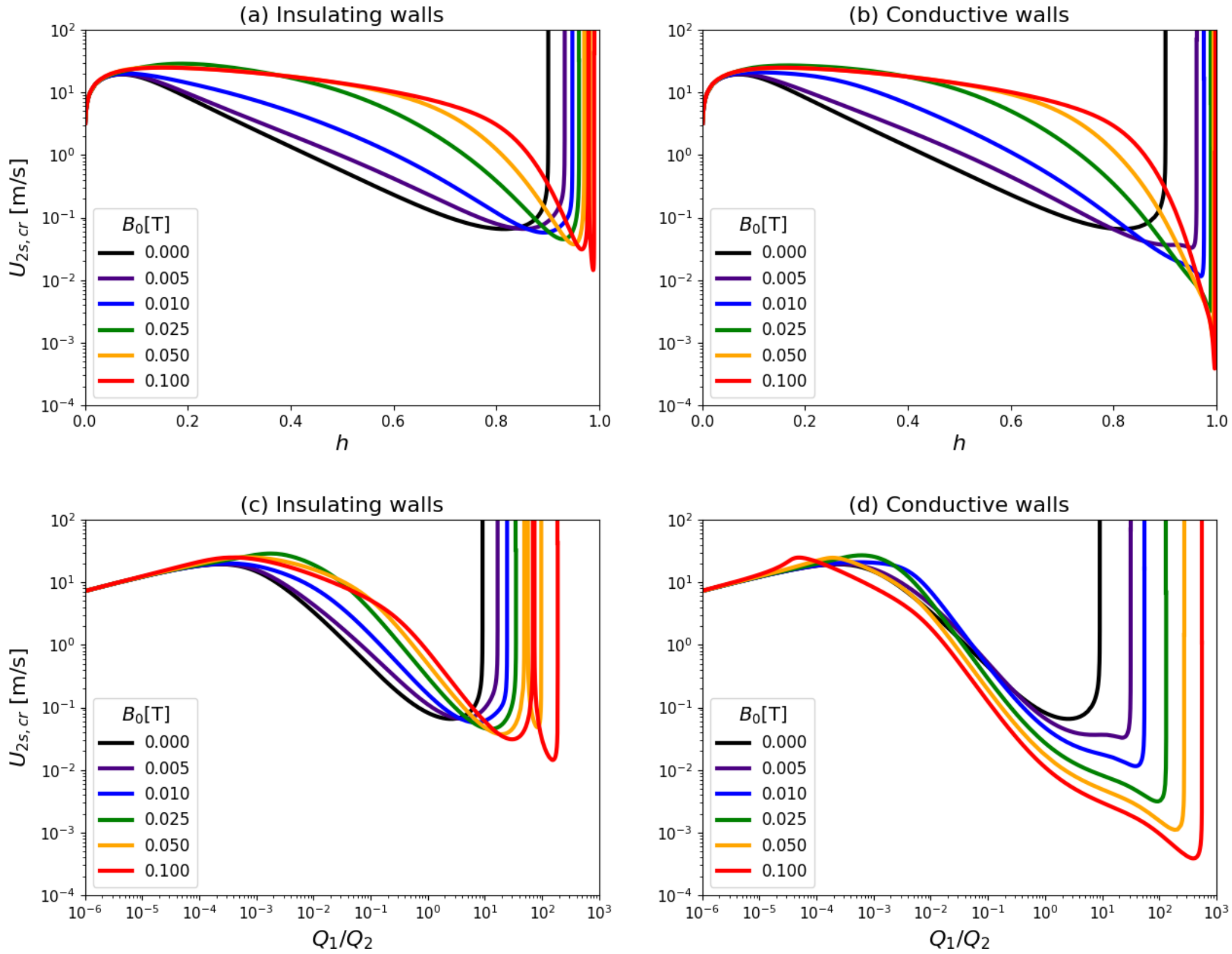

*Figure 13. Critical superficial velocity $U_{2s,cr}$ versus holdup h (a,b) and flux ratio $Q_1/Q_2$ (c,d) for different values of the magnetic field $B_0$ in a channel with insulating (a,c) and perfectly conducting (b,d) lower wall. [mercury-air]*

The non-monotonic variation of the critical superficial velocities observed in previous figures becomes more comprehensible in light of Figure 14, which presents stability maps in a plane $(U_{2s}, U_{1s})$ for different values of the magnetic field strengths. The presence of gravity (for $r > 1$) expands the stability region of long waves compared to the zero-gravity case. At high superficial velocities, the effect of gravity becomes negligible, and the neutral stability boundary approaches that predicted at zero gravity conditions (see Figure 9). This behavior is clearly observed in the case of insulating lower wall (Figure 14a). As a result, for sufficiently low $U_{2s}$, the flow remains stable for any value of $U_{1s}$. For a conducting lower wall (Figure 14b) this stable region is reached at even lower $U_{2s}$, which in the figure's scales is observed only for low $B_0$. Generally, with an insulating lower wall (Fig. 14a), the magnetic field tends to extend the gravity stabilized region for long wave disturbances. In the range of intermediate $U_{2s}$, the critical $U_{1s}$ increases with $B_0$. In this range, where the mercury occupies a large portion of the flow cross section (see Figure 12), instability of the liquid-gas interface is known to result in transition to slug flow and is associated with long wave instability (e.g., Barmak et al., 2016), which is accompanied with a steep increase in the pressure drop. Thus, applying a magnetic field can possibly postpone this transition to larger flow rates of the conductive liquid. However, for low $U_{1s}$ (i.e., low holdups), where instability of the interface is known to result in transition from stratified-smooth to the stratified-wavy pattern, the critical $U_{2s}$ is almost unaffected by the magnetic field strength. This behavior supports the above hypothesis that for thin mercury layers, the onset of instability is predominantly governed by the flow dynamics of the non-conductive air layer. On the other hand, in the case of conductive walls (Fig. 14b), only a weak magnetic field enlarges the gravity stabilized region for long wave disturbances. Further increase of $B_0$ leads to destabilization.

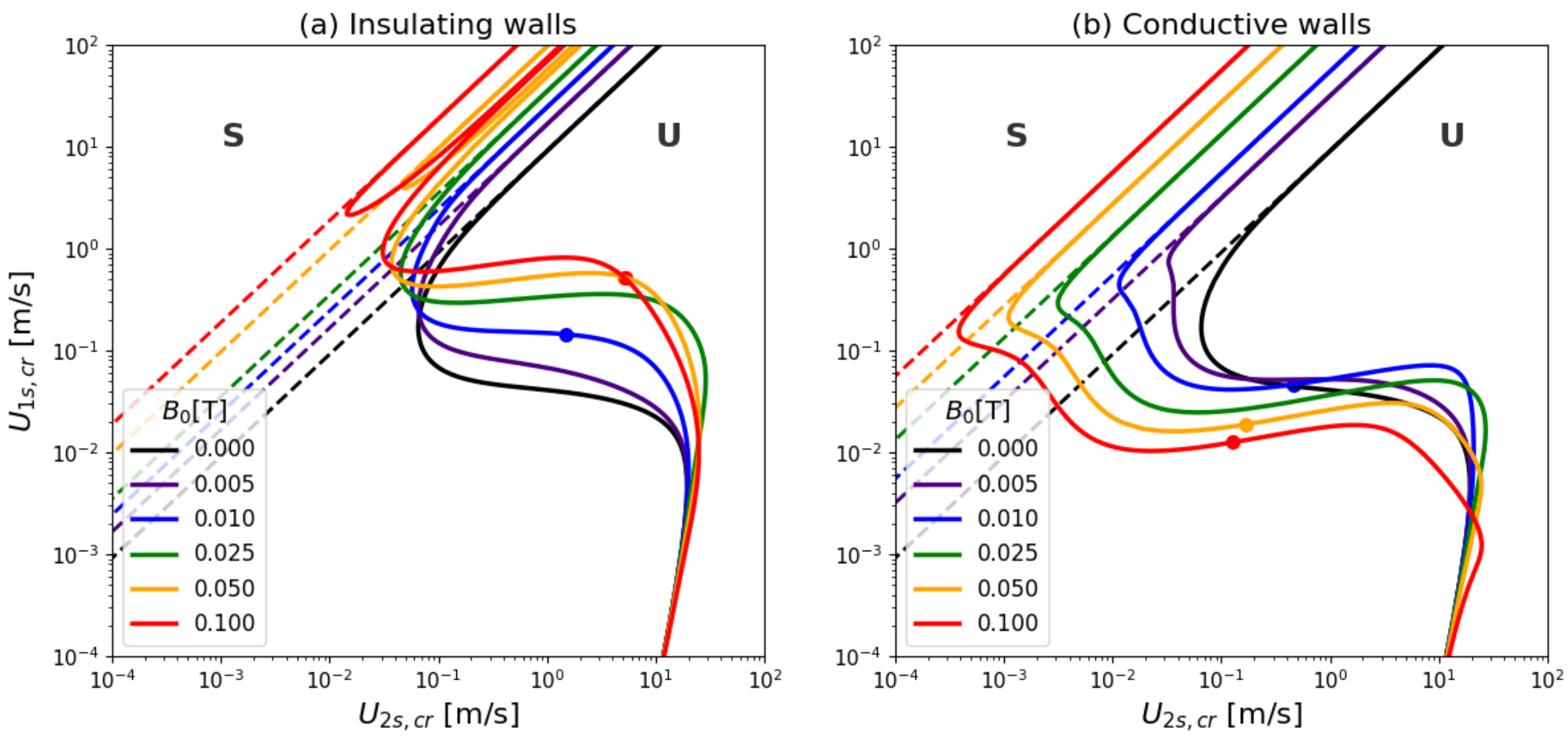

*Figure 14. Stability maps of superficial critical velocities $U_{1s}, U_{2s}$ for fixed values of magnetic field $B_0$ in a channel with insulating (a) and perfectly conducting (b) lower wall; zero-gravity boundaries are denoted with dashed lines; [mercury-air]. The dots correspond to $Q_1/Q_2 = 0.1$, for which perturbation patterns are reported below.*

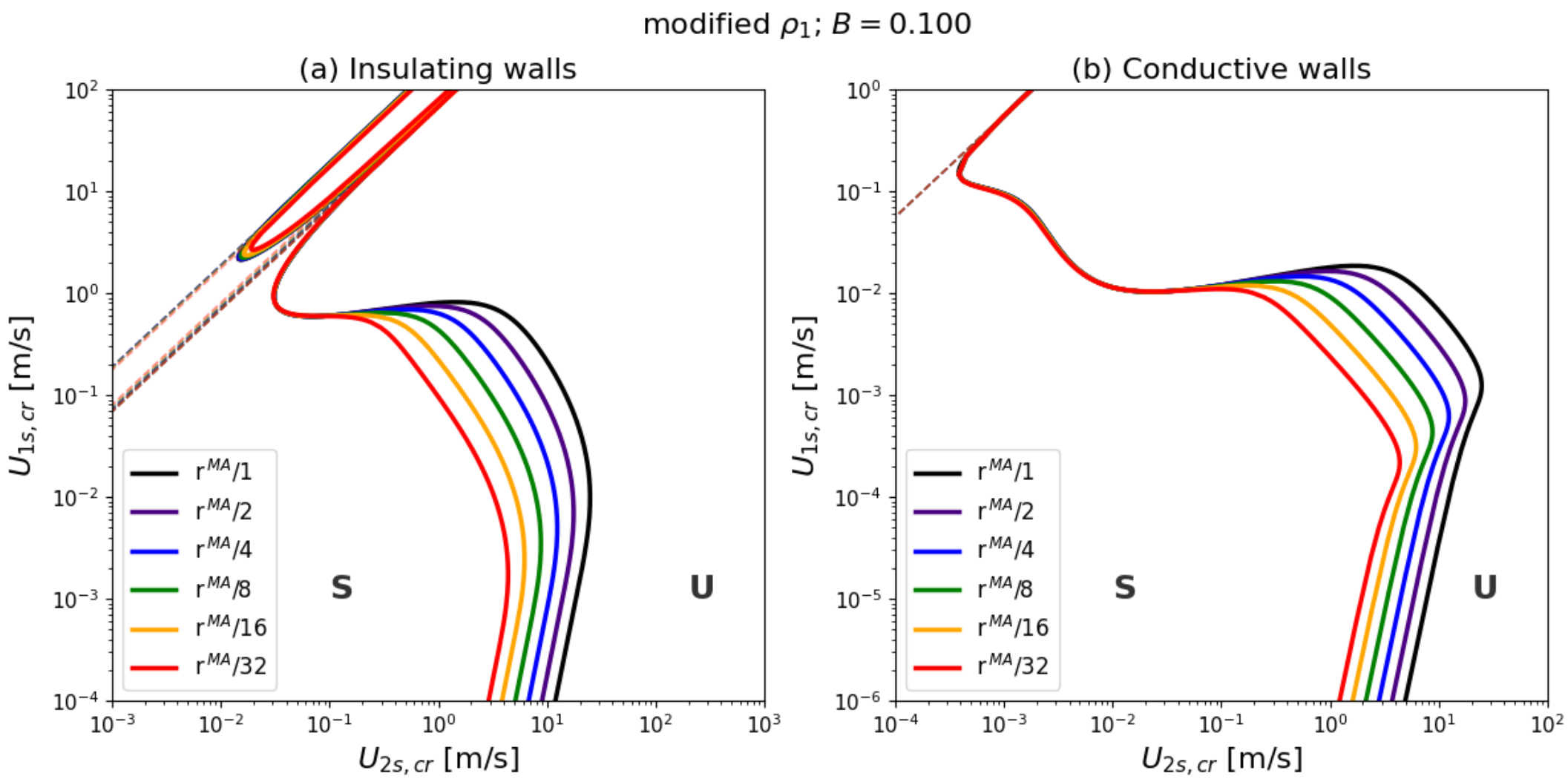

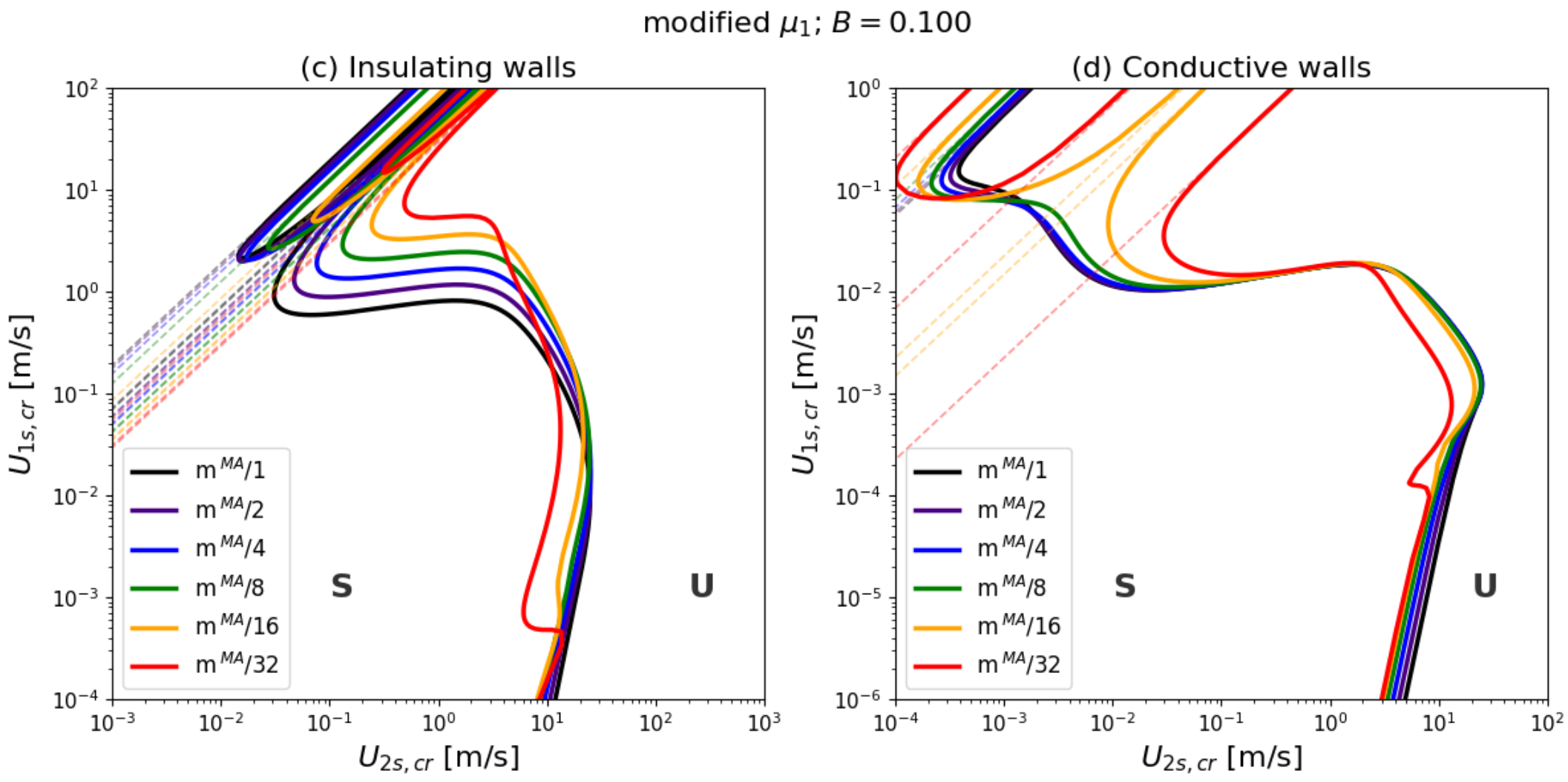

*Figure 15. Stability maps of superficial critical velocities $U_{1s}, U_{2s}$ for modified values of $\rho_1$ (a,b) and $\mu_1$ (c,d) with $B_0 = 0.1\,T$ in a channel with insulating (a) and perfectly conducting (b) lower wall. The reference r$^{MA}$ and m$^{MA}$ values refer to mercury-air system (see Table 2)*

As mentioned above, the large number of flow-defining parameters complicates any meaningful parametric study. Nevertheless, Figure 15 shows how stability boundaries of the considered mercury-air flow at $B_0 = 0.1\,T$ change when the conductive liquid density $\rho_1$ (frames a,b) and viscosity $\mu_1$ (frames c,d) are varied. Note that this is only an illustrative example that does not correspond to any realistic fluids. Stability boundaries for modified $\rho_1$ (and accordingly modified $r$) are shown in Fig.15a,b. Because the base-flow solution is independent of the fluids' density, it remains unchanged when the density ratio is varied, and therefore the general shape of the stability boundaries is preserved. As expected, the stable flow region decreases as $r$ is reduced, since the stabilizing effect of gravity (discussed above with reference to Figure 14) weakens when the density difference between the phases is smaller. The shape and location of the stability boundary change slightly owing to the values of $Re_{2,cr} = \sqrt{-c_{1,-1}/c_{1,1}}$. The value of $c_{1,-1}$ is proportional to $(r-1)$ (Eq. 3.45), while $c_{1,1}$ consists of several terms (Eqs. 3.47-3.49), one of which is not proportional to $r$ and vanishes at large $Q_1/Q_2$. Therefore, for larger $Q_1/Q_2$, both $c_{1,-1}$ and $c_{1,1}$ scale proportionally to $r$, causing the stability boundaries for different $r$ to coincide.

Changes in the viscosity of the conductive fluid affect both $m$ and $Ha_1$, and therefore, also alter the base-flow solution. Although the stability boundaries shown in Fig.15c,d remain qualitatively similar, several effects of varying $\mu_1$ are noted. In an insulating channel and for intermediate $U_{2s}$, the stable region extends to higher critical $U_{1s}$ when $\mu_1$ is reduced. Comparison with Fig. 14a suggests that, in this range of $U_{2s}$, the potential for stabilizing of the stratified flow configuration by applying a magnetic field is greater when the conductive fluid has a lower viscosity. However, the $U_{2s}$ range where magnetic-field induced stabilization occurs decreases as $\mu_1$ is reduced. In particular, for low $U_{1s}$ (i.e., low $Q_1/Q_2$ and holdups), the stable region is reduced when the conductive fluid is less viscous both in insulating and conductive channels. Additionally, in the conducting channel, an extra stable region an extra stable region (corresponding to $c_{1,1} < 0$) emerges at high $U_{1s}$ and reduced $\mu_1$ (frame d: $m^{MA}/16$ and $m^{MA}/32$ for $Q_1/Q_2 \approx 10$). For the mercury–air system, such a region appeared within the studied flow-rate range only in the insulating channel. However, stability analyses of gas–liquid flow in the absence of a magnetic field that consider all wavenumbers (Barmak et al., 2016) indicate that, in the region of high $U_{1s}$(i.e., high $Q_1/Q_2$and high holdups), instability is triggered by short waves rather than long waves. Therefore, the practical significance of the long-wave stability boundaries in this region, and of their variation with system parameters, may be limited.

### 4.4. Disturbance profiles

Further insight into the impact of the magnetic field strength on the onset of instability can be obtained by examining the disturbance profiles. Recalling that the streamwise part of the disturbance is $u_j = \phi_j'$, in the first approximation, when $k \to 0$, it has the form: $u_j = u_{j;0} + ik\left(u_{j;1,1} \cdot Re_2 + u_{j;1,-1}/Re_2\right)$. Note, that both $u_{j;0}$ and the expression in brackets are real functions. For the long wave disturbances, $u_j \sim u_{j;0}$.

Figures 16 and 17 show profiles of $u_{j;0}$ for different magnetic field strengths and fixed values of $h$ and $Q_1/Q_2$, respectively. Since the amplitude of the disturbance is arbitrary, the values presented are scaled by $max|u_{j;0}| = 1$. Note that $u_{1;0}(0) - u_{2;0}(0) = \phi_{1;0}'(0) - \phi_{2;0}'(0) = \left(U_2'(0) - U_1'(0)\right)\eta$, hence the streamwise disturbance undergoes a jump at the interface. The results show that the largest absolute value of the disturbance is at the air-side of the interface ($u_{2;0}$), but there is an additional local extremum in the bulk of the air layer. The effect of the lower wall conductivity on the streamwise velocity perturbation profiles for a fixed holdup seems to be negligible for all holdups. For fixed values of flow-rate ratio, the effect of the lower wall conductivity on the shape of the $u_{j;0}$ profile is mainly due to different holdups of the conducting phase for the cases of conducting and insulating lower wall (see Figure 17). In all cases, the largest disturbance amplitudes are located either at the interface or in the bulk of the air layer. This suggests that the long wave instability sets either as an instability of the shear flow in the gas layer, or results from an interfacial instability. This observation explains why the shape of axial velocity perturbation profiles is practically independent on the wall conductivity. Note also, that with increasing the magnetic field, the axial velocity disturbances in the conducting phase are almost completely damped compared to the disturbance at the interface (Fig. 16).

The stream function of the velocity disturbance is given by $\psi_0(x, y) = \phi_0(y)\cos kx$, so that $u_0 = \partial\psi_0/\partial x$ and $v = -\partial\psi_0/\partial y$. The disturbance streamlines for $Q_1/Q_2 = 0.1$ and three values of the magnetic field $B_0 = 0.01, 0.05$, and $0.1T$ are shown in Figure 18. All the disturbances are normalized to $\eta = 1$, enabling comparison of the stream function perturbation amplitudes. Distinct vortex pairs are observed in each layer, with the maximum (and corresponding minimum) values of $\psi_0$ at the vortex centers listed in Table 4. The velocity perturbations in the air layer are clearly stronger than those in the mercury layer. In the air, the maximum disturbance stream function increases with $B_0$, particularly when the lower wall is conductive. With a conductive wall, the disturbance in the mercury layer also grows with $B_0$. This is in contrast with an insulating wall for which the disturbance in mercury weakens as the magnetic field increases. These behaviors align with the opposite effects of the magnetic field on the stability boundary for conductive versus insulating walls. As shown in Figure 14, a stronger magnetic field destabilizes the flow when the lower wall is conductive, but stabilizes it when the wall is insulating, with higher critical flow rates (corresponding to $Q_1/Q_2 = 0.1$ ) obtained at larger $B_0$.

*Table 4. Maximal values of $\psi_0$ in both phases for $Q_1/Q_2 = 0.1$ and $\eta = 1$. (mercury-air)*

| Phase | Walls | $B_0[T]$ | | | | | |
|---|---|---|---|---|---|---|---|
| | | 0.000 | 0.005 | 0.010 | 0.025 | 0.050 | 0.100 |
| lower | insulating | 1.782 | 1.109 | 0.556 | 0.255 | 0.195 | 0.142 |
| | conductive | 1.782 | 1.734 | 2.328 | 4.486 | 6.789 | 9.772 |
| upper | insulating | 3.442 | 4.473 | 6.097 | 10.428 | 15.126 | 21.042 |
| | conductive | 3.442 | 4.916 | 7.439 | 13.129 | 18.987 | 26.523 |

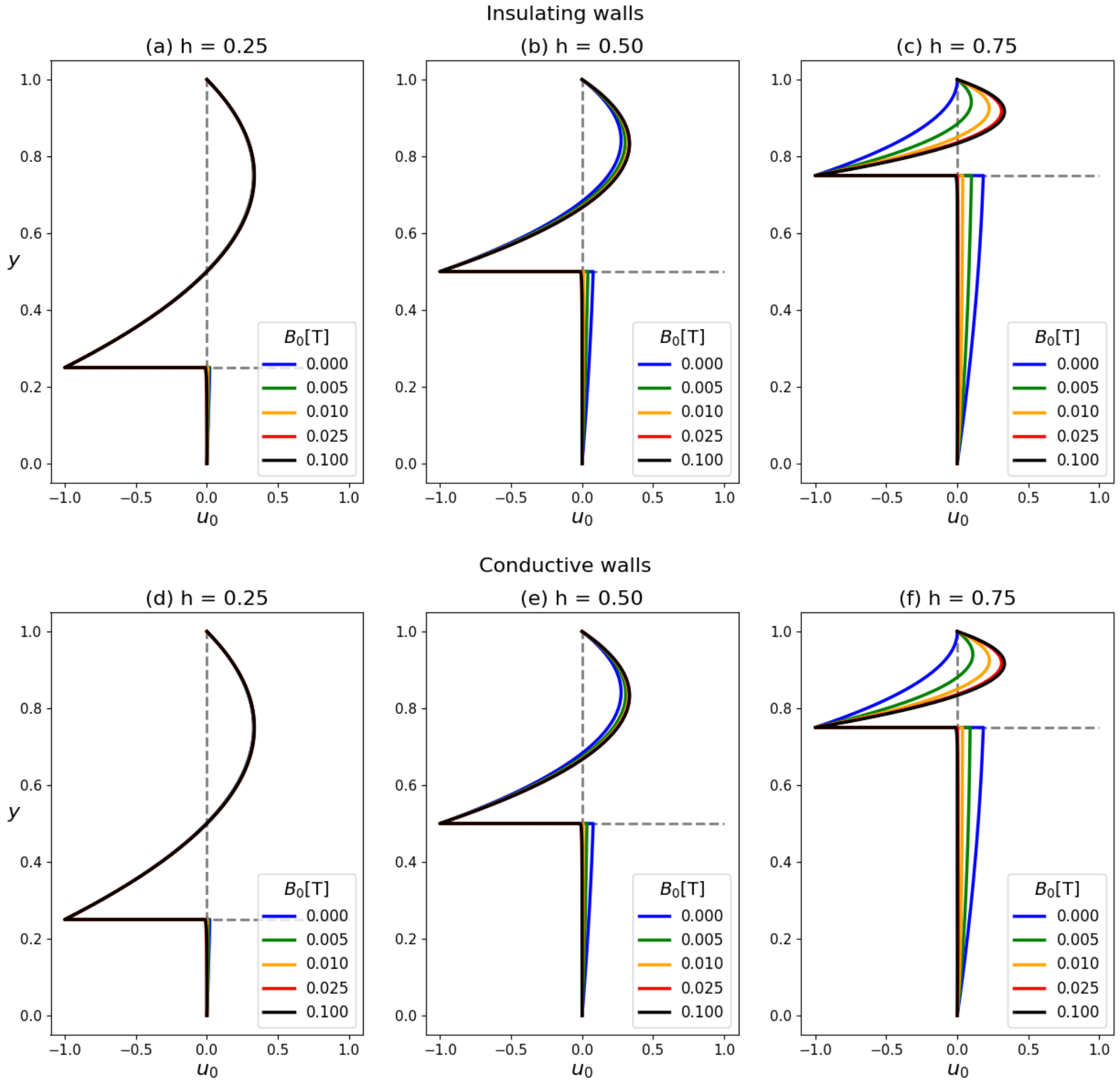

*Figure 16. $u_{j;0}$ (denoted by $u_0$ for both phases, j=1,2, here and in figures below) profiles for different values of magnetic field $B_0$ and fixed values of holdup h in a channel with insulating (a,b,c) and perfectly conducting (d,e,f) lower wall. [mercury-air]*

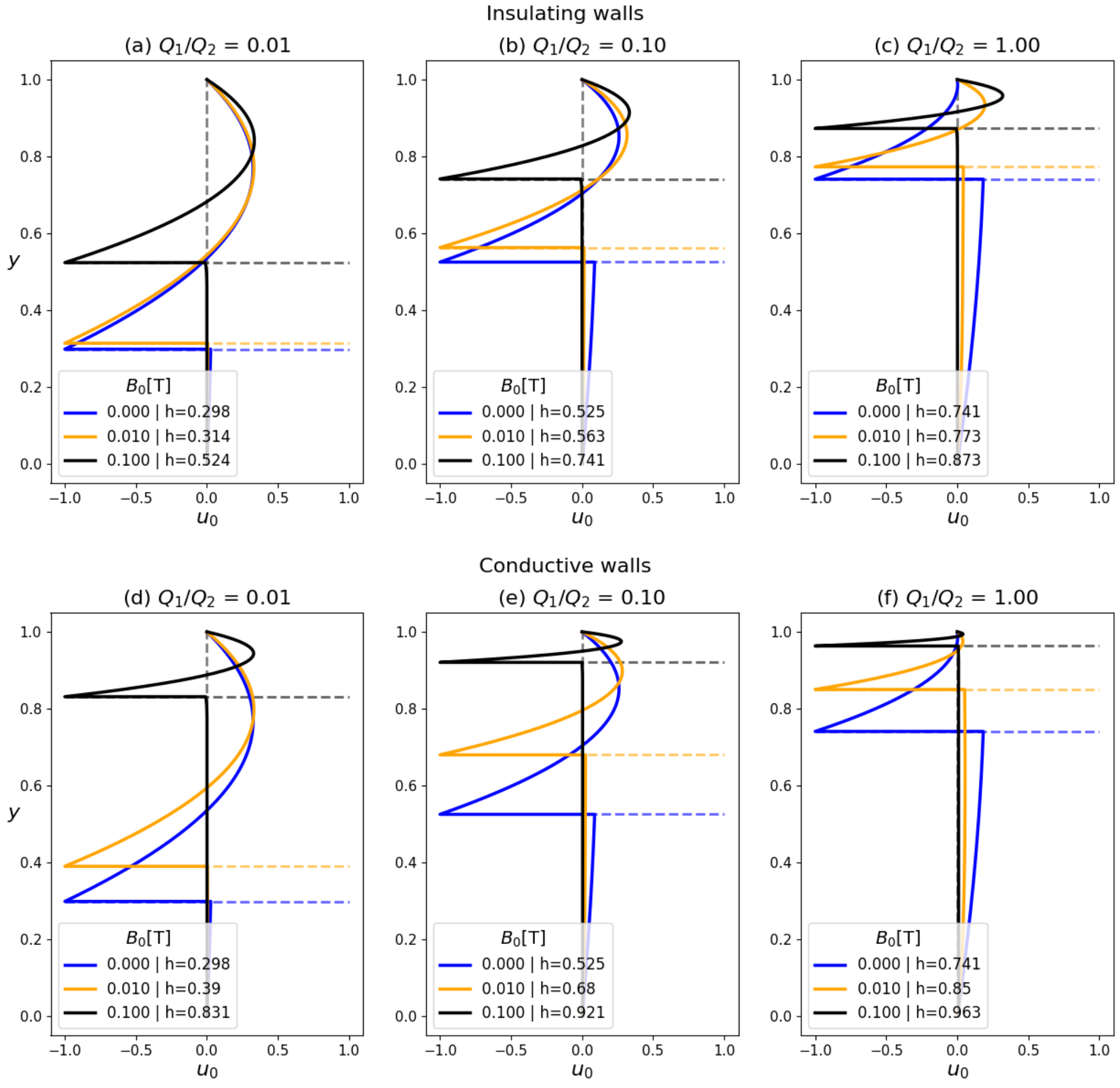

*Figure 17. $u_{j;0}$ profiles for different values of magnetic field $B_0$ and fixed values of the flux ratio $Q_1/Q_2$ in a channel with insulating (a,b,c) and perfectly conducting (d,e,f) lower wall. [mercury-air]*

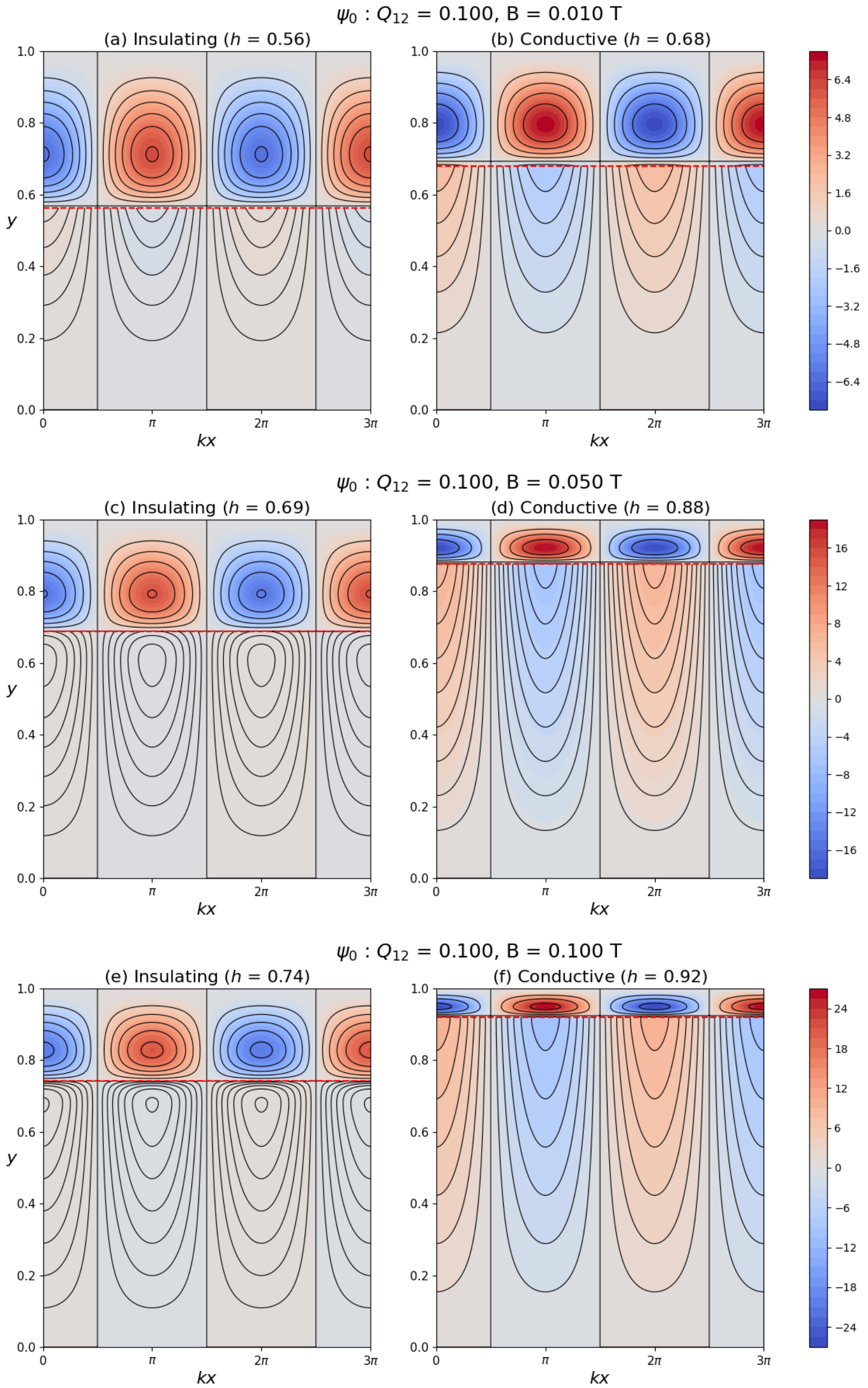

*Figure 18. Disturbances streamlines for $Q_1/Q_2 = 0.1$ and $B = 0.01$ (a,b), $B = 0.05$ (c,d), $B = 0.10$ (e,f) in channels with perfectly insulating (a,c,e) and perfectly conductive (b,d,f) walls. [mercury-air]*

Although for long wave disturbances, $u_j \sim u_{j;0}$, an addition to $u_j$ for a small but finite $k$ is defined by $u_{j;1,cr} = \left(u_{j;1,1} \cdot Re_{2,cr} + u_{j;1,-1} / Re_{2,cr}\right)$, and is presented in Figure 19 at the critical value of $Re_2$ for fixed holdups. Here also the disturbances are scaled to obtain $max|u_{j;1,cr}| = 1$. With increasing $B_0$ the amplitude of $u_{1;1,cr}$ becomes smaller than that of $u_{2;1,cr}$, however, contrarily to $u_{1;0}$, in the conducting phase, this component of the perturbation is not completely damped. This may point to the significance of disturbances in the conducting layer for the short wave instability, which is yet to be studied. Additionally, the shape of profiles of $u_{1;1}$ is strongly affected by the holdup. The profile of $u_{2;1,cr}$ exhibits several local extrema, and at large holdups their amplitude grows with the magnetic field similarly to what was observed above for $u_{j;0}$. Similar behavior of $u_{j;1,cr}$ profiles is observed when they are plotted for fixed $Q_1/Q_2$ (Figure 17). Here too, the shape of profiles is affected by the increase of the holdup with the increasing $B_0$.

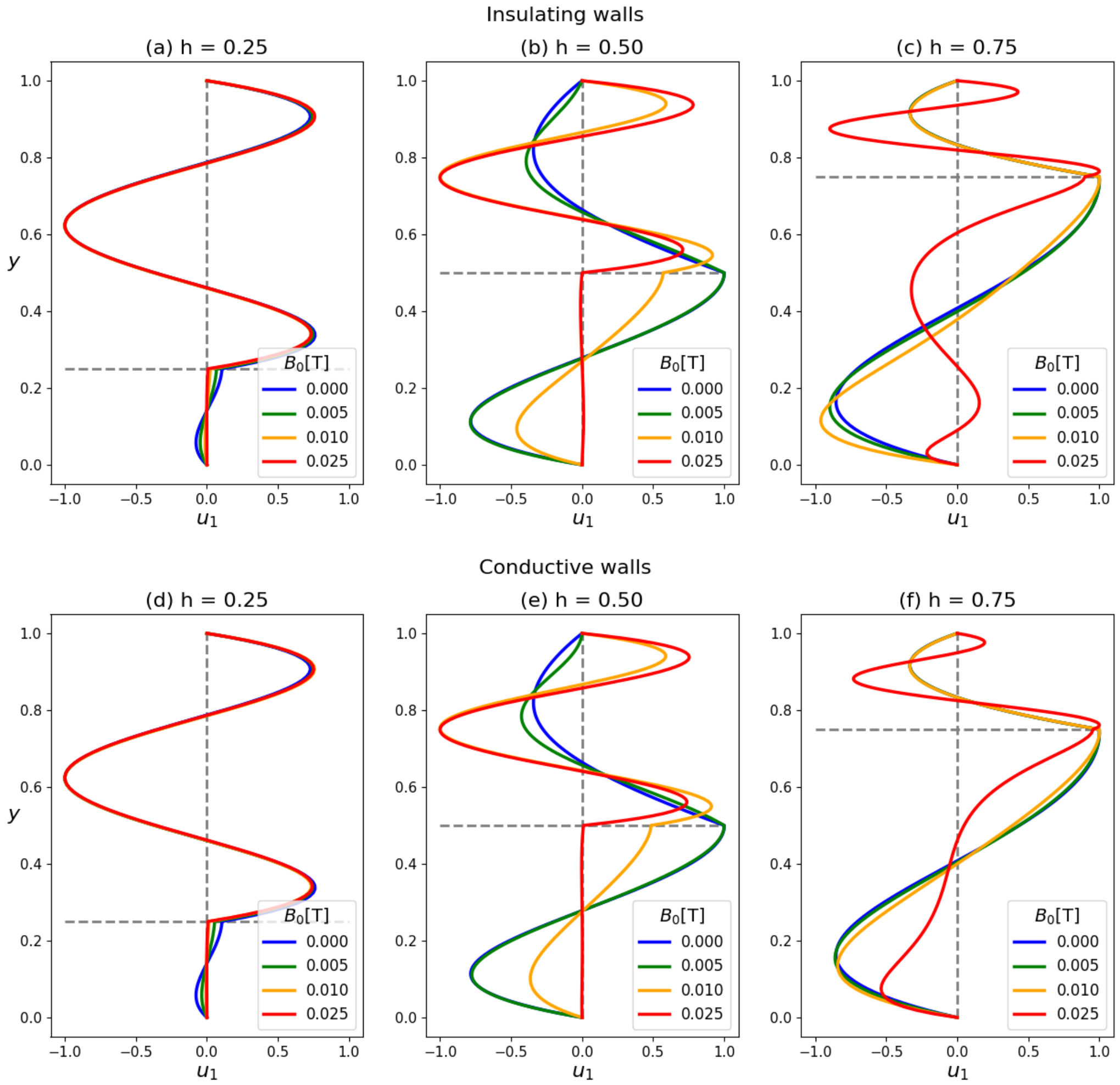

*Figure 19. $u_{j;1,cr}$ profiles for different values of magnetic field $B_0$ and fixed values of holdup $h$ in a channel with insulating (a,b,c) and perfectly conducting (d,e,f) lower wall. [mercury-air]*

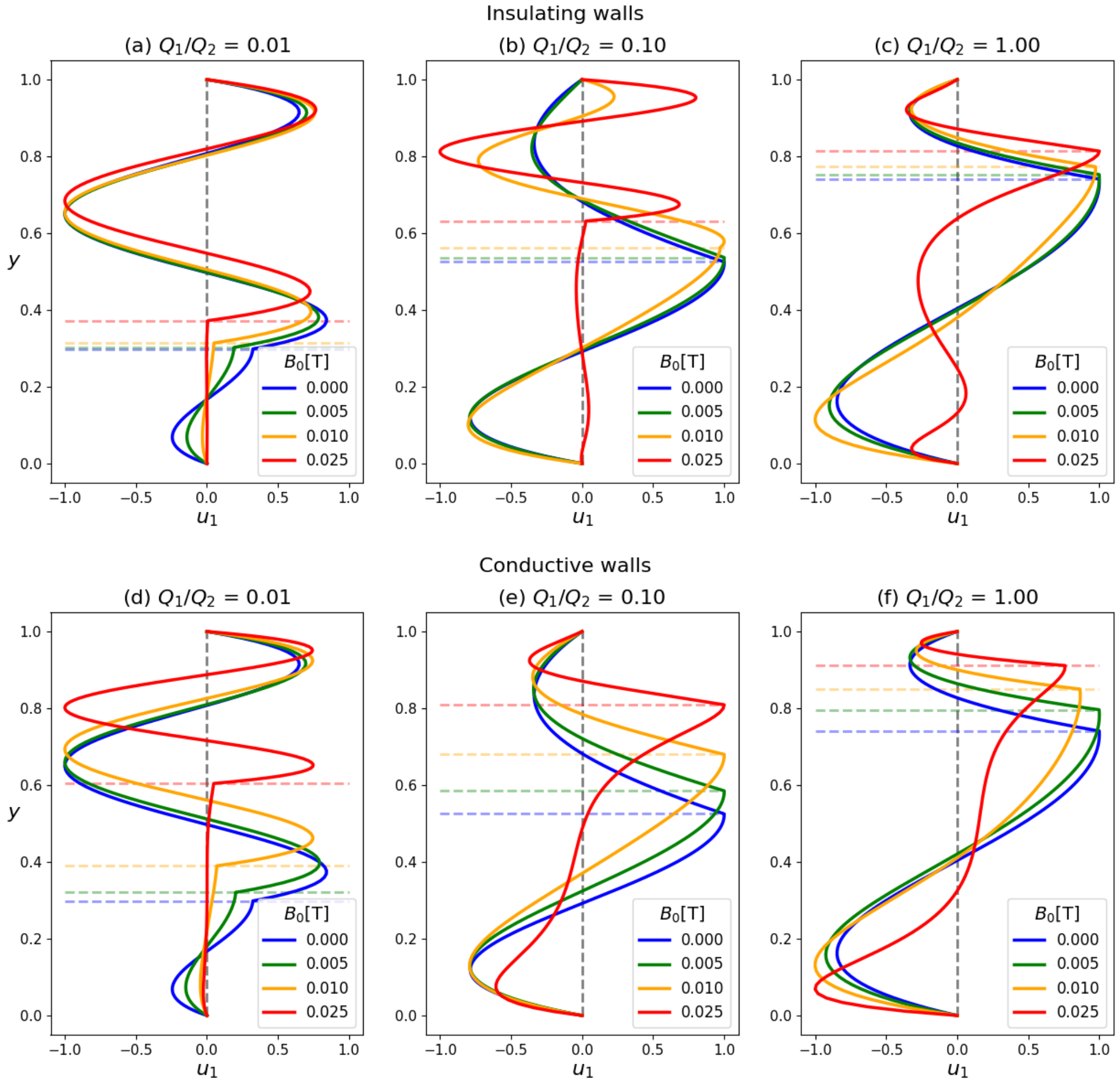

*Figure 20. $u_{j;1,cr}$ profiles for different values of magnetic field $B_0$ and fixed values of flux relation $Q_1/Q_2$ in a channel with insulating (a,b,c) and perfectly conducting (d,e,f) lower wall. [mercury-air]*

In light of the similarity of the disturbance profiles in the two cases of insulating and conducting lower wall observed in Figs. 19 and 20, we reiterate the above conclusion that, in spite of the difference in the holdups and the critical superficial velocities, the long-wave instability in both cases sets in owing to the same instability mechanism.

Figure 21 shows profiles of $|u_j|$ evaluated with the first two terms of the asymptotic expansion, and those computed numerically for the complete problem (2.11)-(2.16), for the case of perfectly conducting lower wall. Here $h = 0.5$ and disturbances are normalized by their maximal value, so that amplitudes of the plotted disturbances are $\max|u_j| = 1$. We observe again (see Table 3) that for $k \leq 10^{-4}$ the asymptotic results can be considered as quantitatively accurate, while for $k = 10^{-3}$ only qualitative similarity can be observed. The asymptotic results can be slightly improved by consideration of more terms in the asymptotic expansion, which is discussed in the next section.

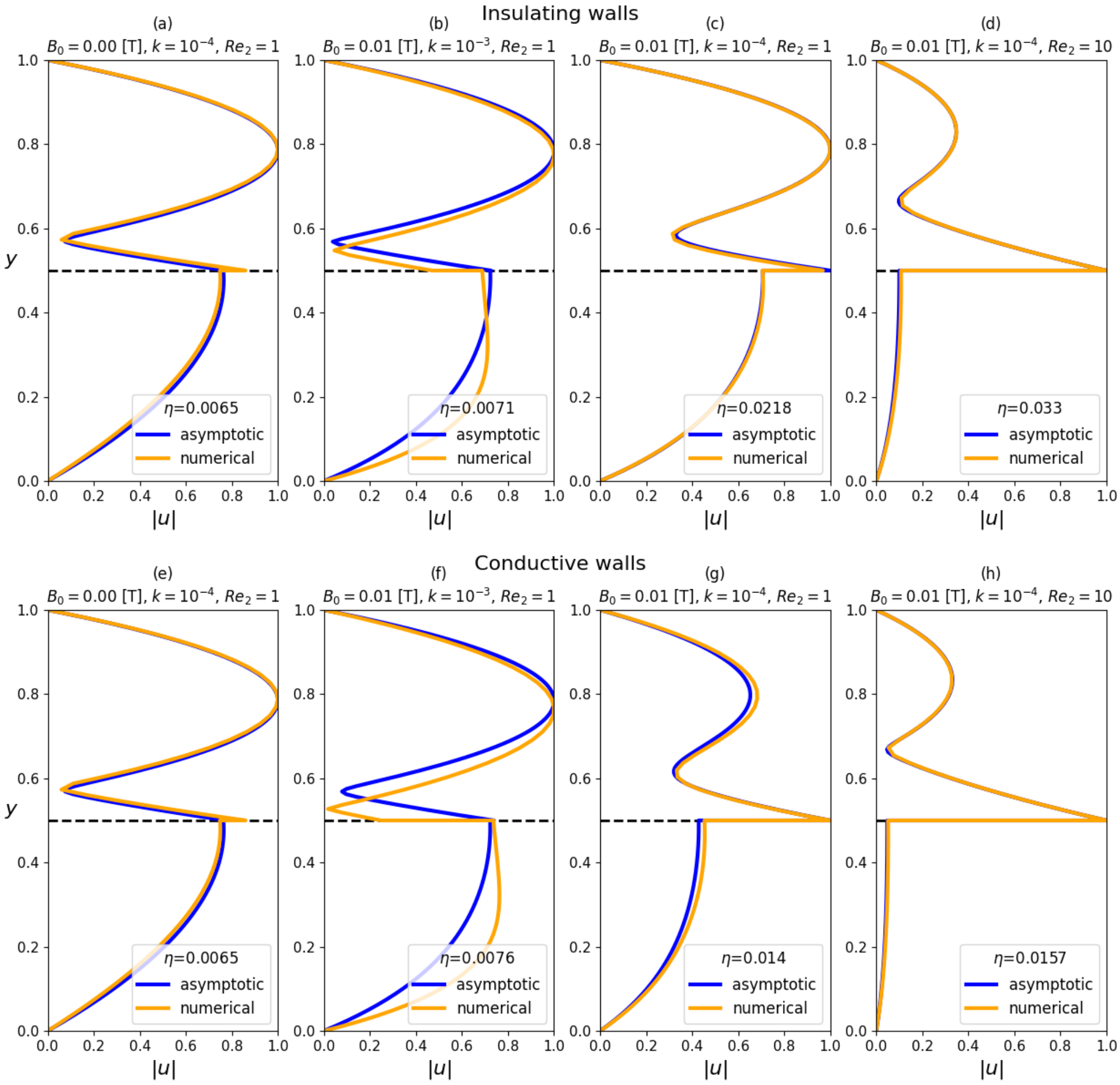

*Figure 21. Comparison of asymptotic and numerical profiles; (a-d) – insulating lower wall, (e-h) – conductive. [mercury-air]*

### 4.5. Additional terms

It is of interest to examine whether computation of additional terms in the expansion (3.29) (see Appendix B) can improve the asymptotic approximation for larger $k$. For this purpose, we represent $c$ as $c = \sum_{M=0}^{\infty} \sum_{N=-1}^{M} (ik)^M \, Re_2^N \, c_{M,N}$. As an example, the first several coefficients of the expansion for the case of $h = 0.5$, $B = 0$ are presented in Table 5.

The value of $|c_{M,0}|$ grows as $\approx 10^{4M}$, and $|c_{M,M} \, Re_{2,cr}^M|$ as $\approx 10^{5M}$, given that $Re_{2,cr} \approx \sqrt{-c_{1,-1}/\, c_{1,1}} \approx 10^2$ for small $k$, and $|c_{M,M}| \approx 10^{2M} .. 10^{3M}$. Note that the values of $r$ and $m$, which are present in the right-hand side of (3.9, 3.10), for the mercury-air flow are $r \approx 10^4$, $m \approx 10^2$.

*Table 5. Coefficients $c_{M,N}$ for $h = 0.5$, $B_0 = 0$ $(mercury - air)$.*

| $c_{M,N}$ | N = −1 | N = 0 | N = 1 | N = 2 | N = 3 | N = 4 | N = 5 | N = 6 |
|---|---|---|---|---|---|---|---|---|
| M = 0 | | 2.66e+00 | | | | | | |
| M = 1 | -1.42e+06 | | 2.08e+02 | | | | | |
| M = 2 | | -3.14e+08 | | 4.59e+04 | | | | |
| M = 3 | 1.14e+14 | | -1.03e+11 | | 1.26e+07 | | | |
| M = 4 | | 7.61e+16 | | -3.76e+13 | | 3.87e+09 | | |
| M = 5 | -1.85e+22 | | 4.18e+19 | | -1.44e+16 | | 1.27e+12 | |
| M = 6 | | -2.05e+25 | | 2.14e+22 | | -5.71e+18 | | 4.40e+14 |

Figure 22 illustrates an example of the asymptotic value predicted for the critical $Re_2$, obtained as the order of terms in the asymptotic expansion increases, for the case $h = 0.5, \mathrm{B}_0 = 0$. The asymptotic prediction is compared with the numerical results. The first-order approximation matches the numerically computed $Re_{2,cr}$ up to $k \approx 10^{-5}$. With the increase of the number of terms, we observe that the asymptotic series matches the numerical result for larger $k$, reaching $k \approx 3 \cdot 10^{-5}$, and diverges afterwards.

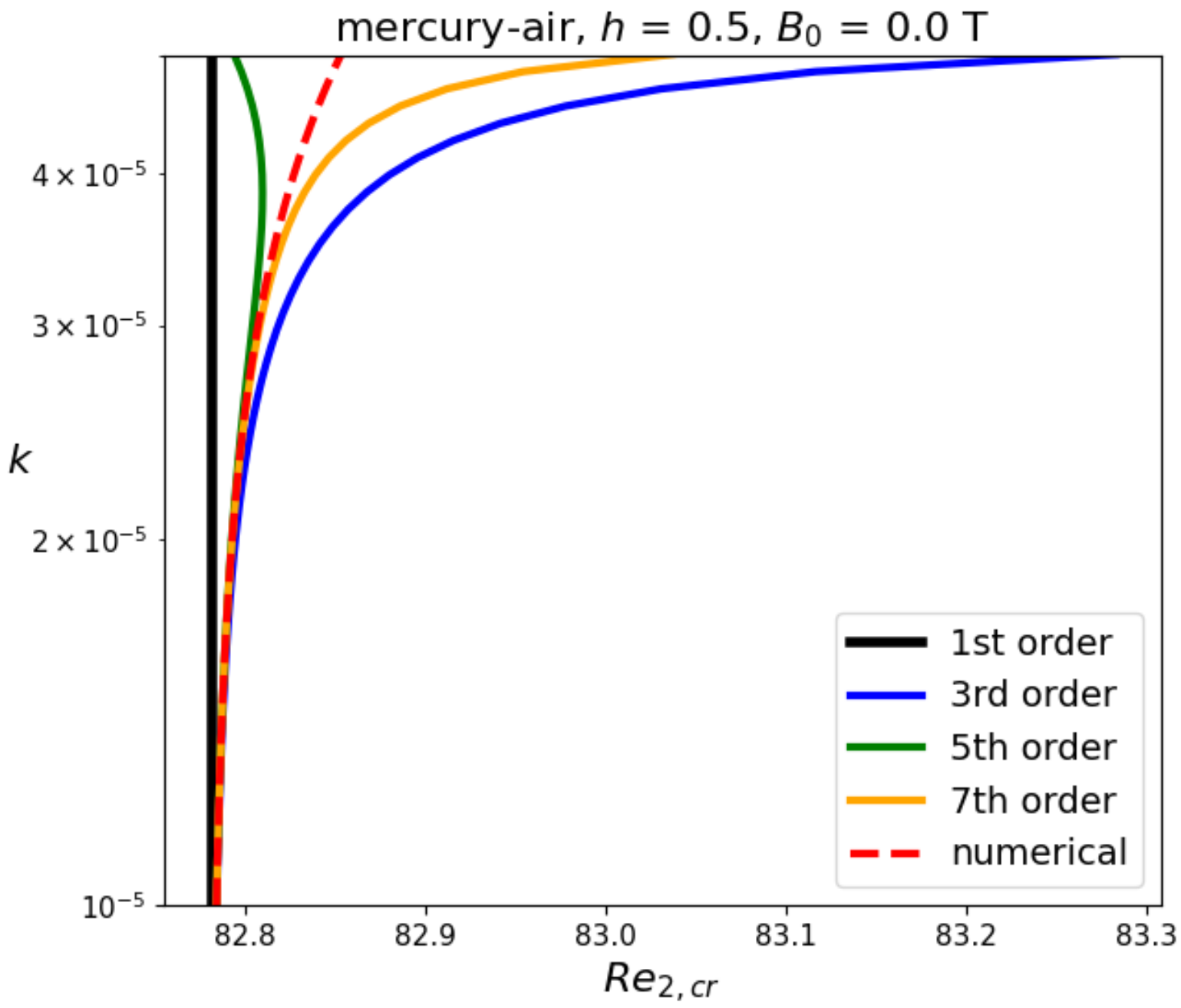

*Figure 22. Comparison of asymptotic and numerical critical $Re_2(k)$. [mercury-air]*

The effect of the increased magnetic field $\mathrm{B}_0 = 0.1$ can be observed in Figure 23; for a fixed holdup of $h = 0.5$ (although the corresponding $Q_1/Q_2$ varies). The plots (a,b) for insulating and conducting wall are similar, with the exception that the corresponding $Re_{2,cr}$ is larger for insulating lower wall and is smaller for conducting lower wall, when they are compared for this flow configuration in the absence of the magnetic field. Frames (c,d) show the values of $Im(c)/k$ obtained for a fixed $Re_2$ vs. $k$ . The selected $Re_2$ is the critical value obtained in the limit of $k \to 0$. As shown, the trends of $Im(c)$ vs $k$ are similar for conductive and insulating wall, however, the disturbance amplification rate is different. These plots illustrate that higher-order approximation can be used to ensure that the long-wave region is reached at certain small wavenumber, so that computations for even smaller $k$ will not change the result.

As shown in Figure 23(a,b), the $Re_{2,cr}$, obtained by the first-order approximation remains accurate up to $k \approx 3 \times 10^{-4}$, whereas by including higher order terms, the validity of the asymptotic expansion extends to $k > 10^{-3}$. Another example is a configuration of $h = 0.85$, $\mathrm{B}_0 = 0$ shown in Figure 24. While the stability curves have different shapes from those in Figure 22, the higher order approximation also matches the numerical result up to $k \approx 3 \times 10^{-5}$.

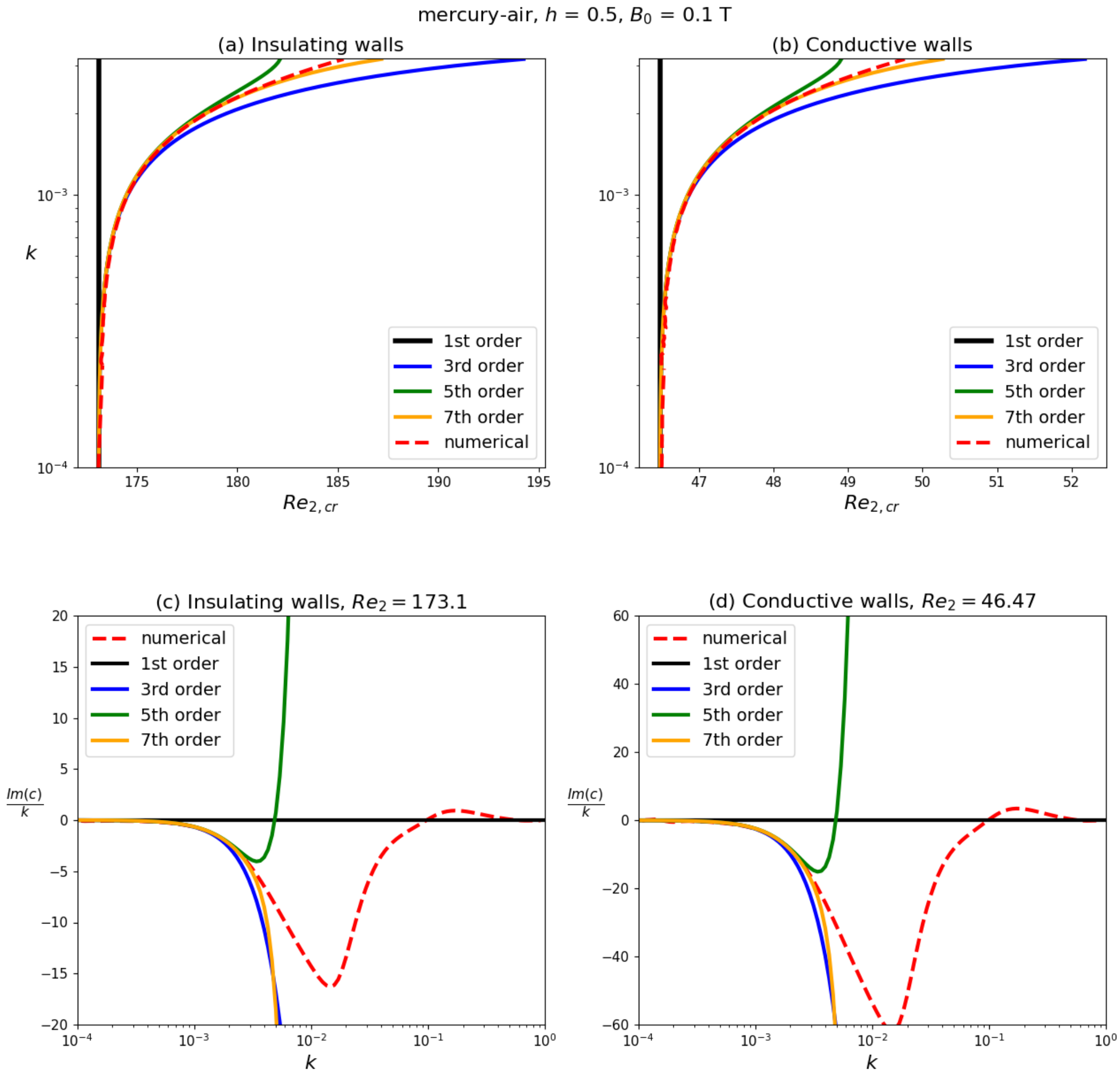

*Figure 23. Comparison of asymptotic and numerical critical $Re_2(k)$ (a,b) and $Im(c)$ (c,d); $h = 0.5,\ B_0 = 0.1T$. [mercury-air]*

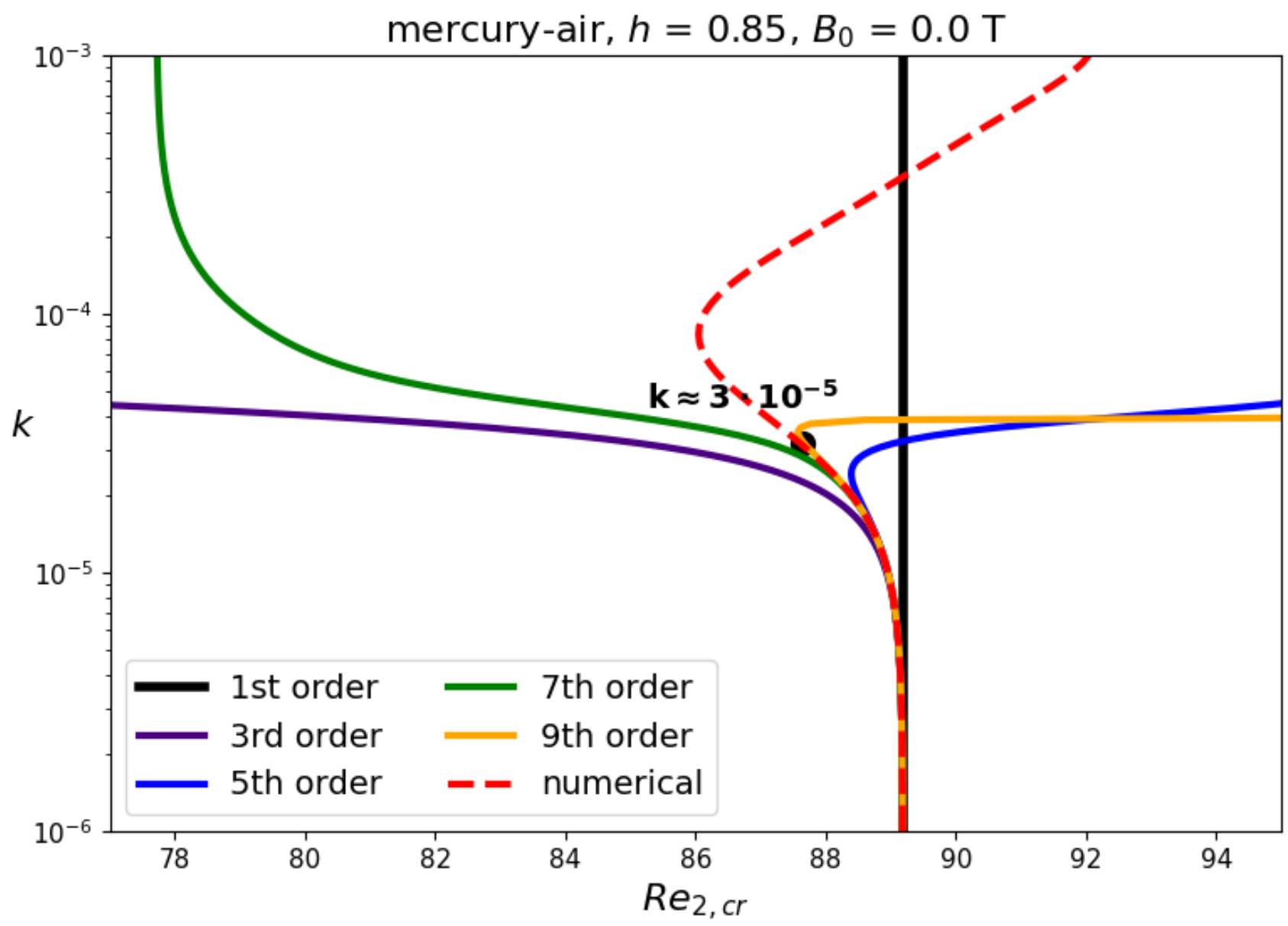

*Figure 24. Comparison of asymptotic and numerical critical $Re_2(k)$, $h = 0.85$. [mercury-air]*

In rare cases (e.g., see Fig. 24), additional terms of the asymptotic expansion indicate the existence of local minimum of $Re_{2,cr}(k)$ at small $k$, but their rapid divergence does not allow one to make a definite conclusion. On the other hand, the asymptotic expansions yield correct independent results at small but finite wavenumbers, which can be used for validation of numerical solutions.

As can be seen from Figures 22-24, the range of $k$, for which first-order approximation can be considered accurate, depends on the problem parameters such as $B, h, m, r$. For example, for the case of water-oil flow (Barmak et al., 2016) with $r = 1.25$, $m = 2.0$, accurate results can be obtained up to $k \approx 10^{-3}$ for $h = 0.2$ and up to $k \approx 10^{-4}$ for $h = 0.5$. In most cases considered here, additional terms in the asymptotic expansion allowed us to increase $k$ in ≈5 times compared with its value permitting for an accurate result with the first order terms only.

## 5. Conclusion

The long-wave instability of two-phase stratified MHD flow was analytically studied for the first time using an asymptotic expansion in a power series of the small wavenumber. Proof that the Squire's transformation holds also for two-phase MHD flows, presented in this study, allows us to restrict the analysis to two-dimensional disturbances. Extending the technique of asymptotic expansion in the series of a small wavenumber, we examined whether evaluation of the higher-order terms can extend validity of the asymptotic solution. As a result, with the increase of the number of expansion terms, the asymptotic approximation agreed with the findings of the full linear stability analysis at larger wavenumbers, which we could increase within an order of magnitude.

We were able to extend the validity of the analytical solutions to larger wavenumbers—sometimes up to the order of $10^{-3}$, whereas previous long-wave stability results for non-conducting fluids were typically limited to wavenumbers $k \leq 10^{-5}$. These analytical results can be used, in particular, to validate numerical solutions, and represent the upper bound of stability diagram, which has yet to be fully determined within the linear stability study that accounts for all possible infinitesimal disturbances.

To narrow the set of parameters to be varied, we chose a particular representative case of mercury and air flowing in a channel of 0.02m height and considered two limiting cases of perfectly insulating and perfectly conducting lower boundary. We argue that in such two-phase flow, conductivity of the lower boundary qualitatively affects the basic flow configuration, which necessarily affects its stability properties. The stability results for the two cases are indeed different, which is seen in the stability diagrams, and the variation of the wave speed with the magnetic field strength. At the same time, the velocity disturbance profiles remain similar for the conducting or insulating lower boundary. We concluded that in both cases, the long wave instability sets in owing to the same physical reasons, while the critical parameters of this instability are different because they are defined by different base flows. The streamwise velocity perturbations reach their highest amplitudes at the interface and exhibit another peak within the air layer. Therefore, the onset of instability is driven by an interaction between interfacial and shear destabilizing effects.

Examining the wave speed $c_0$, scaled by the interfacial velocity, reveals a non-monotonic dependence on both the flux ratio $Q_1/Q_2$ and the magnetic field strength, $B_0$, with minima below unity and maxima above unity. For an insulating wall, the wave speed is of the order of the interfacial velocity over wide range of $Q_1/Q_2$ and $B_0$. In contrast, with a conducting lower wall, the maxima and minima are much more pronounced exhibiting over an order-of-magnitude variation with the magnetic field strength.

The perturbation growth rates also display a non-monotonic trend, reaching their maximum at moderate flux ratios. A notable difference is observed between wall types: for insulating walls, the growth rate decreases with increasing magnetic field strength, as expected, whereas for a conductive lower boundary, the growth rate unexpectedly increases. Thus, while a magnetic field can stabilize the stratified flow, which is associated with relatively low pressure losses, when the channel walls are insulating, conductive walls instead promote destabilization and can trigger a transition in the flow pattern.

The most unexpected result is the non-monotonic dependence of the critical flow rates on the increasing the magnetic field, so that even within the simplified long wave model the magnetic field effect can be stabilizing or destabilizing. It is argued that at large magnetic fields, when the mercury flow is strongly damped, perturbations of the interface still can cause the flow destabilization.

## Acknowledgment

This research was supported by Israel Science Foundation (ISF) grant No 1363/23 .

## Appendix A: Linear stability problem

Consider a disturbance of the base flow, whose evaluation is described in Supplementary material 1. Let us denote

$$\boldsymbol{V}_j = U_j(y)\boldsymbol{e}_x + \boldsymbol{v}_j, \qquad \boldsymbol{B}_j = \boldsymbol{e}_y + \frac{Re_{mj}}{Ha_j}\left(\mathfrak{B}_j(y)\boldsymbol{e}_x + \boldsymbol{b}_j\right), \qquad P_j = P_j^{bf} + p_j, \tag{A.0}$$

where $U_j$, $\mathfrak{B}_j$ are the solutions of (S2.1.2-S2.1.11), $P_j^{bf}$ is taken from (S2.1.12), and $\boldsymbol{v}_j, \boldsymbol{b}_j, p_j$ are, respectively, their disturbances. Let also the disturbance of the interface be given by a surface $y = \eta(x, z, t)$.

Assume that the disturbances are small enough, i.e. $\boldsymbol{v}_j, \boldsymbol{b}_j, p_j, \eta = O(\varepsilon)$, and the terms of order $O(\varepsilon^2)$ can be neglected.

### *A.1. Linearized equations*

We substitute eq. (A.0) into eqs. (S1.4.7-S1.4.8) and neglect the higher order terms. This results in equations:

$$div\, \boldsymbol{v}_j = 0, \tag{A.1.1}$$

$$\left[\frac{\partial}{\partial t} + U_j\frac{\partial}{\partial x}\right]\boldsymbol{v}_j + v_{j,y}U_j'\boldsymbol{e}_x + \frac{\rho_2}{\rho_j}\nabla p_j = \frac{1}{Re_j}\Delta\boldsymbol{v}_j + \\ + \frac{Ha_j}{Re_j} rot\, \boldsymbol{b}_j \times \boldsymbol{e}_y + \frac{Re_{mj}}{Re_j}\left[rot\, \boldsymbol{b}_j \times \mathfrak{B}\boldsymbol{e}_x - \mathfrak{B}_j'\boldsymbol{e}_z \times \boldsymbol{b}_j\right], \tag{A.1.2}$$

and the equations for $\boldsymbol{b}_j$ are:

$$div\, \boldsymbol{b}_j = 0, \tag{A.1.3}$$

$$\frac{Re_{mj}}{Ha_j}\left[\frac{\partial}{\partial t} + U_j\frac{\partial}{\partial x}\right]\boldsymbol{b}_j + \frac{Re_{mj}}{Ha_j}v_{j,y}\mathfrak{B}_j'\boldsymbol{e}_x = \frac{1}{Ha_j}\Delta\boldsymbol{b}_j + \frac{\partial\boldsymbol{v}_j}{\partial y} + \frac{Re_{mj}}{Ha_j}\left(\mathfrak{B}_j\frac{\partial\boldsymbol{v}_j}{\partial x} + b_{j,y}U_j'\boldsymbol{e}_x\right). \tag{A.1.4}$$

Boundary conditions at the walls (S1.4.9-S1.4.12) become:

$$\boldsymbol{v}_1 \underset{y=-n}{=} 0, \qquad \boldsymbol{v}_2 \underset{y=1}{=} 0; \tag{A.1.5}$$

$$\frac{\partial b_{1,x}}{\partial y} - \frac{1}{C_1^w}b_{1,x} \underset{y=-n}{=} \left(1 - \frac{\tau_w^2}{C_1^w M_1^w}\right)\frac{\partial b_{1,y}}{\partial x}, \qquad \frac{\partial b_{2,x}}{\partial y} + \frac{1}{C_2^w}b_{2,x} \underset{y=1}{=} \left(1 - \frac{\tau_w^2}{C_2^w M_2^w}\right)\frac{\partial b_{2,y}}{\partial x}, \tag{A.1.6}$$

$$\frac{\partial b_{1,z}}{\partial y} - \frac{1}{C_1^w}b_{1,z} \underset{y=-n}{=} \left(1 - \frac{\tau_w^2}{C_1^w M_1^w}\right)\frac{\partial b_{1,y}}{\partial z}, \qquad \frac{\partial b_{2,z}}{\partial y} + \frac{1}{C_2^w}b_{2,z} \underset{y=1}{=} \left(1 - \frac{\tau_w^2}{C_2^w M_2^w}\right)\frac{\partial b_{2,y}}{\partial z}, \tag{A.1.7}$$

$$b_{1,y} \underset{y=-n}{=} -\frac{M_1^w}{2}\left(\frac{\partial b_{1,x}}{\partial x} + \frac{\partial b_{1,z}}{\partial z}\right), \qquad b_{2,y} \underset{y=1}{=} \frac{M_2^w}{2}\left(\frac{\partial b_{2,x}}{\partial x} + \frac{\partial b_{2,z}}{\partial z}\right). \tag{A.1.8}$$

At the interface, defined by $y = \eta(x, z, t)$, the unit normal vector is parallel to $\nabla(y - \eta)$, and is given by:

$$\boldsymbol{n} = \frac{1}{\sqrt{1 + (\eta_x')^2 + (\eta_z')^2}}(-\eta_x' \quad 1 \quad -\eta_z')^T. \tag{A.1.9}$$

Neglecting terms of the order of $O(\varepsilon^2)$ and higher,

$$\boldsymbol{n} \approx (-\eta'_x \quad 1 \quad -\eta'_z)^T, \qquad -div\,\boldsymbol{n} \approx \eta''_{xx} + \eta''_{zz}, \tag{A.1.10}$$

$$\boldsymbol{\tau}_x \approx (1 \quad \eta'_x \quad 0)^T,, \qquad \boldsymbol{\tau}_z \approx (0 \quad \eta'_z \quad 1)^T, \tag{A.1.11}$$

where $\boldsymbol{\tau}_x, \boldsymbol{\tau}_z$ are the vectors tangent to the interface.

The conditions at the interface can be evaluated for $y = 0$ using the Taylor expansion, e.g., $U_j(\eta) \approx U_j(0) + \eta U'_j(0)$. Then the continuity of velocity at the interface, defined by eq. (S1.4.13), becomes:

$$v_{1,x} + U'_1\eta \underset{y=0}{=} v_{2,x} + U'_2\eta, \qquad v_{1,y} \underset{y=0}{=} v_{2,y}, \qquad v_{1,z} \underset{y=0}{=} v_{2,z}; \tag{A.1.12}$$

the continuity of the shear stresses (S1.4.14) reduces to:

$$m\left(U''_1\eta + \frac{\partial v_{1,x}}{\partial y} + \frac{\partial v_{1,y}}{\partial x}\right) \underset{y=0}{=} \left(U''_2\eta + \frac{\partial v_{2,x}}{\partial y} + \frac{\partial v_{2,y}}{\partial x}\right), \tag{A.1.13}$$

$$m\left(\frac{\partial v_{1,z}}{\partial y} + \frac{\partial v_{1,y}}{\partial z}\right) \underset{y=0}{=} \left(\frac{\partial v_{2,z}}{\partial y} + \frac{\partial v_{2,y}}{\partial z}\right); \tag{A.1.14}$$

and the jump of the normal stress (S1.4.15) is:

$$(p_1 - p_2) - \frac{2}{Re_2}\left(m\frac{\partial v_{1,y}}{\partial y} - \frac{\partial v_{2,y}}{\partial y}\right) \underset{y=0}{=} \left(\frac{\partial P_2^{bf}}{\partial y} - \frac{\partial P_1^{bf}}{\partial y}\right)\eta - \frac{\eta''_{xx} + \eta''_{zz}}{We}; \tag{A.1.15}$$

for the normal components of magnetic field (S1.4.17) we have:

$$\frac{Re_{m1}}{Ha_1}\left(b_{1,y} - \eta'_x\mathfrak{B}_1\right) \underset{y=0}{=} \frac{Re_{m2}}{Ha_2}\left(b_{2,y} - \eta'_x\mathfrak{B}_2\right), \tag{A.1.16}$$

and for the tangential components (S1.4.16),

$$mHa_1(b_{1,x} + \mathfrak{B}'_1\eta) + \ m\frac{Ha_1^2}{Re_{m1}}\eta'_x \underset{y=0}{=} Ha_2(b_{2,x} + \mathfrak{B}'_2\eta) + \frac{Ha_2^2}{Re_{m2}}\eta'_x, \tag{A.1.17}$$

$$mHa_1 b_{1,z} + \ m\frac{Ha_1^2}{Re_{m1}}\eta'_z \underset{y=0}{=} Ha_2 b_{2,z} + \frac{Ha_2^2}{Re_{m2}}\eta'_z. \tag{A.1.18}$$

Eq. (S1.4.18) yields (here we also use eqs. (A.1.12), (A.1.16)):

$$\frac{1}{Ha_1}\left(\frac{\partial b_{1,y}}{\partial z} - \frac{\partial b_{1,z}}{\partial y}\right) - \frac{1}{Ha_2}\left(\frac{\partial b_{2,y}}{\partial z} - \frac{\partial b_{2,z}}{\partial y}\right) \underset{y=0}{=} 0, \tag{A.1.19}$$

$$\begin{aligned} \frac{1}{Ha_1}\left(\frac{\partial b_{1,x}}{\partial y} - \frac{\partial b_{1,y}}{\partial x} + \mathfrak{B}''_1\eta\right) - \frac{1}{Ha_2}\left(\frac{\partial b_{2,x}}{\partial y} - \frac{\partial b_{2,y}}{\partial x} + \mathfrak{B}''_2\eta\right) \underset{y=0}{=} \\ \underset{y=0}{=} \left[\frac{Re_{m1}}{Ha_1}\mathfrak{B}_1 - \frac{Re_{m2}}{Ha_2}\mathfrak{B}_2\right](U_1\eta'_x - v_{y,1}). \end{aligned} \tag{A.1.20}$$

Finally, the kinematic condition is obtained from $\frac{d}{dt}(y - \eta) = 0$, which yields

$$\eta'_t + U_j\eta'_x \underset{y=0}{=} v_{j,y}. \tag{A.1.21}$$

## *A.2. Squire's transformation*

Let us assume that the most unstable disturbances in the $x$-$z$ plane are spatially periodic, as well as periodic in time:

$$\begin{pmatrix} \boldsymbol{v}_j(x,y,z,t) \\ \boldsymbol{b}_j(x,y,z,t) \\ p_j(x,y,z,t) \\ \eta(x,z,t) \end{pmatrix} = \begin{pmatrix} \boldsymbol{v}_j(y) \\ \boldsymbol{b}_j(y) \\ p_j(y) \\ \eta \end{pmatrix} \exp[ik_x(x-ct)+ik_z z], \qquad (A.2.1)$$

where $c \in \mathbb{C}$, $k_x, k_z \in \mathbb{R}$, $k_x > 0$. Also, denote $k^2 = k_x^2 + k_z^2$, $k > 0$. Then the system (A.1.1-A.1.4) becomes:

$$v_y' + ik_x v_x + ik_z v_z = 0, \qquad (A.2.2)$$

$$ik_x(U-c)v_x + U'v_y + ik_x p = \frac{1}{Re}(v_x'' - k^2 v_x) + \frac{Ha}{Re}\left(b_x' - ik_x b_y\right) + \frac{Re_m}{Re}\mathfrak{B}' b_y, \qquad (A.2.3a)$$

$$ik_x(U-c)v_y + p' = \frac{1}{Re}\left(v_y'' - k^2 v_y\right) + \frac{Re_m}{Re}\left[\mathfrak{B}\left(ik_x b_y - b_x'\right) - \mathfrak{B}' b_x\right], \qquad (A.2.3b)$$

$$ik_x(U-c)v_z + ik_z p = \frac{1}{Re}(v_z'' - k^2 v_z) + \frac{Ha}{Re}\left(b_z' - ik_z b_y\right) + \frac{Re_m}{Re}\mathfrak{B}(ik_x b_z - ik_z b_x); \qquad (A.2.3c)$$

$$b_y' + ik_x b_x + ik_z b_z = 0, \qquad (A.2.4)$$

$$ik_x(U-c)b_x + \mathfrak{B}' v_y = \frac{1}{Re_m}(b_x'' - k^2 b_x + Ha\, v_x') + \left(ik_x \mathfrak{B} v_x + U' b_y\right), \qquad (A.2.5a)$$

$$ik_x(U-c)b_y = \frac{1}{Re_m}\left(b_y'' - k^2 b_y + Ha\, v_y'\right) + ik_x \mathfrak{B} v_y, \qquad (A.2.5b)$$

$$ik_x(U-c)b_z = \frac{1}{Re_m}(b_z'' - k^2 b_z + Ha\, v_z') + ik_x \mathfrak{B} v_z. \qquad (A.2.5c)$$

For brevity, we temporarily omit index $j$. Let us perform the following transformation:

$$kv_n = k_x v_x + k_z v_z, \qquad kv_t = k_x v_z - k_z v_x; \qquad (A.2.6a)$$

$$kb_n = k_x b_x + k_z b_z, \qquad kb_t = k_x b_z - k_z b_x; \qquad (A.2.6b)$$

$$\widehat{Re} = k_x/k\, Re, \qquad \widehat{Re}_m = k_x/k\, Re_m, \qquad \hat{p} = k/k_x\, p; \qquad (A.2.6c)$$

and denote $\alpha = k_z/k_x$. We add equations (A.2.3a) and (A.2.3c), multiplying them by the coefficients $(1, \alpha)$ and $(-\alpha, 1)$, respectively, and do the same for equations (A.2.5a) and (A.2.5c). Note that $kb_x = (k_x b_n - k_z b_t)$. The system rewritten for new variables is:

$$v_y' + ikv_n = 0, \qquad (A.2.7)$$

$$ik(U-c)v_n + U'v_y + ik\hat{p} = \frac{1}{\widehat{Re}}(v_n'' - k^2 v_n) + \frac{Ha}{\widehat{Re}}\left(b_n' - ikb_y\right) + \frac{\widehat{Re}_m}{\widehat{Re}}\left(\mathfrak{B}' b_y + ik\alpha \mathfrak{B} b_t\right), \qquad (A.2.8a)$$

$$ik(U-c)v_y + \hat{p}' = \frac{1}{\widehat{Re}}\left(v_y'' - k^2 v_y\right) + \frac{\widehat{Re}_m}{\widehat{Re}}\left(ik\mathfrak{B} b_y - [\mathfrak{B} \cdot (b_n - \alpha b_t)]'\right), \qquad (A.2.8b)$$

$$ik(U-c)v_t - \alpha U' v_y = \frac{1}{\widehat{Re}}(v_t'' - k^2 v_t) + \frac{Ha}{\widehat{Re}} b_t' + \frac{\widehat{Re}_m}{\widehat{Re}}\left(ik\mathfrak{B} b_t - \alpha \mathfrak{B}' b_y\right); \qquad (A.2.8c)$$

$$b_y' + ikb_n = 0, \qquad (A.2.9)$$

$$ik(U-c)b_n + \left(\mathfrak{B}'v_y - U'b_y\right) = \frac{1}{\widehat{Re}_m}\left(b_n'' - k^2 b_n + Ha\, v_n'\right) + ik\mathfrak{B}v_n, \qquad (A.2.10a)$$

$$ik(U-c)b_y = \frac{1}{\widehat{Re}_m}\left(b_y'' - k^2 b_y + Ha\, v_y'\right) + ik\mathfrak{B}v_y, \qquad (A.2.10b)$$

$$ik(U-c)b_t - \alpha\left(\mathfrak{B}'v_y - U'b_y\right) = \frac{1}{\widehat{Re}_m}\left(b_t'' - k^2 b_t + Ha\, v_t'\right) + ik\mathfrak{B}v_t. \qquad (A.2.10c)$$

Note that equation (A.2.10a) can be obtained by taking derivative of (A.2.10b) and using (A.2.7), (A.2.9).

We can see that condition (A.1.15) at the interface includes $p$, the value of which is taken from (A.2.8a), and therefore depends on $b_t$. The latter is connected to $v_t$. So the Squire's transformation can not be used.

However, if we follow (Lock, 1955; Chandrasekhar, 1953; Stuart, 1954) and neglect the terms of the magnetic Prandtl number order, $O(Pr_m) = O(Re_m/Re)$, substituting (A.2.9, A.2.10a) into (A.2.8a) yields:

$$v_{j,y}' + ikv_{j,n} = 0, \qquad (A.2.11a)$$

$$ik\left(U_j - c\right)v_{j,n} + U_j' v_{j,y} + ik\frac{\rho_2}{\rho_j}\hat{p}_j = \frac{1}{\widehat{Re}_j}\left(v_{j,n}'' - k^2 v_{j,n} - Ha_j^2 v_{j,n}\right), \qquad (A.2.11b)$$

$$ik\left(U_j - c\right)v_{j,y} + \frac{\rho_2}{\rho_j}\hat{p}_j' = \frac{1}{\widehat{Re}_j}\left(v_{j,y}'' - k^2 v_{j,y}\right). \qquad (A.2.11c)$$

Let us also denote $\hat{\eta} = k_x/k\ \eta$, $K_F = Re_2^2/Fr_2$ and $K_W = Re_2^2/We_2$; $K_F$ and $K_W$ are dimensionless parameters that do not depend on the flow velocity scale $U_*$. Then the conditions on the interface are:

$$ik\left(U_j - c\right)\hat{\eta} \underset{y=0}{=} v_{j,y}; \qquad (A.2.12a)$$

$$v_{1,n} + U_1'\hat{\eta} \underset{y=0}{=} v_{2,n} + U_2'\hat{\eta}, \qquad v_{1,y} \underset{y=0}{=} v_{2,y}; \qquad (A.2.12b)$$

$$m\left(v_{1,n}' + ikv_{1,y} + U_1''\hat{\eta}\right) \underset{y=0}{=} \left(v_{2,n}' + ikv_{2,y} + U_2''\hat{\eta}\right), \qquad (A.2.12c)$$

$$(\hat{p}_1 - \hat{p}_2) - \frac{2}{\widehat{Re}_2}\left(mv_{1,y}' - v_{2,y}'\right) = \frac{1}{\widehat{Re}_2^2}\left((r-1)K_F + k^2 K_W\right)\hat{\eta}, \qquad (A.2.12d)$$

and conditions at the walls are:

$$v_{1,n}(-n) = v_{1,y}(-n) = 0, \qquad v_{2,n}(1) = v_{2,y}(1) = 0. \qquad (A.2.13)$$

The problem (A.2.11-A.2.13) for $(v_y, v_n, \hat{p}, \hat{\eta})$ can be solved separately from other variables, and coincides with the original problem in case of two-dimensional disturbance, i.e. $k_z = 0$ and $k = k_x$, with smaller Reynold's number $\widehat{Re}_2 = k_x/k\, Re_2 \le Re_2$. In other words, if some 3-dimensional disturbance with wavenumbers $(k_x, k_z)$, and with either of the variables $(v_y, v_n, \hat{p}, \hat{\eta}) \ne 0$, is unstable for some $Re_2$, then there is a two-dimensional disturbance with wavenumber $k = \sqrt{k_x^2 + k_z^2}$ that is unstable for a smaller Reynold's number $\widehat{Re}_2$. Below, we prove that all disturbances with $(v_y, v_n, \hat{p}, \hat{\eta}) = 0$ and $\boldsymbol{b} \ne 0$ are stable, and, therefore, for the stability analysis it is sufficient to consider only two-dimensional disturbances of the velocity, pressure and interface deformation.

### *A.3. Squire's theorem*

Let us prove that for the stability analysis it is sufficient to consider only the two-dimensional sub-problem (A.2.11-A.2.13), i.e. any eigenvalue $c$ of the whole problem that corresponds to an unstable disturbance ($Im\, c > 0$) also must be an eigenvalue of the sub-problem. Note that if $c$ is not an eigenvalue of (A.2.11-A.2.13), then, by definition, there is no non-trivial solution, and $v_y, v_n, \hat{p}, \hat{\eta}$ must be zero.

Consider a disturbance with $(v_y, v_n, \hat{p}, \hat{\eta}) = 0$. From (A.2.8a), (A.2.9), (A.2.10b), and conditions (A.1.6-A.1.8), also $b_n, b_y = 0$ (as we neglected $Re_m/Re$, (A.2.8a) gives $b_n' - ikb_y = 0$; combining it with (A.2.9) yields $b_y'' - k^2 b_y = 0$). Therefore we have a following problem for $v_t$, $b_t$:

$$b_{j,t}'' - k^2 b_{j,t} + Ha\, v_{j,t}' = 0, \qquad (A.4.1)$$

$$ik(U_j - c)v_{j,t} = \frac{1}{Re_j}(v_t'' - k^2 v_t) + \frac{Ha_j}{Re_j} b_{j,t}'; \qquad (A.4.2)$$

the boundary conditions are:

$$v_{1,t} \underset{y=-n}{=} 0, \qquad v_{1,t} \underset{y=1}{=} 0; \qquad (A.4.3)$$

$$C_1^w b_{1,t}' - b_{1,t} \underset{y=-n}{=} 0, \qquad C_2^w b_{2,t}' + b_{2,t} \underset{y=1}{=} 0, \qquad (A.4.4)$$

and the conditions at the interface are:

$$v_{1,t} \underset{y=0}{=} v_{2,t}, \qquad (A.4.5)$$

$$m\, v_{1,t}' \underset{y=0}{=} v_{2,t}', \qquad (A.4.6)$$

$$m\, Ha_1 b_{1,t} \underset{y=0}{=} Ha_2 b_{2,t}, \qquad (A.4.7)$$

$$b_{1,t}'/Ha_1 \underset{y=0}{=} b_{2,t}'/Ha_2. \qquad (A.4.8)$$

The equation (A.4.1) and conditions (A.3.3-A.4.8) hold also for complex conjugates $b_{j,t}^*$ and $v_{j,t}^*$.

Consider integrals

$$\int_0^1 v_{2,t}'' v_{2,t}^*\, dy = v_{2,t}'(1) v_{2,t}^*(1) - v_{2,t}'(0) v_{2,t}^*(0) - \int_0^1 \left|v_{2,t}'\right| dy\,; \qquad (A.4.9a)$$

$$Ha_2 \int_0^1 b_{2,t}'\, v_{2,t}^* dy = Ha_2 \left(b_{2,t}(1) v_{2,t}^*(1) - b_{2,t}(0) v_{2,t}^*(0)\right) - \int_0^1 b_{2,t} \cdot \left(k^2\, b_{2,t} - b_{2,t}''\right)^* dy\,; \quad (A.4.9b)$$

$$\int_0^1 b_{2,t} \left(b_{2,t}''\right)^* dy = b_{2,t}(1) b_{2,t}'^*(1) - b_{2,t}(0) b_{2,t}'^*(0) - \int_0^1 \left|b_{2,t}'\right|^2 dy\,. \qquad (A.4.9c)$$

Similar evaluations can be carried out for the lower phase. Taking this into account, we multiply (A.4.2) by $mv_{1,t}^*$ and $v_{2,t}^*$, and integrate from $-n$ to 0 and from 0 to 1 respectively, and then take the real part:

$$Re_j\, Im(c)\, \|v_t\|^2 = -k^2 \|v_t\|^2 - \|v_t'\|^2 - k^2 \|b_t\|^2 - \|b_t'\|^2 + \left[b_{2,t}(1) b_{2,t}'^*(1) - b_{1,t}(-n) b_{1,t}'^*(-n)\right], (A.4.10)$$

where $\|f\|^2 = m \int_{-n}^0 |f_1|^2 dy + \int_0^1 |f_2|^2 dy$.

In the cases of insulating walls ($C^w = 0$) and perfectly conductive walls ($C^w = \infty$) either $b_t$ or $b_t'$ equals 0 on the boundary, hence $b_t b_t' = 0$; otherwise, from (A.4.4):

$$b_{2,t}(1) b_{2,t}'^*(1) - b_{1,t}(-n) b_{1,t}'^*(-n) = -\left(C_2^w \left|b_{2,t}'(1)\right|^2 + C_1^w \left|b_{1,t}'(1)\right|^2\right) \le 0. \qquad (A.4.11)$$

Then the right-hand side of (A.4.10) is $\le 0$ (and $< 0$ if $v_t \ne 0$), therefore $Im(c) \le 0$, so that the above disturbances cannot be unstable.

*A.4. Orr-Sommerfeld equation*

Considering a two-dimensional disturbance ($k_z = 0$, $k_x = k$), the Orr-Sommerfeld equation is re-derived taking into account the Lorenz force. We introduce the stream function $\phi_j$ whereby:

$$v_{j,x} = \phi_j', \qquad v_{j,y} = -ik\phi_j; \qquad (A.3.1)$$

Equation (A.2.11a) is automatically satisfied, and the boundary conditions become:

$$\phi_1(-n) = \phi_1'(-n) = 0, \qquad \phi_2(1) = \phi_2'(1) = 0; \qquad (A.3.2)$$

The conditions (A.2.12a-A.2.12.d) are rewritten as:

$$\phi_j \underset{y=0}{=} (c - U_j)\eta; \qquad (A.3.3)$$

$$\phi_1 \underset{y=0}{=} \phi_2, \qquad (A.3.4)$$

$$\phi_1' + U_1'\eta \underset{y=0}{=} \phi_2' + U_2'\eta, \qquad (A.3.5)$$

$$m[\phi_1'' + U_1''\eta + k^2\phi_1] \underset{y=0}{=} \phi_2'' + U_2''\eta + k^2\phi_2; \qquad (A.3.6)$$

The expression for $p_j$ is derived from eq. (A.2.11b):

$$p_j = \frac{\rho_j}{\rho_2}\left[U_j'\phi_j + (c - U_j)\phi_j'\right] + \frac{\mu_j/\mu_2}{ikRe_2}\left(\phi_j''' - k^2\phi_j' - Ha_j^2\phi_j'\right); \qquad (A.3.7)$$

Substituting this into (A.2.12d), and using (A.3.3), yields the condition on normal stress:

$$\begin{gathered} m(\phi_1''' - 3k^2\phi_1' - Ha_1^2\phi_1') - (\phi_2''' - 3k^2\phi_2' - Ha_1^2\phi_2') + \\ +ikRe_2(c - U_j)\left[r(\phi_j' + U_j'\eta) - (\phi_j' + U_j'\eta)\right] = ikRe_2\left[\frac{r-1}{Fr_2} + \frac{k^2}{We_2}\right]\eta; \end{gathered} \qquad (A.3.8)$$

Finally, the Orr-Sommerfeld equation is obtained by substituting (A.3.7) into (A.2.11c):

$$(U_j - c)(\phi_j'' - k^2\phi_j) - U_j''\phi_j = \frac{1}{ikRe_j}\left(\phi_j'''' - k^2\phi_j'' + k^4\phi_j - Ha_j^2\phi_j''\right). \qquad (A.3.9)$$

## Appendix B: Higher order terms

Additional terms of the asymptotic expansion (3.29-3.30) can be calculated, which may help to estimate $c, \phi_j$ for small, but finite, $k$. It is also possible to evaluate an expansion by powers of $Re_2$, which can be more convenient if estimations are needed to be made for a wider range of superficial velocities. The coefficients are computed once for given holdup (therefore for fixed phases' flow rate ratio), magnetic field strength and phases' properties and then can be used for different wave numbers, $k$. The phase velocity and the stream function are represented as

$$c = \sum_{M,N} c_{M,N} \, (ik)^M \, Re_2^N \,, \qquad \phi_j = \sum_{M,N} (ik)^M \, Re_2^N \, \phi_{j;M,N} \qquad (B.1)$$

where $M, N \in Z$, and we assume $c_{M,N} = 0$, $\phi_{j;M,N} = 0$ for $M < 0$.

Substituting the expansions (B.1) into the equations (3.5-3.11) and collecting terms with $(ik)^M \, Re_2^N$ yields the problem for $\phi_{j;M,N}$:

$$\phi''''_{j;M,N} - Ha_j^2 \phi''_{j;M,N} = F_{j;M,N}(y), \qquad (B.2a)$$

$$\phi_{1;M,N}(-n) = \phi'_{1;M,N}(-n) = 0, \qquad \phi_{2;M,N}(1) = \phi'_{2;M,N}(1) = 0, \qquad (B.2b)$$

$$\phi_{1;M,N} - \phi_{2;M,N} \underset{y=0}{=} 0, \qquad (B.2c)$$

$$\phi'_{1;M,N} - \phi'_{2;M,N} \underset{y=0}{=} G^v_{M,N}, \qquad (B.2d)$$

$$m\phi''_{1;M,N} - \phi''_{2;M,N} \underset{y=0}{=} G^s_{M,N}, \qquad (B.2e)$$

$$m\,(\phi_1''' - Ha_1^2\, \phi_1') - (\phi_2''' - Ha_2^2\, \phi_2') \underset{y=0}{=} G^n_{M,N}, \qquad (B.2f)$$

where the right-hand sides are defined as:

$$\begin{aligned} F_{j;M,N}(y) &\stackrel{\text{def}}{=} \vartheta_j\big( U_j \phi''_{j;M-1,N-1} + U_j \phi_{j;M-3,N-1} - U_j'' \phi_{j;M-1,N-1} \big) - \\ &-\vartheta_j \sum_{I+I'=M-1} \sum_{J+J'=N-1} c_{I,J} \left( \phi''_{j;I',J'} + \phi_{j;I'-2,J'} \right) - \left( 2\phi''_{j;M-2,N} + \phi_{j;M-4,N} \right), \end{aligned} \qquad (B.3a)$$

$$G^v_{M,N} \stackrel{\text{def}}{=} \delta_{M,0}\delta_{N,0}(m-1)U_1'; \qquad (B.3b)$$

$$G^s_{M,N} \stackrel{\text{def}}{=} \delta_{M,0}\delta_{N,0}(U_2'' - m\,U_1'') + (m-1)\big(c_{M-2,N} - \delta_{M,2}\delta_{N,0}\big), \qquad (B.3c)$$

$$\begin{aligned} G^n_{M,N} &\stackrel{\text{def}}{=} \delta_{N,-1}\big(\delta_{M,1}(r-1)K_F - \delta_{M,3}K_W\big) - 3(m-1)\big(\phi'_{1;M-2,N} + \delta_{M,2}\delta_{N,0}U_1'\big) - \\ &-(r-1) \sum_{I+I'=M-1} \sum_{J+J'=N-1} \big(c_{I,J} - \delta_{I,0}\delta_{J,0}\big)\Big(\phi'_{1;I',J'} + \delta_{I',0}\delta_{J',0}U_1'\Big). \end{aligned} \qquad (B.3d)$$

Here $\delta_{I,J}$ is the Kronecker symbol (i.e., $\delta_{I,J} = 1$ when $I = J$ and $\delta_{I,J} = 0$ otherwise). The coefficients $c_{M,N}$ are found from the relation:

$$c_{M,N} = \delta_{M,0}\delta_{N,0} + \phi_{1;M,N}(0). \qquad (B.4)$$

It can be seen that the right hand side is zero for $N < -1$ and for $N > M$, so that we have to consider only $N$ varying from $-1$ to $M$. As before, in (3.20), we represent $\phi_{j;M,N}$ using $\xi_j$ defined in eqs. (3.12-3.13):

$$\phi_j = A_{j;M,N}\xi_j + B_{j;M,N}\xi_j' + C_{j;M,N}\xi_j'' + D_{j;M,N}\xi_j''' + f_{j;M,N}(y); \qquad (B.5)$$

where

$$f_{j;M,N}(y) = A^f_{j;M,N}\xi_j + B^f_{j;M,N}\xi'_j + I_{j;M,N}(y), \qquad (B.6a)$$

$$I_{j;M,N}(y) = \int_0^y \xi_j(y - s) \cdot F_{j;M,N}(s)ds\,; \qquad (B.6b)$$

and coefficients $A^f_{j;M,N}, B^f_{j;M,N}$ are chosen so that

$$f_{j;M,N}(0) = f'_{j;M,N}(0) = 0, \qquad f_{1;M,N}(-n) = f'_{1;M,N}(-n) = 0, \qquad f_{2;M,N}(1) = f'_{2;M,N}(1) = 0. \quad (B.6c)$$

Then the coefficients $\left(A_{1;M,N}, B_{1;M,N}, C_{1;M,N}, D_{1;M,N}, A_{2;M,N}, B_{2;M,N}, C_{2;M,N}, D_{2;M,N}\right)^T \stackrel{\text{def}}{=} \boldsymbol{a}_{M,N}$ are expressed using $\boldsymbol{a}^v$, $\boldsymbol{a}^s$, $\boldsymbol{a}^n$ from (3.28) as:

$$\boldsymbol{a}_{M,N} = G^v_{M,N}\boldsymbol{a}^v + \left(G^s_{M,N} + B^f_{2;M,N} - mB^f_{1;M,N}\right)\boldsymbol{a}^s + \left(G^n_{M,N} + A^f_{2;M,N} - mA^f_{1;M,N}\right)\boldsymbol{a}^n. \qquad (B.7)$$

In this representation, $c_{M,N} = \delta_{M,0}\delta_{N,0} + D_{1;M,N}$.

Alternatively, instead of representation (A.5), we can write

$$\phi_{j;M,N} = G^v_{M,N}\varphi^v_j + \left(G^s_{M,N} + B^f_{2;M,N} - mB^f_{1;M,N}\right)\varphi^s_j + \left(G^n_{M,N} + A^f_{2;M,N} - mA^f_{1;M,N}\right)\varphi^n_j + f_{j;M,N}, \quad (B.8)$$

where

$$\varphi^\alpha_j(y) = \Xi_j(y) \cdot \boldsymbol{a}^\alpha_j, \qquad \alpha \in \{\text{"v"}, \text{"s"}, \text{"n"}\}; \qquad (B.9a)$$

$$\Xi_1 = \xi_1\boldsymbol{e}_1 + \xi'_1\boldsymbol{e}_2 + \xi''_1\boldsymbol{e}_3 + \xi'''_1\boldsymbol{e}_4, \qquad \Xi_2 = \xi_2\boldsymbol{e}_5 + \xi'_2\boldsymbol{e}_6 + \xi''_2\boldsymbol{e}_7 + \xi'''_2\boldsymbol{e}_8. \qquad (B.9b)$$

The algorithm for calculation of the terms of the expansions (B.1) can be summarized as follows:

- **Compute $\boldsymbol{\xi_j, \xi'_j, \xi''_j, \xi'''_j}$ ($\boldsymbol{j = 1, 2}$).** *Either directly from (3.19), or by solving problem (3.12-3.13).*
- **Compute base flow $\boldsymbol{U_j}$.** *It can be expressed using $\xi_j$ as shown in Supplementary material 2, or found from the problem (2.1-2.8) by any other means.*
- **Compute matrix $\boldsymbol{S}$ from (3.28) and vectors $\boldsymbol{a^v}$, $\boldsymbol{a^s}$, $\boldsymbol{a^n}$ from (3.28).** *Alternatively, it is possible to use functions from (B.9a) instead of vectors.*
- **Do for M from 0 to a desired power of $\boldsymbol{k}$, and for N from -1 to M.** *It can also be shown, that $c_{M,N}, \phi_{j,M,N}$ are zero when $M + N$ is odd, so that it is enough to compute only for $N$ having the same parity as M.*
    - **Compute $\boldsymbol{F_{j;M,N}(y), G^v_{M,N}, G^s_{M,N}, G^n_{M,N}}$.**
    - **Compute $\boldsymbol{I^v_{j;M,N}(y)}$.** *Requires integration.*
    - **Compute $\boldsymbol{f^v_{j;M,N}(y)}$.** *Coefficients $A^f_{j;M,N}, B^f_{j;M,N}$ are found from a $2 \times 2$ linear system. Since it uses the same matrix for all $M, N$, it can be inverted in advance.*
    - **Calculate $\boldsymbol{a_{M,N}}$ from (B.7), $\boldsymbol{\phi_{j;M,N}}$ from (B.5) and $\boldsymbol{c_{M,N}}$ from (B.7).** *Alternatively, if the functions from (B.9a) are used, $\phi_{j;M,N}$ is found from (B.8), and $c_{M,N}$ from (B.4).*

Implementation of the algorithm requires to store functions $\phi_{j;M,N}$ and involves addition, multiplication, integration and differentiation of these functions. Note that all the functions used in the algorithm ($U_j$, $\phi_{j;M,N}, I_{j;M,N}, f_{j;M,N}$) are contained in the subspace spanned by the with basis $\{x^p e^{qx} \mid p, q \in Z, p \geq 0\}$ (or simply $\{x^p\}$ when $Ha_j = 0$), and all the required operations do not extend outside this subspace. Therefore, the algorithm can be implemented using symbolic computations; for instance, the results in the present article are obtained by a procedure programmed in Maple. Although, symbolic computations allow to search for $c_{M,N}$ in the form of expressions containing $n, m, r, Ha_j$ e.t.c., however in practice, the complexity of such expressions grows rapidly and exhausts computational resources. Instead, $c_{M,N}$ can be calculated as floating point numbers.

Another option is to represent functions as vectors containing coefficients of their expansion in the basis $\{x^p e^{qx}\}$. Then the integration and differentiation operators can be represented via the matrix algebra, while multiplication of the functions is performed as vector convolution.

Finally, the program can operate with approximate functions using some numerical method (such as Chebyshev collocation method). In that case, $I_{j;N,M}(y)$ can be calculated either using numerical integration, or by solving the Cauchy problem (3.18) or, alternatively, $f_{j;N,M}(y)$ can be found directly from the boundary value problem.